\documentclass[12pt]{article}
\usepackage{mheck}

\def\bee{\mathbb{E}}

\date{March, 2015}

\institution{SISSA}{Scuola Internazionale Superiore di Studi Avanzati, via Bonomea 265,  34100 Trieste, ITALY}
\institution{harvard}{Jefferson Physical Laboratory, Harvard University, Cambridge, MA 02138, USA}

\title{Galois covers of $\cn=2$
BPS spectra\\
and
quantum monodromy}

\authors{Sergio Cecotti \worksat{\SISSA} \footnote{e-mail: {\tt cecotti@sissa.it}} and 
Michele Del Zotto \worksat{\harvard} \footnote{e-mail: {\tt eledelz@gmail.com}}}

\abstract{
The BPS spectrum of many 4d $\cn=2$ theories may be seen as the (categorical) Galois cover of
the BPS spectrum of a different 4d $\cn=2$ model.
The Galois group $\mathbb{G}$ acts as a physical symmetry of the covering $\cn=2$ model. The simplest instance is $SU(2)$ SQCD with $N_f=2$ quarks, whose BPS spectrum is a $\Z_2$--cover of 
the BPS spectrum of pure SYM. More generally, $\cn=2$ SYM with simply--laced gauge group $G$ admits $\Z_k$--covers for all $k\in\mathbb{N}$; e.g.\! the $\Z_2$--cover of $SO(8)$ SYM is $SO(8)$ SYM coupled to two copies of the $E_6$ Minahan--Nemeshanski SCFT.
Galois covers simplify considerably the computation of the BPS spectrum at $\mathbb{G}$--symmetric points, in both finite and infinite chambers. \\
\indent When the covering and quotient QFTs admit a geometric engineering, say for class $\cs$ models,
the categorical spectral cover may be realized as a covering map in the geometry. A particularly nice instance is when the spectral Galois cover is induced by a modular cover of principal modular curves, $X(NM)\to X(M)$, or, more generally, by  \emph{regular} Grothendieck's \emph{dessins d'enfants;} the BPS spectra of the corresponding $\cn=2$ QFTs have magic properties.\\ 
\indent The Galois covers
allow to study effectively the action of 
the quantum (half)monodromy $\mathbb{K}(q)$ of 4d $\cn=2$ QFTs. We present several examples 
and applications of the spectral covering philosophy.}

\begin{document}

\maketitle

\tableofcontents
\newpage

%\tableofcontents

%\newpage

\section{Introduction and Outlook}\label{intt}

The original elegant proof of the Kontsevich--Soibelman wall crossing formula (WCF) \cite{KS1}
for $\cn=2$ SYM with $G=SU(2)$ --- due to Gaiotto, Moore, and Neitzke (GMN) \cite{GMN:2008} ---  was based on the observation that the BPS spectrum of 
$\cn=2$ SQCD with gauge group $SU(2)$ and two   flavors is a `double cover' of the BPS spectrum of pure SYM\footnote{ We stress that this holds in \emph{all} BPS chambers of $SU(2)$ SYM.}, so  the WCF for SYM is obtained by `pushing down' the simpler  WCF for SQCD with $N_f=2$. 

There are several (related) senses in which $\cn=2$ SQCD with $N_c=N_f=2$ is a `double cover' of its $N_f=0$ cousin. Geometrically,
seeing both theories as class $\cs[A_1]$ models
\cite{GMN:2009,Gaiotto} defined by quadratic differentials $\phi_2^{(N_f)}$ on the sphere, 
we have $\phi_2^{(2)}=\xi^* \phi_2^{(0)}$ where
$\xi\colon \mathbb{P}^1\to \mathbb{P}^1$ is a certain double cover (a meromorphic function of degree 2).
Thus the GMN spectral `double cover' is induced from the Galois double cover of the field of rational functions $\C(z)$ by its algebraic extension $\C(\sqrt{z})$. From the point of view of the 2d/4d correspondence \cite{CNV}, the 2d (2,2) model associated to $N_f=2$ SQCD is a double cover, in the sense of $tt^*$ \cite{ttstar,CV92}, of the
 (2,2) theory corresponding to pure SYM.\footnote{ These 2d models are (2,2) Landau--Ginzburg theories with superpotentials $W_{N_f=2}(X)=e^{2X}+e^{-2X}$ and $W_{N_f=0}(Y)=e^{Y}+e^{-Y}$ \cite{CV11}. } All $tt^*$ quantities (metric \cite{ttstar}, CFIV index \cite{Cecotti:1992qh}, and brane amplitudes \cite{branes}) for $N_f=0$ SQCD are the $tt^*$ push--down of the corresponding quantities for $N_f=2$. 

Since the 4d BPS spectrum may  be computed geometrically \cite{GMN:2009,Gaiotto:2010be,Gaiotto:2011tf,Gaiotto:2012rg,Gaiotto:2012db}, the equality
$\phi_2^{(2)}=\xi^* \phi_2^{(0)}$ implies that,
in a sense, the BPS spectrum of $N_f=2$ SQCD `doubly covers' the one 
of pure SYM, as observed in \cite{GMN:2008}.
On the other hand, the covering property of the corresponding 2d (2,2) models
may be used to construct \emph{new} 4d $\cn=2$ QFTs
(generically without Lagrangian formulation)
as multiple covers of known theories, as was done in \cite{infty,CDZG}.

The purpose of this paper is to make more precise the
notion of a $\cn=2$ QFT having a BPS spectrum which is a Galois cover of the BPS spectrum of
another $\cn=2$ theory,
 introduce some technical tools to work effectively with such
coverings, and discuss some generalization and new applications.

The basic observation is that the map between the BPS spectra of $N_f=2$ and $N_f=0$ $SU(2)$ SQCD is an elementary
instance of a math gadget known as
the \emph{Bongartz--Gabriel push--down functor} \cite{gal1,gal2}.  This functor is one of the most efficient tools in Representation Theory, and hence a powerful technique to compute BPS spectra of 
general 4d $\cn=2$ theories.

The connection is as follows (see section 2 for details): to most 4d $\cn=2$ theories we may associate a bounded
$\C$--category $\cc$ such that
its BPS states are identified with continuous families of stable linear
functors $X\colon \cc\to \mathsf{mod}\,\C$ (the category of finite--dimensional $\C$--spaces);
we write $\mathsf{mod}\,\cc$ for the (Abelian) category of such functors.
The BPS spectrum of a $\cn=2$ model is a `cover' of the spectrum of another $\cn=2$ model when the category $\cc_1$ of the first theory is a \emph{Galois covering} of the category $\cc_2$ of the second one. 

\subsubsection*{$N_f=2\longrightarrow N_f=0$ revisited}
As an illustration of our strategy, in this introduction we rephrase the GMN computation in the categorical language, leaving the general story and more advanced applications for the main body of the paper.
To SQCD with $N_c=N_f=2$  it is associated the unique $\C$--category $\cc^{(2)}$ on
four objects $\co_a$, $a=1,2,3,4$, and morphism spaces such that
\begin{equation}
\dim \cc^{(2)}(\co_a,\co_b)=\begin{cases}1 &\text{if }a=b\\
1 &\text{if $a$ even and $b$ odd}\\
0 & \text{otherwise}.
\end{cases}
\end{equation}
The category $\cc^{(2)}$ has a  $\Z_2$ group of automorphisms acting freely on the objects $\co_a$, i.e.\! $\co_1\leftrightarrow \co_3, \co_2\leftrightarrow\co_4$; then it makes sense to consider the \emph{orbit category} $\cc^{(0)}\equiv \cc^{(2)}/\Z_2$. One says that the category $\cc^{(2)}$ is
the \emph{Galois cover} of the category $\cc^{(0)}$ with \emph{Galois group} $\Z_2$. In this example the
orbit category $\cc^{(0)}$ coincides with the linear category
for pure $SU(2)$ SYM. Hence
$N_f=2$ SQCD is the $\Z_2$ Galois cover
of pure SYM, as originally found by GMN. The canonical covering functor $F\colon\cc^{(2)}\to \cc^{(0)}$ induces two
functors between the functor categories
$\mathsf{mod}\,\cc^{(0)}$, $\mathsf{mod}\,\cc^{(2)}$:
the \emph{pull up} functor $F^\lambda$, and 
the \emph{push down} functor $F_\lambda$
\begin{equation}\label{updown1}
F^\lambda\colon \mathsf{mod}\, \cc^{(0)}\to \mathsf{mod}\,\cc^{(2)},\qquad F_\lambda\colon \mathsf{mod}\, \cc^{(2)}\to \mathsf{mod}\,\cc^{(0)}
\end{equation}
which are each other adjoint. With respect to the 
Euler form,
\begin{equation}
\langle X, Y\rangle_\cc= \sum_{k=0}^\infty (-1)^k\,\dim\mathrm{Ext}_{\mathsf{mod}\,\cc}^k(X,Y),\qquad X,Y\in\mathsf{mod}\,\cc,
\end{equation} 
we have
\begin{equation}\label{ajoint}
\langle X, F^\lambda(Y)\rangle_{\cc^{(2)}}=\langle F_\lambda(X), Y\rangle_{\cc^{(0)}},\qquad X\in\mathsf{mod}\,\cc^{(2)},\ Y\in\mathsf{mod}\,\cc^{(0)}.
\end{equation}
The $W$ boson of $N_f=2$ SQCD is the pull--back of the pure SYM $W$ boson. Then the adjoint formula \eqref{ajoint} says that the push forward $F_\lambda$ preserves the magnetic charge of the states. $F_\lambda$ restricted to the magnetically charged BPS states is then
a two--to--one
correspondence; this is the original GMN trick.

The cover $\cc^{(2)}\to\cc^{(0)}$ has a nice Lie algebraic interpretation. We known \cite{CV11,ACCERV1,ACCERV2,cattoy} that the indecomposables of $\mathsf{mod}\,\cc^{(2)}$ are in correspondence with the positive roots of the affine Lie algebra ${A}_{3}^{(1)}$ which has a $\Z_2$ outer automorphism subgroup acting freely on the simple roots. We consider the $\Z_2$--invariant Lie subalgebra $(A_3^{(1)})^{\;\Z_2}\cong {A}_1^{(1)}$. The indecomposables of 
$\mathsf{mod}\,\cc^{(0)}$ are in correspondence with the positive roots of the invariant Lie subalgebra
${A}_1^{(1)}$.

We may endow the category $\mathsf{mod}\,\cc^{(2)}$ with a stability function (i.e.\! $\cn=2$ central charge) which is the pull--back of the one
for $\mathsf{mod}\,\cc^{(0)}$. In this case
we may use the functors $F^\lambda$, $F_\lambda$ to relate the stability of functors in $\mathsf{mod}\,\cc^{(2)}$ and $\mathsf{mod}\,\cc^{(0)}$, that is, to compare the BPS spectra of the two theories in 
corresponding chambers. The fact that the respective $W$ bosons are related by pull back, implies that Yang--Mills
weak (strong) coupling pulls back to weak (strong) YM coupling.

The Galois cover of BPS spectra
leads to relations between the quantum monodromies
\cite{CNV} of the cover and quotient $\cn=2$ QFTs.
In the simple case of $N_f=2$ SQCD covering $N_f=0$, this reduces to the functional identity
of quantum operators (here $q=e^{2\pi i\tau}$ is a complex number in the unit disk)
\begin{equation}
\mathbb{M}^{(N_f=2)}(q)=\mathbb{M}^{(N_f=0)}(q^2),
\end{equation}
where equality holds under a natural identification of the Hilbert spaces on which the two operators act.

\subsubsection*{A general strategy to compute
(symmetric) BPS spectra}
Going to the general case, the previous observation leads to a new strategy to compute 
the BPS spectrum of the covering theory
from the one of the (usually simpler) quotient QFT.
This strategy applies to all
special points where the
central charge $Z_\mathrm{cover}$ 
is the pull back of a
central charge for the quotient theory, i.e.\!
$Z_\mathrm{cover}=F^\lambda Z_\mathrm{quot}$.  
These
special points are precisely the ones which
are invariant under the Galois group\footnote{ In the general case, the Galois group may be any group $\mathbb{G}$ of automorphisms
of the covering category $\cc$ which acts freely
on the objects (and has finitely many orbits).}
$\mathbb{G}$: i.e.\! they are points of \emph{enhanced symmetry} for the 4d QFT.
The existence of an additional 
symmetry makes technically natural 
\cite{'tHooft:1979bh}
to
focus on such points (and chambers) in parameter space, 
even if they
are highly non--generic. 

{Non--generic}
points may be intersection points of
walls of marginal stability. If this is the case, one must be careful with the physical interpretation of the mathematical results. In the examples, we find the following pattern which we expect to be rather general.
Consider a generic point in parameter space at distance $\epsilon$ from the given $\mathbb{G}$--invariant point; for small $\epsilon$, the BPS spectrum consists of two sectors:
\begin{itemize}
\item[1)] the sector of particles one gets from the covering approach at the invariant point, which form complete $\mathbb{G}$--representations and have bounded $\mathbb{G}$--invariant masses (up to $O(\epsilon)$ corrections) as $\epsilon\to 0$; 
\item[2)] if the $\mathbb{G}$--invariant point belongs to wall(s) of marginal stability, in addition we have BPS particles which do not form complete $\mathbb{G}$--representations, and have masses of order $O(1/\epsilon)$ unbounded as $\epsilon\to 0$. This sector of the spectrum depends on the particular perturbation away from the symmetric point (i.e.\! on which side of the wall(s) we are), and
is also hard to compute. 
\end{itemize}
The `hard' sector 2) is physically irrelevant when we are very close to the $\mathbb{G}$--invariant point, and the `pull back' spectrum 1) is asymptotically exact in the limit $\epsilon\to 0$. 

\subsubsection*{A sample of results from the
covering technique}

There are many examples of Galois coverings between pairs of 4d $\cn=2$ QFTs; all of them may be used to compute BPS spectra at points in parameter space where
the Galois group $\mathbb{G}$ is realized
as a physical symmetry.
In this paper we focus on three large classes of examples. The first two are generalizations of
$N_f=2\to N_f=0$ in different directions.

\subparagraph{A) Coverings between class $\cs[A_1]$ models.} We consider models
defined by quadratic differentials on surfaces
with several punctures (regular and irregular) \cite{Gaiotto,CV11}. We describe the BPS spectrum in detail for class $\cs[A_1]$ models associated to surfaces of genus zero and one. A large subclass of them covers $SU(2)$ SQCD or $SU(2)$ $\cn=2^*$.
Since the BPS spectra of these quotient theories have an elegant description in terms of the root system of Lie algebras, the BPS spectra
of the covering theories are also described by root systems of a class of Lie algebras known as
\emph{extended affine}. One may also characterize the allowed quantum numbers of the BPS states in terms of roots of Kac--Moody algebras, see section 6.
 
\subparagraph{B) Coverings of pure SYM by SYM coupled to `matter'.} $\cn=2$ SYM with
simply--laced gauge group $G$ admits, for all $k\in\mathbb{N}$, a $\Z_k$
Galois cover by
SYM with the same gauge group coupled to two copies of a certain superconformal `matter' system
which depends on $G$ and $k$ (this reduces to two copies of the quark for $G=SU(2)$ and $k=2$).  
A typical example is
the double cover of pure SYM with gauge group
$SO(8)$ by SYM with the same group coupled to two copies of the Minahan--Nemeshanski theory of type $E_6$ \cite{MN1,MN2}. Other examples
involve in the `matter' systems the Minahan--Nemeshanski theories of types $E_7$, $E_8$,
and other strongly--coupled SCFTs with or without a Lagrangian formulation.

Despite the apparent complication of the coupled theory, the fact that --- at $\Z_k$--symmetric points --- it has a description as a Galois cover of pure SYM, allows to reduce the computations to the simpler ones for the uncoupled  model.
This simplification is particularly dramatic in the action of the monodromy.

\subparagraph{C) Modular covers of $\cn=2$ theories, \emph{dessins d'enfants.}} When the Galois cover of the geometry
(and hence of the linear category) arises
from a \emph{modular covering} --- so that
$\mathbb{G}$ is a sub--quotient of the
modular group $PSL(2,\Z)$ --- the covering
$\cn=2$ QFTs are of extraordinary complexity:
 their gauge groups have a huge number of
simple factors. For instance, the \emph{minimal}
example of \S.\ref{minexample} has a gauge group of the form
\begin{equation}
G_\mathrm{gauge}= \prod_{i=1}^{2304} G_i\qquad\quad G_i\ \text{simple,}
\end{equation}
and thousands  of different matter sectors.
 Nevertheless the magic
of the modular Galois covers allows to describe
their BPS spectra (at $\mathbb{G}$--symmetric points) very easily. 

Even more general examples may be constructed using regular Grothendieck's \emph{dessins d'enfants,} see section 4 for details.

\subsubsection*{Organization of the paper}
 
Section 2 is a detailed catalogue of mathematical techniques we shall use in the subsequent sections to compute BPS spectra. \S.2.1 is a review of basic facts mainly intended to establish the notation. \S.2.2, \S.2.3, and \S.2.4
contain both reviews of mathematical tools not generally known in the physics community, and new materials which are specific to the algebras which appear in physics. Section 3 contains the first examples and applications. Section 4 describes the modular paradise and the relation with Grothendieck theory of the \emph{dessins d'enfants.} Section 5 discusses the relation between Galois covers in the sense of \S.2.4 and the (quantum) monodromy operator \cite{CNV}. Section 6 contains some computations of  BPS spectra for class $\cs[A_1]$ theories with genus one Gaiotto surfaces and describes their relations with the roots systems of the extended affine (GIM) Lie algebras. Several appendices contain additional examples, as well as technicalities and
 reviews of specific math tools.   

\section{BPS spectra, GIM Lie algebras,
and Galois covers}

In this section we collect the technical tools we shall
use to study various $\cn=2$ models in later sections. \S\S.\ref{revrev}, \ref{biquivers}, \ref{revLie}, \ref{emalas} and \ref{sec:coveringfunctors} are reviews of
more or less known facts.

\subsection{Review of basic facts}\label{revrev}

\subsubsection{BPS states and Representation Theory}\label{rtreview}

We quickly review the framework of \cite{ACCERV2,cattoy}. 
We consider the Coulomb branch of a 4d $\cn=2$ gauge theory. Its 
charges
 take value in a  lattice $\Gamma$. 
 Dirac quantization endows $\Gamma$  with a skew-symmetric pairing $\langle \cdot , \cdot \rangle_D\colon \Gamma\times\Gamma \to \mathbb{Z}$. 
In a chosen Coulomb vacuum, the central charge of $\cn=2$ \textsc{susy} defines a linear map $Z\colon\Gamma \to \C$. 
$Z$ encodes the stability condition of 
 BPS states. 
Varying the moduli,  the stability condition changes, and the BPS spectra jump discontinuously (wall--crossing).
 A BPS chamber is a region in parameter space with constant BPS spectrum; BPS spectra in contiguous chambers are related by the KS formula \cite{KS1,GMN:2008,wallcrossingtopstrings,guk1,guk2}.
The $\cn=2$ model has a BPS--quiver if there exist
stable BPS hypermultiplets of charge vectors $e_i\in \Gamma$ satisfying two conditions \cite{ACCERV2,CV11}: 
$Z(e_i)\in H$ for some half--space $H\subset \C$, and all BPS states have charge vectors of the form $\pm \sum_i \Z_+\,e_i$.
Then the integral matrix\footnote{ $B_{ij}$ is the \emph{exchange matrix} of the quiver. $B_{ij}$ is the net
number of arrows from $e_i$ to $e_j$. For generic $\cw$  all 1 and 2-loops can be integrated out, and we remain with
the 2-acyclic quiver specified by $B$.} $B_{ij} \equiv \langle e_i, e_j \rangle_D$ defines the model's (BPS--)quiver $Q$: 
its nodes are in 1--to--1 correspondence with the simple charges $e_i$, and we draw $B_{ij}$ oriented arrows from $e_i$ to $e_j$.
 The BPS--quiver encodes a four--supercharges quiver 
quantum mechanics (SQM), with gauge group $\prod_{i=1}^D U(x_i)$ and bifundamental arrows,  that captures the dynamics on
 the world-line of the BPS particle of charge $\boldsymbol{x} = \sum_{i=1}^D x_i e_i\in\Gamma$ \cite{Denef}; its  superpotential $\cw$ is a linear combination of oriented cycles in $Q$. 
 The BPS states with charge $\boldsymbol{x}\in\Gamma$ correspond to the 
 \textsc{susy} vacua of the quiver SQM given by the cohomology of its vacuum moduli. Standard GIT arguments allow to trade $D$--flatness for stability: then a (classical) vacuum configuration of the SQM  describing particles of charge $\boldsymbol{x}\in\Gamma$ is identified with an isoclass of \emph{stable}
modules $X$ of the Jacobian algebra $\mathscr{J}\equiv \C Q / (\partial \cw)$ 
(the path algebra $\C Q$ bounded by the bilateral ideal generated by $\partial \cw$)
with $\mathbf{dim}\,X=\boldsymbol{x}$.  
Rotating $H$ to the upper--half plane, the stability condition defined by the central charge  $Z\colon\Gamma \to \C$ becomes
\be
X \text{ is stable } \, \Longleftrightarrow \, \forall\; \text{proper non--zero submodule $Y$ of $X$:}\ \ \arg Z(Y) < \arg Z(X).
\ee
Isoclasses of stable modules appear in continuous families parametrized by K\"ahlerian moduli spaces $\cm_X$.
The corresponding BPS $\cn=2$ supermultiplets have a Clifford vacuum in a $SU(2)_{\text{spin}} \times SU(2)_R$  representation determined by the Lefshetz/Hodge decompositions
 of $H^{p,q}(\cm_X)$ \cite{DZS},
 so the maximal spin occurring is $\dim_{\C} \cm_X / 2$. Hypermultiplets correspond to rigid 
representations with $\cm_X$ a point.  A necessary condition for a $\mathscr{J}$--module $X$ to be stable in \emph{some} chamber is that $X$ is a \emph{brick}, namely $\mathrm{End}\,X=\C$ \cite{cattoy}. 

The BPS spectral problem is then reduced to the Representation Theory (RT) of the Jacobian algebra $\mathscr{J}$. 
A $\cn=2$ model has different quiver descriptions: indeed there are many possible choices of the half--plane $H$.
Different choices produce different, but physically equivalent, algebras $\mathscr{J}$ \cite{ACCERV1,ACCERV2}.  
The corresponding SQMs are 
related by 1d Seiberg dualities a.k.a.\! quiver mutations
\cite{zele}.

\subsubsection{Review of pure $SU(2)$ SYM}\label{purereview}

We shall need some results in the Representation Theory of the algebra associated to $SU(2)$ SYM, i.e.\! the Kronecker algebra
$\mathsf{Kr}$ $\equiv$ the path algebra of the
Kronecker quiver
\begin{equation}
\xymatrix{\circ\ar@<0.4ex>[rr]^A\ar@<-0.4ex>[rr]_B && \bullet}
\end{equation}
The Abelian category of
$\mathsf{Kr}$--modules has a separating decomposition
into three linear categories (the \emph{regular} category, $\mathcal{T}$, is an Abelian category in its own right)
\cite{ringel,cbrq,bluebook2} (see also \cite{cattoy}) 
\begin{equation}\label{ringel}
\mathsf{mod}\,\mathsf{Kr}=\mathcal{P}\bigvee \mathcal{T}\bigvee\mathcal{Q},
\end{equation}
where the notation means that a $\mathsf{Kr}$--module can be (uniquely) written as 
\begin{equation}
p\oplus t\oplus q\qquad
\text{with }p\in\mathcal{P},\; t\in\mathcal{T},\;
q\in\mathcal{Q}.\end{equation}
 \emph{Separating} means that in the \textsc{rhs} of \eqref{ringel} morphisms go from left to right, i.e.\! 
\begin{equation}
\mathrm{Hom}(\mathcal{Q},\ct)=\mathrm{Hom}(\cq,\cp)=\mathrm{Hom}(\ct,\cp)=0.
\end{equation}
Objects of $\cp$ (resp.\! $\cq$) are called
\emph{preprojective modules} (resp.\! \emph{preinjective modules}). Duality interchanges
$\cp$ and $\cq$.
Each object $p$, $q$, and $t$ is, in turn, the direct sum of
 indecomposable modules in $\cp$, $\cq$, and $\ct$, respectively. They are characterized by the Ringel defect \cite{ringel,cbrq}
 \begin{equation}\label{defect}
 m(X)=\dim X_\circ-\dim X_\bullet:
 \end{equation}
\begin{itemize}
\item[\textit{a)}] An indecomposable $\mathsf{Kr}$--module $X$ is preprojective iff $m(X)<0$. In this case the arrows $A$, $B$ are injective non--iso. Indecomposable preprojective modules are rigid and have dimension
$\mathbf{dim}\,P_n=(n-1,n)$, $n\in\mathbb{N}$; 
\item[\textit{b)}] An indecomposable $\mathsf{Kr}$--module $X$ is preinjective iff $m(X)>0$. In this case the arrows $A$, $B$ are surjective non--iso. Indecomposable injective modules are rigid and have dimension
$\mathbf{dim}\,Q_n=(n,n-1)$, $n\in\mathbb{N}$;
\item[\textit{c)}] An indecomposable $\mathsf{Kr}$--module $X$ is regular iff $m(X)=0$.  Regular indecomposable modules appear in families parametrized by $\mathbb{P}^1$ and have dimension
$\mathbf{dim}\,R_n(\lambda)=(n,n)$, with $n\in\mathbb{N}$ and $\lambda\in\mathbb{P}^1$.
The indecomposable $R_n(\lambda)$ has the form
$\xymatrix{\C^n\ar@<0.4ex>[r]^{J(\lambda,n)}\ar@<-0.4ex>[r]_{\mathrm{Id}}& \C^n}$ with
$J(\lambda,n)$ the Jordan block of size $n$ and eigenvalue $\lambda$.
\end{itemize}

An indecomposable $\mathsf{Kr}$--module $X$ is a brick if either
\textit{a)} $m(X)\neq0$, or \textit{b)} $m(X)=0$ and $\mathbf{dim}\,X=\delta\equiv (1,1)$ \cite{cbrq}.
We know that being a brick is a necessary condition for being stable in \emph{some} chamber. In the $\mathsf{Kr}$ case, all bricks are simultaneously stable
under a separating central charge, i.e.\! a central charge whose phase is ordered as the \textsc{rhs}
of \eqref{ringel}
\begin{equation}\label{separating}
\arg Z(\cp)<\arg Z(\ct) < \arg Z(\cq).
\end{equation}

The physical interpretation of these RT facts is \cite{cattoy}: \textit{a)} the separating central charge gives the weakly coupled chamber;
\textit{b)} the Ringel defect is the magnetic charge;
\textit{c)} the weak--coupling BPS spectrum consists of a single vector multiplet of zero magnetic charge and electric charge $2$, associated to the $\mathbb{P}^1$ family of regular bricks, together with two infinite towers of dyonic hypermultiplets, $P_n$ of magnetic charge $-1$ and electric charge $2n$,
and $Q_n$ of magnetic charge $+1$ and electric charge $2n+2$. 

In terms of $Z_\bullet\equiv Z(e_\bullet)$, $Z_\circ\equiv Z(e_\circ)$, eqn.\eqref{separating} is just
$\arg Z_\bullet < \arg Z_\circ$. We note that all
regular modules $R\in\ct$ have the same phase
$\arg Z(R)=\theta_r\equiv\arg(Z_\circ+Z_\bullet)$ while
\begin{equation}
\arg Z(P_{n+1})< \theta_r,\qquad
\arg Z(Q_{n+1})\big)> \theta_r.
\end{equation}
%\paragraph{Aside: the Auslander--Reiten translation.} For the more mathematically oriented reader we recall the notion of Auslander--Reiten translation $\tau$, and its inverse $\tau^-$ \cite{ringel,cbrq,bluebook2}, which will be used to provide alternative proofs
%of some results. If $X$ is a non--projective (resp.\! non--injective) indecomposable, $\tau X$ (resp.\! $\tau^-X$) is a non--injective (resp.\! non--projective) indecomposable such that $\tau^-\tau X\cong X$
%(resp.\! $\tau\tau^-X\cong X$). By definition,
%an indecomposable $X$ is preprojective (resp.\! preinjective) iff $X\cong \tau^{-j}P_i$ (resp.\! $\cong\tau^jQ_i$) where $j\in\mathbb{N}_0$
%and $P_i$ (resp.\! $Q_i$) is the projective cover (resp.\!
%injective envelope) of the simple module $S_i$ with support at the $i$--th node. For all Euclidean algebras (i.e.\! $\C Q$ with $Q$ an acyclic orientation of an affine Dynkin graph) the separating decomposition in the \textsc{rhs} of \eqref{ringel} holds, and $\tau, \tau^-$ are inverse automorphisms of the (Abelian) regular category $\ct$. In this case the (unique) $W$ boson coincides with to the family of bricks $Y$ which are
%fixed by the translation, $\tau Y\cong Y$.
%

\subsection{Triangular factors  of Jacobian algebras}

\subsubsection{Triangular factors and Tits forms}\label{biquivers}

Let $Q$ be a $2$--acyclic quiver with (non degenerate, reduced) superpotential $\cw$.
We consider subquivers $\widetilde{Q}\subset Q$ which are \emph{acyclic} and not full; we take $\widetilde{Q}$ to be maximal in the sense that all nodes of $Q$ are nodes of $\widetilde{Q}$ and there are no acyclic subquiver $Q^\prime$ with
%\begin{equation}
$\widetilde{Q}\subsetneq Q^\prime\subsetneq Q$.
%\end{equation}
Then, given two nodes $i$ and $j$, either $\widetilde{Q}$ contains all the arrows of $Q$ between $i$ and $j$, or does not contain any such arrow. Moreover, adding to $\widetilde{Q}$ any omitted arrow $\alpha\colon i\rightarrow j$ will close a cycle. We represent the pair $(Q,\widetilde{Q})$ by a \emph{bi--quiver} where the arrows of $\widetilde{Q}$ are solid and those of
$Q\setminus \widetilde{Q}$ are dashed. The underlying \textit{bi--graph} is obtained by ignoring the orientation of the arrows but keeping the distinction dashed/solid. 

We write $(\mathsf{dash})$ for the two--sided 
ideal in the Jacobian algebra $\mathscr{J}=\C Q/(\partial\cw)$ generated by the dashed arrows.
Let $\mathscr{T}$ be the factor\footnote{ Below we shall
refer to $\mathscr{T}$ simply as a \emph{factor} of $\mathscr{J}$.} algebra $\mathscr{J}/(\mathsf{dash})$ of $\mathscr{J}$ and $\pi\colon \mathscr{J}\to
\mathscr{T}$ the canonical quotient.
A module of $\mathscr{T}$ is naturally a module of
$\mathscr{J}$; as a representation of the quiver,
it is just a representation in which the dashed arrows happen to vanish.

We give an alternative description of $\mathscr{T}$.
Let $I$ be the (bilateral) ideal of $\C \widetilde{Q}$
\begin{equation}
 I:= \Big(\partial\cw\big|\Big),
\end{equation}
where $(\cdot)\big|$ means that the dashed arrows in $Q\setminus\widetilde{Q}$ are set to zero. Since $\widetilde{Q}$ is acyclic, while each dashed arrow $\alpha$ closes a cycle in $Q$, $I$ is generated by the relations
\begin{equation}
 \left\{\partial_\alpha\cw\big|=0,\ \Big|\ \alpha\in Q\setminus \widetilde{Q}\right\}.
\end{equation}
The relations  $\partial_\alpha\cw\big|$ are not necessarily independent; we choose a minimal set of generators of $I$, $R$, and write $b_{ij}$ for the number of elements in $R$ with source $i$ and target $j$. Obviously,
\begin{equation}\label{xxxx42}
 0\leq b_{ij} \leq a_{ij}\equiv \#\big\{\text{dashed arrows }j\rightarrow i\big\}. 
\end{equation}
One has $\mathscr{T}=\C \widetilde{Q}/I$.
We have a functor
\begin{equation}\label{functor}
 F\colon\quad \mathsf{mod}\, \C\widetilde{Q}/I\rightarrow \mathsf{mod}\, \C Q/(\partial\cw)
\end{equation}
whose image are the modules of $\mathscr{J}\equiv \C Q/(\partial\cw)$ having vanishing arrows in $Q\setminus \widetilde{Q}$. 
$F$ preserves indecomposables and isoclasses.
The algebra $\mathscr{T}\equiv \C\widetilde{Q}/I$ is \emph{triangular}.\footnote{\ An algebra is \emph{triangular} if it has the form $\C \widetilde{Q}/I$ with $\widetilde{Q}$ acyclic; $\mathscr{T}$ was \emph{constructed} to be triangular. Other authors use instead the adjective \emph{directed} to denote the same class of algebras. We prefer to avoid this term to prevent confusion with representation--directed algebras.} Let $n$ be the number of nodes of the quiver $\widetilde{Q}$. With respect to this quiver, we define the three $n\times n$ symmetric matrices $A_{ij}$, $B_{ij}$, and $C_{ij}$ as
\begin{equation}\label{whatAAA}
\begin{aligned}
A_{ij} &= C_{ij}+a_{ij}+a_{ji},\qquad\qquad
B_{ij} = C_{ij}+b_{ij}+b_{ji},\\
C_{ij} &= 2\,\delta_{ij}-\#\big\{\text{arrows }i\rightarrow j\big\}-\#\big\{\text{arrows }j\rightarrow i\big\}.
\end{aligned}\end{equation}
$C$ is the Cartan matrix of $\widetilde{Q}$ in the sense of Kac \cite{Kac1,Kac2,Kac3}; $B$ is the Brenner matrix of the triangular algebra $\C\widetilde{Q}/I$\cite{dPlects}.
The corresponding integral quadratic forms $\mathbb{Z}^n\rightarrow \Z$ will be denoted as 
\begin{equation}
 q_A(\boldsymbol{x})= \tfrac{1}{2}\,\boldsymbol{x}^tA\boldsymbol{x},\qquad q_B(\boldsymbol{x})=\tfrac{1}{2}\,\boldsymbol{x}^tB\boldsymbol{x},\qquad q_C(\boldsymbol{x})=\tfrac{1}{2}\,\boldsymbol{x}^tC\boldsymbol{x}.
\end{equation}
%\begin{itemize}
$q_A(\cdot)$ is the quadratic form of the underlying bi--graph of $(Q,\widetilde{Q})$,
$q_B(\cdot)$ is the \textit{Tits form} of the triangular algebra $\C \widetilde{Q}/I$, and
$q_C(\cdot)$ is the \textit{Tits form} of the hereditary algebra $\C \widetilde{Q}$. 
%\end{itemize}
In the examples of interest $q_A(\cdot)\equiv q_B(\cdot)$; we shall write the Tits form $q_A(\cdot)$ simply as $q(\cdot)$. Two Tits forms that differs by a linear isometry defined over $\Z$ are said to be $\Z$--equivalent.

Counting dimensions, one sees that, if $X$ is an
indecomposable module of $\mathscr{T}$,
\begin{equation}\label{zdim}
\dim \cm_X= 1-q(\mathbf{dim}\,X).
\end{equation}
In particular $q(\mathbf{dim}\,X)\leq 1$, that is, the dimension vector of an decomposable $\mathscr{T}$--module is \emph{a root} of the Tits form $q(\cdot)$. \emph{Real roots} $\boldsymbol{x}$, i.e.\! 
$\boldsymbol{x}\in\Gamma$ with $q(\boldsymbol{x})=1$
(and connected support in $\widetilde{Q}$), correspond to dimension vectors of rigid indecomposables which, if stable, yield hypermultiplets. \emph{Imaginary} roots of $q(\cdot)$,
$q(\boldsymbol{x})\leq 0$, correspond to higher spin 
BPS multiplets: the maximal spin of a BPS
particle of charge vector $\boldsymbol{x}\in\Gamma$ is
\be\label{spintits}
\text{max spin}(\boldsymbol{x}) =  \big(1 - q(\boldsymbol{x})\big)/2.
\ee
The dimension $\boldsymbol{x}$ of a stable module of a
triangular algebra is not merely a root of $q(\cdot)$ which is just the dimension of an indecomposable $X$ (i.e.\! $\mathrm{End}\,X$ is a local ring). $\boldsymbol{x}$ satisfies the stronger condition of being a \emph{Schur root} i.e.\! the dimension of a \emph{brick} $X$ (i.e.\! $\mathrm{End}\,X=\C$).

%The Tits form of the triangular subalgebras of a complete
%$\mathcal{N}=2$ theory \cite{CV11} is weakly semi--definite positive, i.e.\! $q(\boldsymbol{x})\geq 0$ $\forall\,\boldsymbol{x}\in\mathbb{N}^D$.
%The BPS spectrum of a complete $\cn=2$ model consists only of hypermultiplets and vector multiplets. Conversely, a non--complete $\cn=2$ theory necessarily has chambers with BPS particles of arbitrary high spin \cite{cattoy}. 
\def\scj{\mathscr{J}}
\def\sct{\mathscr{T}}

\subsubsection{Triangular factors and BPS sectors}

A BPS particle 
 corresponds to a (family of) stable module of the Jacobian algebra $\mathscr{J}\equiv \C Q/(\partial\cw)$ of the quiver with superpotential.
Any module of a factor 
$\scj/I$ is naturally a module of the 
original algebra $\scj$; so
a module of $\scj/I$, if stable,
corresponds to a BPS particle. To each
factor algebra $\scj/I$ is thus associated a
  \emph{sector} of the BPS spectrum\footnote{ The sectors overlap: the sectors of $\scj/I_1$ and $\scj/I_2$ both contain the sectors associated to the common factor $\scj/(I_1,I_2)$ of $\scj/I_1$ and $\scj/I_2$.}.
If the Representation Theory of the factor $\scj/I$ is elementary, the BPS particles in the corresponding sector are easily determined.  
This observation applies, in particular, to
BPS sectors described by triangular factors  $\sct$ of $\scj$, which we call
\emph{triangular sectors}.
The states of a triangular sector are controlled by the Tits form of the triangular algebra $\sct$.
Many previous computations of BPS spectra
may be rephrased in the language of triangular sectors. The simplest instance
is when there is a single triangular factor 
$\sct$ whose sector happens to be the \emph{full} BPS spectrum; a slightly more general case is when   
there is a family of triangular
factors $\sct_A$ such that the union of their sectors is the full BPS spectrum. When this happens we say that the BPS chamber \emph{is triangular}.
Obviously, all chambers are triangular when
the algebra $\mathscr{J}$ itself is triangular, as in $ADE$ Argyres--Douglas systems or
$SU(2)$ SQCD with $N_f\leq 3$. Strong coupling finite chambers for $\cn=2$ SYM and SQCD with simply--laced gauge groups are also examples of
triangular chambers \cite{ACCERV2}, as are the `canonical' finite chambers of $(G,G^\prime)$ models \cite{CNV}, and Arnold models \cite{Arnold1,Arnold2}.    
A further example of a triangular chamber is the 
finite chamber of $E_6$ Minahan--Nemeshanski \cite{MN3}.
In all these cases the BPS spectrum is controlled by the Tits forms, and is described by the root systems of their associated Lie algebras as we are going to describe. 

\subsection{Slodowy--GIM Lie algebras.
Extended affine Lie algebras }\label{GIM-EALA}

\subsubsection{Slodowy GIM Lie algebras}\label{revLie}
The usual Kac--Moody construction \cite{KacMoody2} allows to associate to each graph (without loops) a Lie algebra. The construction may be generalized to bi--graphs \cite{Slodowy1,Slodowy2}. Let $A=\{a_{ij}\}$ be $n\times n$ matrix defined in \S.\ref{biquivers}, 
namely, $a_{ij}$ is the symmetric\footnote{\ Our definitions are not the most general ones; they correspond to the symmetric (as contrasted to symmetrizable) case.} integral $n\times n$ matrix,  with diagonal components $a_{ii}=2$ for all $i=1,2,\dots, n$,
where nodes $i$, $j$ are connected by $|a_{ij}|$ undashed (resp.\! dashed) edges if $a_{ij}<0$ (resp.\! $a_{ij}>0$).
To the matrix $A$ we attach the GIM (Generalized Intersection Matrix) Lie algebra $\mathfrak{L}_A$
\cite{Slodowy1,Slodowy2} generated by elements
\begin{equation}\{e_i, f_i, h_i\}_{i=1,2,\dots, n}\end{equation}
and satisfying the generalized Serre relations
\begin{align}\label{serre1}
 &[h_i,h_j]=0,\hskip 2.6cm [h_i,e_j]=a_{ij}\,e_j\\
&[h_i,f_j]=-a_{ij}\, f_j,\hskip 1.61cm [e_i,f_i]=h_i\\
& \text{if }i\neq j\ \text{and }a_{ij}\leq 0\hskip 1cm \left\{\begin{aligned} &[e_i,f_j]=0\\ &(\mathrm{ad}\,e_i)^{1-a_{ij}}\,e_j=
                                                  (\mathrm{ad}\,f_i)^{1-a_{ij}}\,f_j=0
                                                   \end{aligned}\right.\\
& \text{if }i\neq j\ \text{and }a_{ij}> 0\hskip1cm \left\{\begin{aligned} &[e_i,e_j]=[f_i,f_j]=0\\ &(\mathrm{ad}\,e_i)^{1+a_{ij}}\,f_j=
                                                  (\mathrm{ad}\,f_i)^{1+a_{ij}}\,e_j=0.
                                                   \end{aligned}\right.
\label{serren}\end{align}
The corresponding bi--graph is called the \emph{Dynkin} bi--graph of $\mathfrak{L}_A$. The graphical conventions are the same as in the theory of the quadratic integral (Tits) forms \cite{dPlects}; the relevant Tits form being
$q(\boldsymbol{x})\equiv \tfrac{1}{2} \boldsymbol{x}^tA\boldsymbol{x}$. A given $\boldsymbol{x} \in \Z^n$ is a \emph{root} of $\mathfrak{L}_A$ iff $q(\boldsymbol{x}) \leq 1$. 
A given root $\boldsymbol{x}$ is \emph{Schur} (for a chosen orientation of the bi--graph) iff it is the dimension vector of a brick of $\C \widetilde{Q}/I$. 
Two Tits forms which differ by a linear isometry defined over $\Z$ are said to be $\Z$--equivalent: 
$\Z$--equivalent Tits forms correspond to the same Slodowy--GIM Lie algebra. In particular, an algebra $\mathfrak{L}_A$ is Kac--Moody
 if and only if the Dynkin bi--graph is $\Z$--equivalent to one with no dashed edges.
 
In this paper we are interested in the GIM Lie algebras specified by the bi--graph underlying the bi--quiver $(Q,\widetilde{Q})$ 
of a triangular factor of the Jacobian algebra of a $\cn=2$ model. Not all GIM Lie algebras may arise from a
$\cn=2$ QFT. A (conjectural) necessary condition in order for a GIM Lie algebra to arise from a QFT is provided by the $2d/4d$ correspondence \cite{CNV,CV11}.
A Slodowy GIM Lie algebra $\mathfrak{L}_A$ satisfies this condition iff 
there is an unipotent integral matrix $S$ with
\begin{equation}\label{ggggtt}
A=S+S^t,
\end{equation}
such that the following two properties hold:
\begin{enumerate}
\item[\emph{i.})] the quiver $Q_S$ with incidence matrix
$B=S-S^t$
and the bi--graph $G_A$ associated to the symmetric matrix $A$ have the same underling graph, and the set of arrows $Q_S$ which correspond
 to solid edges of $G_A$ form
 an acyclic subquiver $\widetilde{Q}_S$;
\item[\emph{ii.})] the \textit{spectral radius} of the matrix $M\equiv-(S^{-1})^tS$ is $1$, and the Jordan blocks of $M$ have at most size\footnote{ For complete theories \cite{CNV}, the blocks of size 2 can be associated only to the eigenvalue $+1$.} 2, and no size 2 block corresponds to the eigenvalue $-1$.
\end{enumerate}
This condition just says that $S$ is a solution to the Diophantine equations of the 2d $(2,2)$ classification program \cite{CV92} with $\hat c<2$.
A GIM Lie algebra $\mathfrak{L}_A$ which satisfies the condition is called \emph{2d/4d admissible.} 
The arguments of \cite{CV92} give
\medskip

\noindent\textbf{Fact 1.} \textit{If the Tits form $q$ is positive semi--definite then $\mathfrak{L}_A$ is $2d/4d$ admissible.}
\medskip

\textsc{Proof.} Choose a decomposition of the matrix $A$ satisfying \textit{i.)}.  
Let\footnote{ We recall that the \emph{radical} of a quadratic form $q\colon \Gamma\to\Z$ is the sublattice $\mathrm{rad}\,q\subset\Gamma$  of
elements $v\in\Gamma$ annihilated by the matrix $A$ of $q$, i.e.\! such that $Av=0$. } $v\in\mathrm{rad}\,q$; then
\begin{equation}
0=-(S^{-1})^t \, A \, v=(M-1)v
\end{equation}
so $\mathrm{rad}\,q$ is invariant under $M$. $M$ descends to a linear map $\overline{M}$ on
 $\mathbb{Q}^n\big/\mathrm{rad}\,q\otimes \mathbb{Q}$. The quadratic form $q$ descends to a positive--definite form
 $\overline{q}$ on $\mathbb{Q}^n\big/\mathrm{rad}\,q\otimes \mathbb{Q}$. Now \cite{CV92}
\begin{equation}
M^tAM=S^tS^{-1}(S+S^t)(S^{-1})^tS=S+S^t=A,
\end{equation}
so $\overline{M}$ is an orthogonal transformation with respect to the positive definite inner product corresponding to the form $\overline{q}$. Then $\overline{M}$ has spectral radius $1$ acting on 
$\mathbb{Q}^n\big/\mathrm{rad}\,q\otimes \mathbb{Q}$; hence $M$ has spectral radius $1$ acting on $\mathbb{Q}^n$. Moreover, $\overline{M}$ is semisimple, so $M$ may have Jordan blocks only associated to the eigenvalue $+1$ and the size of the Jordan blocks is at most $2$.

\subsubsection{Extended affine Lie algebras}\label{emalas}

Slodowy GIM Lie algebras with semi--definite Tits form $q\colon\mathbb{Z}^n \to \mathbb{Z}$ have special interest. We already saw that they are automatically
2d/4d admissible. A GIM Lie algebra $\mathfrak{L}_A$ having the Tits form semi--definite with a radical of rank $\kappa$,
is called an \emph{extended affine Lie algebra of nullity $\kappa$} \cite{gim3,gim4}.
\medskip

Consider the quadratic form $\overline{q}$ induced by $q$ on the lattice $\Z^n/\mathrm{rad}\, q$; it is positive definite. 
A well--known theorem \cite{ringel} says that a (connected) positive--definite form is $\Z$--{equivalent}
to an $ADE$ Tits form.
We say that $\mathfrak{L}_A$ \textit{is an extended affine Lie algebra of type $\mathfrak{g}_r$}, if the corresponding induced form 
$\overline{q}$ is equivalent to the Tits form of the finite--dimensional (simply--laced) Lie algebra $\mathfrak{g}_r$. 

The extended affine Lie algebras of nullity $\kappa$ and type $\mathfrak{g}_r$ are constructed
 recursively from the same type Lie algebras of nullity $\kappa-1$  by
\emph{`affinization'}, that is, as a central extension of their loop algebra. 
At nullity zero we have the finite--type Lie algebra $\mathfrak{g}_r$, 
and at nullity 1 we have its affinization $\mathfrak{g}^{(1)}_r$, which is the usual affine Kac--Moody algebra 
\cite{KacMoody2}. At nullity 2 we have the affinization 
of the Kac--Moody algebra, $\mathfrak{g}^{(1,1)}_r\equiv (\mathfrak{g}_r^{(1)})^{(1)}$, which is a \emph{toroidal Lie algebra} \cite{toroidal}. In general
\begin{equation}\label{recconst}
 \mathfrak{g}_r^{(\overbrace{\begin{scriptsize}\text{1,1,...,1}\end{scriptsize}}^{\kappa\ \text{times}})}\equiv 
\Big(\mathfrak{g}_r^{(\overbrace{\begin{scriptsize}\text{1,1,...,1}\end{scriptsize}}^{\kappa-1\ \text{times}})}\Big)^{(1)}.
\end{equation}
Equivalently, an extended affine Lie algebra of nullity $\kappa$ and type $\mathfrak{g}_r$ is a central extension of the Lie algebra of maps
$(S^1)^\kappa\rightarrow\mathfrak{g}_r$. The central extension is not unique \cite{gim4} and we have a large family of inequivalent Lie algebras.
However, the root system does not depend on the particular central extension but only on $\kappa$ and $\mathfrak{g}_r$, that is,
 only on the Tits form $q$.
If $v \in \mathbb{Z}^n$ is a \emph{root} of an extended affine Lie algebra we say that it is \emph{real} iff $q(v)=1$, 
and \emph{imaginary} iff $q(v)=0$. 
It is clear that the set of real roots is given by $\{\alpha+\delta\}$ where $\alpha$ is a root
 of the finite--type Lie algebra of type $\mathfrak{g}_r$  and $\delta$ is an element of the lattice of imaginary roots $\mathrm{rad}\,q$.
Equivalently, we may choose a generator $\delta\in \mathrm{rad}\,q$ and write the roots in the 
form $\{\hat \alpha+ \rho\}$ where $\hat \alpha=\alpha+n\delta$ is a root of the affine
Lie algebra $\mathfrak{g}_r^{(1)}$ and $\rho\in \mathrm{rad}\, q/\Z\delta$. We shall find convenient to represent
(non--canonically) the
root lattice of an extended Lie algebra as the direct sum of the root lattice of $\mathfrak{g}_r$ (resp.\! $\mathfrak{g}^{(1)}_r$), 
which we shall call the \emph{reduced charge lattice} $\Gamma_\mathrm{red.}$,
 and the lattice $\mathrm{rad}\, q$ (resp.\! $\mathrm{rad}\,q/\Z \delta$).

The extended affine Lie algebra of nullity $\kappa$ and type $\mathfrak{g}_r$
will be written simply as $\mathfrak{g}_{r}^{[\kappa]}$.

\subsubsection{Triangular factors of class $\cs[A_1]$ algebras and their Lie algebras}\label{classA1tits}

The quiver with superpotential of a theory of class $\cs[A_1]$ is determined by an ideal triangulation of the corresponding ultraviolet curve \cite{triangulation1,LF1,LF2}, 
ideal arc flips being in correspondence with quiver mutations \cite{triangulation1}. To make the correspondence complete, it is necessary to allow ideal triangulations with self--folded triangles
\cite{triangulation1} (even if all surfaces admit triangulations without self--folded triangles).

In this subsection we describe the structure of
the Tits form of the triangular factors of the Jacobian algebras for class $\cs[A_1]$ models (\emph{class $\cs[A_1]$ algebras} for short). 
In view of \S.\ref{revLie}, we
adopt the Lie theoretical language.
\smallskip

Let $Q$ be the quiver arising from an ideal triangulation of a surface with punctures and marked points on the boundaries, 
and let $\widetilde{Q}$ be a maximal acyclic subquiver of $Q$. We write $n$ for the number of nodes of $Q$. Since the algebras of a $\cs[A_1]$ model are tame \cite{GLBS}, for all the associated bi--quivers $(Q,\widetilde{Q})$ their Tits form $q\colon \Z^n\rightarrow \Z$
is \textit{weakly} non--negative.
The bi--quivers $(Q, \widetilde{Q})$ such that $q$ is non--negative \emph{definite} will be called \emph{good}.
It is easy to check that for
all surfaces (admitting an ideal triangulation) good bi--quivers always exist.

\medskip

\noindent\textbf{Fact 2.} \textit{$\Sigma$ a surface of genus $g$ with $p$ punctures and $b$ boundaries, with $(p+b)>0$, the $i$--th boundary carrying $\ell_i\geq 1$ marked points. $\Sigma$ has Euler characteristic $\chi=2(1-g)-p-b$,
 while the number of arcs in
any ideal triangulation of $\Sigma$ is $n=-3\chi+\sum_i\ell_i$. Let $Q$ be a triangulation quiver of  
$\Sigma$, and $q$ the Tits form of an associated \emph{good} bi--quiver. Then
\begin{itemize}
\item[1)] if $Q$ corresponds to an ideal triangulation of $\Sigma$ \emph{without self--folded triangles,} $q$ is the Tits form of
an \emph{extended affine Lie algebra} which is either
\begin{equation}
 A^{[-\chi+1]}_{n+\chi-1}\qquad\text{or}\qquad D^{[-\chi]}_{n+\chi};
\end{equation}
\item[2)] if $Q$ corresponds to an ideal triangulation \emph{with at least one self--folded triangle,}
$q$ is the Tits form of
an \emph{extended affine Lie algebra} which is either
\begin{equation}
 A^{[-\chi]}_{n+\chi}\qquad\text{or}\qquad D^{[-\chi]}_{n+\chi}.
 \end{equation}
\end{itemize}
} 
\medskip

The maximal nullity $\kappa$ vanishes only for $g=0$, $b=1$, $p=0$, corresponding to the $A_n$ AD models, and $g=0$, $b=1$, $p=1$ which corresponds to the $D_n$ AD models.\medskip

We sketch the proof of the above claims leaving the details to appendix \ref{ppprof}.
A quiver $Q$ is a triangulation quiver of some $\Sigma$ if and only if it can be decomposed into
the following blocks of Types I -- V
(ref.\!\cite{triangulation1} see also \cite{CV11})
\begin{equation}
\begin{gathered}
\xymatrix{\circ\ar[d]\\\circ}\\
\mathrm{I}
\end{gathered}\quad
\begin{gathered}
\xymatrix{\circ\ar[d]\\
\circ\ar[r] & \circ\ar[ul]}\\
\mathrm{II}
\end{gathered}\quad
\begin{gathered}
\xymatrix{\bullet\\
\circ\ar[u]\ar[r] & \bullet}\\
\mathrm{IIIa}
\end{gathered}\quad
\begin{gathered}
\xymatrix{\bullet\ar[d]\\
\circ & \bullet\ar[l]}\\
\mathrm{IIIb}
\end{gathered}\quad
\begin{gathered}
\xymatrix{\bullet \ar[r] & \circ\ar[d]\\
\circ\ar[u]\ar[ur] & \bullet\ar[l]}\\
\mathrm{IV}
\end{gathered}\quad
\begin{gathered}
\xymatrix{\bullet \ar[r] & \circ\ar[r]\ar[dl] & \bullet\ar[d]\ar@/_1pc/[ll]\\
\bullet\ar[u]\ar[rr] && \bullet\ar[ul]}\\
\mathrm{V}
\end{gathered}
\end{equation}
which have two kinds of nodes \emph{black} $\bullet$ and \emph{white} $\circ$.
All triangulation quivers are constructed by
taking a finite collection of blocks of various types
and gluing them together by identifying pairs of
white nodes; gluing nodes belonging to the
same block is forbidden. Some white node may remain unpaired. For our present purposes,
it is convenient to introduce an additional block, of Type $0$, which consists of a single white
node $\circ$; we reinterpret the unpaired white nodes as being paired with  Type $0$ blocks, 
so that in our conventions \emph{all} white nodes of the building
blocks of Types 0--V should be identified in pairs. 
Finally, all opposite pairs of arrows  $\leftrightarrows$ (between the same nodes),  produced in the gluing process, should be deleted. We recall the geometric meaning of the block decomposition 
\cite{triangulation1}: each block Type corresponds to an ideally triangulated, possibly punctured, oriented $n$--gon as in table \ref{puzzle}.
White nodes correspond to the sides of the
$n$--gon which are not boundary segments, thus
internal ideal arcs along which two distinct $n$--gons
are glued.

\begin{table}
\centering
\caption{\label{puzzle} The geometric meaning of block Types}\vskip 9pt
\begin{tabular}{c l}\hline\hline
block Type & \phantom{--------------------}
$n$--gon \\\hline
0 & triangle with two sides on the boundary\\
I & triangle with one side on the boundary\\
II & triangle\\
III & punctured $2$--gon with one side on the boundary\\
IV & punctured $2$--gon\\
V & doubly punctured $1$--gon\\\hline\hline
\end{tabular}
\end{table}

\begin{figure}
\begin{align*}
&0^*:\hskip 1.9cm x_1 &&p_0(x_1)=\frac{1}{2}x_1^2\\
&I^*:\quad\begin{gathered}
           \xymatrix{x_1\ar[rr] &&x_2}
          \end{gathered}&& p_{I}(x_i)=\frac{1}{2}(x_1-x_2)^2\\
&I^\flat:\quad\begin{gathered}
           \xymatrix{x_1\ar@{..>}[rr] &&x_2}
          \end{gathered}&& p_{I^\flat}(x_i)=\frac{1}{2}(x_1+x_2)^2\\
&II^*:\quad\begin{gathered}\xymatrix{& x_2\ar@{..>}[dr]\\ x_1\ar[ur]&& x_3\ar[ll]}\end{gathered}
&& p_{II}(x_i)\equiv \frac{1}{2}(x_1-x_2-x_3)^2\\
&II^\flat:\quad\begin{gathered}\xymatrix{& x_2\ar@{..>}[dr]\\ x_1\ar@{..>}[ur]&& x_3\ar@{..>}[ll]}\end{gathered}
&& p_{II^\flat}(x_i)\equiv \frac{1}{2}(x_1+x_2+x_3)^2\\
&II^\sharp:\quad\begin{gathered}\xymatrix{& x_2\ar[dr]\\ x_1\ar@{..>}[ur]&& x_3\ar@{..>}[ll]}\end{gathered}
&& p_{II^\sharp}(x_i)\equiv \frac{1}{2}(x_1+x_2-x_3)^2+2\,x_1x_3\\
&III^*a:\quad\begin{gathered}\xymatrix{& \text{\fbox{$x_2$}}\\ x_1\ar[ur]\ar[dr]&\\
& \text{\fbox{$x_3$}}}\end{gathered}
&& p_{III}(x_i)\equiv \frac{1}{4}(x_1-2x_2)^2+\frac{1}{4}(x_1-2x_3)^2\\
&IV^*:\quad\begin{gathered}\xymatrix{& \text{\fbox{$x_3$}}\ar[dr]\\ x_1\ar[ur]\ar[dr]&& x_2\ar@{..>}[ll]\\
& \text{\fbox{$x_4$}}\ar[ur]}\end{gathered}
&& \begin{aligned}&p_{IV^*}(x_i)\equiv \frac{1}{4}(x_1+x_2-2x_3)^2+\frac{1}{4}(x_1+x_2-2x_4)^2\\
    &p_{IV^\flat}(x_i)\equiv \frac{1}{4}(x_1-x_2-2x_3)^2+\frac{1}{4}(x_1-x_2-2x_4)^2\\
    &p_{IV^\sharp}(x_i)\equiv \frac{1}{4}(-x_1+x_2-2x_3)^2+\frac{1}{4}(-x_1+x_2-2x_4)^2
   \end{aligned}\\
&V^*:\quad\begin{gathered}\xymatrix{\text{\fbox{$x_1$}} \ar@{..>}[rr]\ar@{..>}[dd] && \text{\fbox{$x_2$}}\ar[dl]\\ 
& x_0 \ar[lu]\ar[dr]\\
\text{\fbox{$x_3$}}\ar[ur] && \text{\fbox{$x_4$}}\ar@{..>}[ll]\ar@{..>}[uu]}\end{gathered}
&& \begin{aligned}p_{V}(x_i)\equiv &\frac{1}{2}\big(x_0-x_1-x_2-x_3-x_4)^2+\\
   &\ \ +\frac{1}{2}(x_1-x_4)^2+\frac{1}{2}(x_2-x_3)^2,\end{aligned}
\end{align*}
\caption{\label{biblocks}The collection $\mathfrak{B}$ of bi--blocks of triangulation bi--quivers. Boxed labels correspond to the black nodes of blocks.
 The bi--block $IV^\flat$ ($IV^\sharp$) is obtained from $IV^\ast$ by inverting \emph{dashed} $\leftrightarrow$ \emph{solid} all the arrows incident node $x_2$ ($x_1$). The bi--block of type $V$ has other forms, we reduce to form $V^*$ by flipping \emph{dashed} $\leftrightarrow$ \emph{solid} all arrows incident to some black node.
 The bi--block $III^*a$ has a second form $III^*b$ with the arrows in the opposite direction.}
\end{figure}
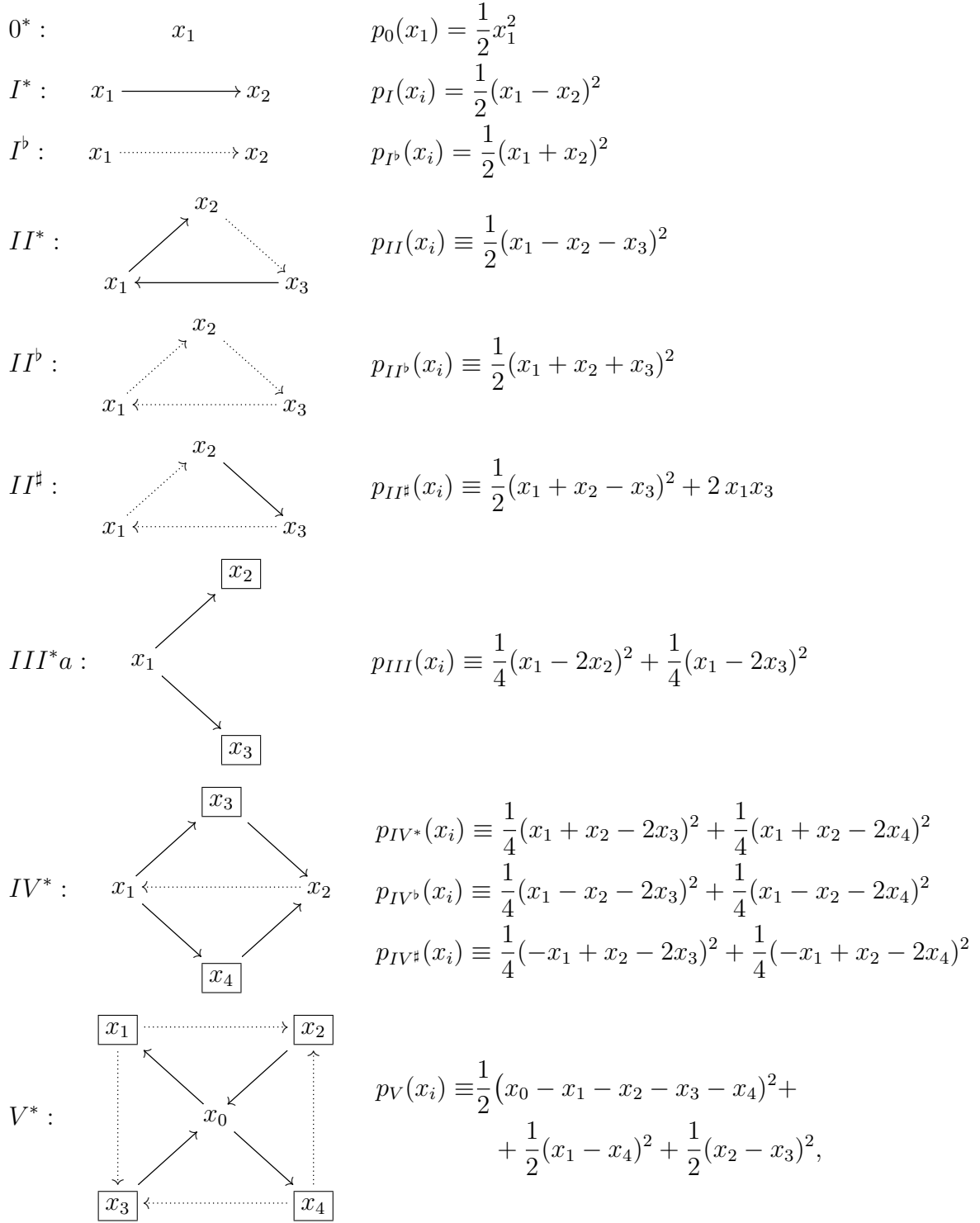

A given block--decomposition of $Q$ induces a decomposition 
of all its associated bi--quivers $(Q,\widetilde{Q})$ in a finite collection $\mathfrak{B}$ of bi--block Types, see figure \ref{biblocks}. By convention, in the bi--block decomposition the deleted pairs of opposite arrows $\leftrightarrows$ are taken to be one solid and one dashed. 

Attaching to the $i$--th node of $Q$ an integer--valued variable $x_i$, the Tits form
 $q(\boldsymbol{x})$ of $(Q,\widetilde{Q})$ is given by the sum over the polynomials $P_\mathfrak{b}(x_i)$ associated to the several bi--blocks
 \begin{equation}\label{titsbiblocks}
 q(\boldsymbol{x})=\sum_{\mathfrak{b}\in\mathfrak{B}}\sum_{\text{bi--blocks}\atop\text{of type $\mathfrak{b}$}}p_\mathfrak{b}(x_i).
 \end{equation} 
With the single exception of $p_{\mathrm{II}^\sharp}(x_i)$, all these polynomials are sums of squares. 
Since $p_{\mathrm{II}^\sharp}(x_i)$ is weakly non--negative, $q(\boldsymbol{x})$ is automatically weakly non--negative, and it is semi--definite iff no Type $\mathrm{II}^\sharp$ bi--block is present. By flipping
\emph{dashed} $\leftrightarrow$ \emph{solid}
all arrows incident to a node of each offending
$\mathrm{II}^\sharp$ bi--block, we get a \emph{good} bi--quiver.
For the computation of the nullity $\kappa$ and
type $\mathfrak{g}_r$ of the Tits form $q(\boldsymbol{x})$, see
appendix \ref{ppprof}.

\subsection{Galois covers of (locally) bounded $\C$--categories}

\subsubsection{Covering functors}\label{sec:coveringfunctors}

We start by recalling some standard definitions \cite{gal1,gal2,gal3,gal4,gal5}
(for a nice survey see \cite{galsurvery}).

A category $\ca$ is called \emph{$\C$--linear} if, for all ordered pairs of objects, $X$, $Y$, the set
of morphisms $\ca(X,Y)$ has a $\C$--vector--space structure and their composition law $\circ$ is bilinear. The category is \emph{connected} if it is not empty nor the disjoint union of two non--empty categories.
A connected $\C$--category $\ca$ is
\emph{locally bounded} if the following three conditions are satisfied:
\begin{itemize}
\item[(a)] distinct objects of $\ca$ are non--isomorphic;
\item[(b)] for all object $X$, $\ca(X,X)$ is a local algebra;
\item[(c)] for all $X\in \ca$,
$\sum\limits_{Y\in\ca}\big(\dim_\C\ca(X,Y)+\dim_\C\ca(Y,X)\big)<\infty$.
\end{itemize}
A \emph{bounded} category is a locally bounded category with finitely many objects. \medskip

A (locally) bounded category $\ca$ may always be written
(uniquely) in the form $\C Q/I$, where $Q$ is a 
(locally) finite quiver and $I$ an admissible ideal
in its path category \cite{gal1,gal2}. The objects of $\ca$ are identified with the nodes of $Q$. The
identity morphism of the $i$--th object $i\in\ca$ is identified with the minimal idempotent $e_i$ of the
algebra $\C Q/I$ given by the lazy path at the $i$--th node of $Q$.
One has 
\begin{equation}
\ca(i,j)= e_j\; \C Q/I\;e_i.
\end{equation}
The bounded category $\ca$ is then canonically identified with the basic associative algebra $\C Q/I$, and we shall use the two languages interchangeably.
Viewing $\ca$ as a category, however, allows
to perform general categorical operations on it, in particular to construct its \emph{covering} and \emph{orbit} categories.
In the categorical language, a (finite dimensional) representation $X$ of the (opposite) quiver bounded by $I$, which is the same as a {right} module of the algebra $\C Q/I$, is seen as
a \emph{linear functor} from the $\C$--category $\ca$ to the $\C$--category of finitely--dimensional $\C$--vector--spaces
\begin{equation}
X\colon \ca\to \mathsf{mod}\,\C.
\end{equation}
Such a functor is called a \emph{module}
of the bounded category $\ca$, and the (Abelian)
category of such functors is denoted as
 $\mathsf{mod}\,\ca$. 
\smallskip

Let $\ca$ be a locally bounded $\C$--category and
$\mathbb{G}$ a group of $\C$--linear automorphisms of $\ca$. 
The group $\mathbb{G}$ acts on 
$\mathsf{mod}\,\ca$ by composition of functors
\begin{equation}
X\longmapsto X^g\equiv X\circ g.
\end{equation}
This action is an autoequivalence and so:
\begin{itemize}
\item[(a)] preserves indecomposable, brick, simple, projective, and injective modules;
\item[(b)] commutes with the Auslander--Reiten translations $\tau$, $\tau^-$.
\end{itemize}
To each $X\in\mathsf{mod}\,\ca$ we associate its
\emph{isotropy subgroup}
 $\mathbb{G}_X\subset \mathbb{G}$
\begin{equation}
\mathbb{G}_X=\Big\{\;g\in\mathbb{G}\; \Big|\; X^g\cong X\;\Big\}.
\end{equation}
Let $\mathbb{H}\subseteq \mathbb{G}$ be a subgroup; we write $\mathsf{mod}^\mathbb{H}\!\ca$ for the full subcategory of $\mathbb{H}$--invariant modules.
\vglue 6pt

An automorphism group $\mathbb{G}$ is said to be \emph{admissible} if it acts freely on the objects $i$ of $\ca$ (i.e.\! on the nodes of its quiver) 
and has finitely many orbits.
If $\mathbb{G}$ is admissible for $\ca$, we 
may consider its \emph{orbit category} $\ca/\mathbb{G}$ which is a bounded $\C$--category.
The objects of the category $\cb\equiv\ca/\mathbb{G}$ are the $\mathbb{G}$--orbits $\mathbb{G}i$ of objects $i\in\ca$, while the morphism spaces are
\begin{equation}\label{a/gmor}
\cb(\mathbb{G}i,\mathbb{G}j)=\bigoplus_{g\in\mathbb{G}}\ca(i,gj).
\end{equation} 
Then we have a canonical \emph{Galois covering}
functor
\begin{equation}
F\colon \ca\longrightarrow \ca/\mathbb{G}\quad\text{given on objects by}\quad i\longmapsto \mathbb{G}i.
\end{equation}
Since $\ca/\mathbb{G}$ is again a bounded
$\C$--category, there is a finite quiver $\overline{Q}$
(whose nodes are the $\mathbb{G}$--orbits of nodes of $Q$) and an admissible ideal $\overline{I}$ such that $\ca/\mathbb{G}=\C \overline{Q}/\overline{I}$. 
\medskip

The process of going from the category $\ca$ and the freely acting automorphism group $\mathbb{G}$
to the orbit category $\ca/\mathbb{G}$ may be reversed. Indeed, the definition \eqref{a/gmor} shows that the morphism spaces of $\ca/\mathbb{G}$ are vector spaces
graded by elements $g\in\mathbb{G}$, the grading being compatible with
compositions. By definition, this means that
$\ca/\mathbb{G}$ is a $\mathbb{G}$--graded $\C$--category. Given a $\mathbb{G}$--graded 
$\C$--category $\cb$ we may construct the \emph{smash product} $\C$--category,
$\cb\# \mathbb{G}$, whose objects
are pairs $(a,g)$, $a\in\cb$, $g\in\mathbb{G}$,
and morphism spaces \cite{galsurvery}
\begin{equation}
\cb\#\mathbb{G}\big((a,g),(b,h)\big)=\cb(a,b)\bigg|_{\text{degree $h^{-1}g$}\atop \text{component}}.
\end{equation} 
It is easy to see that, if $\mathbb{G}$ acts freely on $\ca$,
\begin{equation}\label{wwwz17}
(\ca/\mathbb{G})\# \mathbb{G}\cong \ca.
\end{equation}
%\smallskip

The theory of Galois covers of $\C$--categories has  much the same flavor as the classical
theory of Galois field extensions. A field $k$
(say of zero characteristic)
has an universal extension, namely its algebraic
closure $k^\mathrm{al}$, and universal
Galois group $\mathrm{Gal}(k^\mathrm{al}/k)$.
The fundamental theorem of the classical theory says that
there is an inclusion reversing correspondence
\begin{equation}\label{classGalois0}
\begin{split}
&\Big\{\text{normal subgroups $N\triangleleft \mathrm{Gal}(k^\mathrm{al}/k)$ of finite index}\Big\}\longleftrightarrow\\
&\hskip2cm\longleftrightarrow\quad
\Big\{\text{finite degree Galois extensions of $k$}\Big\}.
\end{split}
\end{equation}
In the same fashion, a (locally bounded)
$\C$--category $\ca$ admits
an \emph{universal Galois cover} $\widetilde{\ca}$
and an universal covering group $\Pi(\ca)$
called the \emph{fundamental group} of the
category $\ca$. The fundamental group of
$\ca$ is explicitly constructed using the
\emph{walks} in the quiver $Q$ of $\ca$;
we shall not use this construction and refer
to \cite{gal3} for details. All locally bounded $\C$--category $\ca$ is automatically
$\Pi(\ca)$--graded. Eqn.\eqref{wwwz17} then gives an explicit formula for its universal cover category
\begin{equation}
\widetilde{\ca}= \ca\# \Pi(\ca).
\end{equation} 
The Galois correspondence then becomes
\begin{gather}\label{classGalois}
\big\{\text{normal subgroups $\mathbb{H}\triangleleft \Pi(\ca)$}\big\}\longleftrightarrow
\big\{\text{Galois coverings $\cc\to \ca$}\big\},
\end{gather}
where the Galois group of $\cc\to \ca$ is $\mathbb{G}\equiv\Pi(\ca)/\mathbb{H}$,
and $\cc= \ca\#(\Pi(\ca)/\mathbb{H})$.
An algebra $\ca$ with $\Pi(\ca)=(1)$
has no non--trivial covers; such an algebra, if connected, is said to be
\emph{simply--connected}.
\smallskip

We are not interested in the Galois cover of $\C$--categories 
$F\colon\ca\to\cb$ \emph{per se}, but rather in the 
functors it induces between the module categories
$\mathsf{mod}\,\ca$ and $\mathsf{mod}\,\cb$.
For simplicity, we assume all categories to be
bounded (so $\mathbb{G}$ is a finite group of order $|\mathbb{G}|$)
and all modules to be finite--dimensional;
more general statements may be found
in the literature \cite{gal1,gal2,gal3,gal4,gal5,galsurvery}.
A Galois cover
$F\colon\ca\to\cb\equiv \ca/\mathbb{G}$ induces two natural functors
between the module categories:
\begin{itemize}
\item the \emph{pull up} functor
$F^\lambda\colon\mathsf{mod}\,\cb\to \mathsf{mod}\,\ca$ defined by composition of functors 
\begin{equation}
F^\lambda\colon X\longmapsto F^\lambda X\equiv X\circ F;
\end{equation}
\item the \emph{push down} functor
$F_\lambda\colon\mathsf{mod}\,\ca\to \mathsf{mod}\,\cb$ is the map which associates to the functor $Y\colon
\ca\to \mathsf{mod}\,\C$  the functor $F_\lambda Y\colon \cb\to\mathsf{mod}\,\C$ 
acting as follows
\begin{align}
&\text{\,$\diamond$ \underline{on objects $\mathbb{G}i$}:}  
\hskip 2.6cm\mathbb{G}i\longmapsto 
F_\lambda Y(\mathbb{G}i)=\bigoplus_{g\in\mathbb{G}}Y(g i)\\ 
&
\begin{aligned}&\text{$\diamond$ \underline{on morphisms $\mathbb{G}i \xrightarrow{\;f\;}\mathbb{G}j$}}:
&&\text{eqn.\eqref{a/gmor}
$\Rightarrow$ $f=\sum_{g\in\mathbb{G}}f_g$
with $f_g\in \ca(i,gj)$}\\
&&&\text{then $F_\lambda Y(f)=\sum_{g\in\mathbb{G}}Y(f_g)$.}
\end{aligned} 
\end{align}
\end{itemize}
\textbf{Properties} \cite{galsurvery}:
\begin{itemize}
\item[1)] the categories $\mathsf{mod}^\mathbb{G}\!\ca$ and $\mathsf{mod}\,\cb$ are equivalent;
\item[2)] for all $X\in\mathsf{mod}\,\ca$ and all
$g\in\mathbb{G}$ we have $F_\lambda X^g\cong X$ and $F^\lambda F_\lambda X\cong \bigoplus_{g\in\mathbb{G}}X^g$;
\item[3)] $F_\lambda$ and $F^\lambda$ are each other right-- and left--adjoints:
\begin{equation*}
\ca(X,F^\lambda Y)\cong \cb(F_\lambda X,Y),\ \  
\ca(F^\lambda Y, X)\cong \cb(Y,F_\lambda X)\ \  \forall\;X\in\mathsf{mod}\,\ca,\; Y\in\mathsf{mod}\,\cb.
\end{equation*}
\end{itemize}
The crucial property for our 
applications is
\medskip

\textbf{Fact 3} \cite{galsurvery}. \textit{$\mathbb{G}$ an admissible group of automorphisms of $\ca$.
Suppose the $\ca$--module $X$ is indecomposable and $\mathbb{G}_X=(1)$. Then $F_\lambda X$
is indecomposable and for all modules $Y$
with $F_\lambda Y\simeq F_\lambda X$ there is $g\in\mathbb{G}$ such that $Y\simeq X^g$.}

\subsubsection{Important special cases}\label{special cases}

A special case, which is relevant for
complete $\cn=2$ theories \cite{CV11},
is when $\ca$ is a \emph{string} algebra.
The indecomposable modules of string algebras
have an explicit construction in terms of strings and bands \cite{stringmod} (see also \cite{cattoy}). 
For convenience of the reader we briefly review their 
construction in appendix \ref{stringbands}. If $C$ is a string (band) we write $M(C)$  (resp.\! $M(C,\mu,n)$ with $\mu\in\C$ and $n\in\mathbb{N}$) for the corresponding module.

Let $\ca$ be a string algebra with a freely acting
automorphism group $\mathbb{G}$, and
let $F\colon\ca\to\cb\equiv\ca/\mathbb{G}$
be the associated Galois cover. The first observation is
that $\cb$ is automatically a string algebra.
Then, if $M(C)\in\mathsf{mod}\,\ca$ is a string module,
its push down $F_\lambda M(C)\in\mathsf{mod}\,\cb$ is a string module
with respect to the image string
\begin{equation}
F_\lambda M(C)= M(F C),
\end{equation}
and hence is automatically indecomposable\footnote{ This may also be seen from \textbf{Fact 3}. Indeed, for all string module $M(C)$, one has $\mathbb{G}_{M(C)}=(1)$.} (but the push down of a brick may be just indecomposable).
On the contrary, the push down of a band is not necessarily a band, since it may be a non--trivial power of a shorter band. In facts, it is a $|\mathbb{G}_C|$
power, where $\mathbb{G}_C\equiv \mathbb{G}_{M(C,\mu,n)}$ is the isotropy subgroup
of the covering band. Thus, a band module $M(C,\mu,n)$ pushes down to an
indecomposable module $M(FC,\mu,n)$ if and only if
its isotropy group is trivial, in agreement with
 \textbf{Fact 3}. 

The pull back of a string of $\cb$ is the disconnected union of $|\mathbb{G}|$ strings
of $\ca$. By \textbf{Properties} 2) this means that the restriction of
$F_\lambda$ to strings is surjective (i.e.\!
all string modules of $\cb$ are push down 
of string modules of $\ca$). In other words, $F_\lambda$
sets a $|\mathbb{G}|$--to--1 correspondence
between string modules of $\ca$ and string modules of $\cb$. 

The pull back of a band is, in general, a disconnected union of bands which forms
one orbit under $\mathbb{G}$.
It produces an indecomposable module $M(C',\mu',n)\in\mathsf{mod}\,\ca$ precisely when
this orbit contains a single element,
that is, $\mathbb{G}_{M(C',\mu',n)}\equiv\mathbb{G}$. In this case
\begin{equation}
F^\lambda M(C,\mu,n)=M(F^{-1}C, \mu^{|\mathbb{G}|},n).
\end{equation}

\paragraph{Bricks.} A band module is a brick if and only if it has the form $M(C,\mu,1)$. Then, if $\mathbb{G}_C=(1)$, the indecomposable module
$F_\lambda M(C,\mu,1)$ is also a brick.
If $M(C)$ is a string module of $\ca$ which is not a brick, the string module $F_\lambda M(C)\in\mathsf{mod}\,\cb$ cannot be a brick.
On the contrary, if $M(C)$ is a brick, $F_\lambda M(C)$ may or may not be a brick. Let $C^\prime=F C$. The module $M(C^\prime)\in\mathsf{mod}\,\cb$ may fail to be a brick only if $C^\prime$ contains a substring $C^{\prime\prime}$ which starts and ends at the same node, which then is a power of a band
$B$. $M(C^\prime)$ is a brick if for all such bands $B$ one of the following two conditions are fulfilled
\begin{equation}
\begin{split}
&\mathbf{1)}\ \ \sum_{i\in\text{sources of }B}(\mathbf{dim}\,M(C^\prime))_i\neq \sum_{i\in\text{sinks of }B}(\mathbf{dim}\,M(C^\prime))_i,\\
&\mathbf{2)}\ \ \mathbf{dim}\,M(C^\prime)\big|_{\mathrm{supp}\,M(B,\mu,1)}\leq \mathbf{dim}\, M(B,\mu,1),
\end{split}
\end{equation}
where sinks and sources are counted with their multiplicities in $B$.

\paragraph{Triangular string algebras.} When $\ca$ (and hence $\cb$)
happens to be, in addition, a triangular algebra,
the dimension vectors of its indecomposable modules are roots of the Tits form $q_\ca$ which, in the string case, is positive semi--definite (string algebras are tame \cite{dPlects,stringmod}). Band modules 
correspond to
imaginary roots of $q_\ca$, while string modules
correspond to real roots \underline{unless} they are in the
projective closure of $\C$--families of 
band modules in which case they are imaginary. 
Hence
\begin{equation}\label{ttttq12}
\begin{aligned}
&q_\ca(\mathbf{dim}\,X)=0 &&\Rightarrow
&&q_\cb(\mathbf{dim}\,F_\lambda X)=0\\
&q_\ca(\mathbf{dim}\,X)=1 &&\Rightarrow
&&q_\cb(\mathbf{dim}\,F_\lambda X)=\begin{cases}1 & F_\lambda X\ \text{rigid: $\mathrm{Ext}^1(X,X)=0$}\\ 
0 & F_\lambda X\ \text{has self--extensions}.
\end{cases}
\end{aligned}
\end{equation}

\subsubsection{Galois covers of dashed categories}

By a \emph{dashed category} we mean a triple
$(\mathscr{J},\mathscr{T},\varpi)$ where $\mathscr{J}$ and $\mathscr{T}$ are bounded $\C$--categories such that, as algebras,
$\mathscr{J}\equiv \C Q/(\partial \cw)$
is the Jacobian algebra
 of a
$2$--acyclic quiver $Q$ with superpotential $\cw$,
while
 $\mathscr{T}$ is its factor defined by an admissible
dashing of $Q$, i.e.\! $\mathscr{T}\equiv \mathscr{J}/(\mathsf{dash})$.
$\varpi\colon \mathscr{J}\to \mathscr{T}$ is then the canonical quotient functor.\footnote{ `Dashed categories' are $\C$--categories. Its objects are the nodes of the quiver $Q$, and the morphisms are
pairs of morphisms of the form $(f,\varpi(f))$. } Clearly one may also consider dashed categories which are only
\emph{locally} bounded.

Let $\mathbb{G}$ be a group of admissible automorphisms of $\mathscr{J}$ 
which preserves the ideal $(\mathsf{dash})$,
that is, which sends dashed (solid) arrows of $Q$ into
dashed (solid) arrows. Then $\mathbb{G}$ is a group of admissible automorphisms of the factor $\mathscr{T}$, and we have a commutative diagram of functors
\begin{equation}
\begin{gathered}
\xymatrix{\mathscr{J} \ar[r]^F\ar[d]_\varpi & \mathscr{J}/\mathbb{G}\ar[d]^{\overline{\varpi}}\\
\mathscr{T}\ar[r]_{\overline{F}} & \mathscr{T}/\mathbb{G} }
\end{gathered}
\end{equation}
The induced functors $\overline{F}^\lambda$, $\overline{F}_\lambda$ relate
corresponding triangular sectors of
$\mathsf{mod}\mathscr{T}$ and 
$\mathsf{mod}\mathscr{T}/\mathbb{G}$.

\subsubsection{Covers of class $\cs[A_1]$ dashed categories}\label{coversofsa1}

We specialize to the case in which
$\mathscr{J}$ is a Jacobian category of a 
class $\cs[A_1]$ $\cn=2$ QFT.
Modulo a handful of exceptions,\footnote{ \label{fool}The argument may be extended to cover the exceptional cases. In facts, for the validity of the argument it suffices that the block decomposition is unique in some \emph{class} of decompositions, e.g.\! the decompositions with minimal or maximal number of pairs of opposite arrows $\leftrightarrows$, or the ones with minimal/maximal number of blocks of a given Type, \emph{etc}. }
we may assume that the corresponding quiver $Q$ has a unique block decomposition without type V blocks\footnote{ All our conclusions remain valid in presence of Type V blocks. However, the assumption that no such block is present allows to shorten our arguments. The straightforward extension to the case in which Type V blocks are present is left to the reader. }; and  that the \emph{good}
 biquiver $(Q,\widetilde{Q})$ associated to the factor $\mathscr{T}$ has no type $II^\flat$ bi--block, cfr.\! \S.\ref{classA1tits}.
Then the Tits form $q_\mathscr{T}$ of $\sct$
is written
(in a unique way) as a sum of squares
associated to the several bi--blocks, eqn.\eqref{titsbiblocks}.

Let $h$ be an automorphism of the bi--quiver.
$h$ maps a (bi)block decomposition into a (bi)block decomposition. Since the decomposition is unique,
$h$ maps each bi--block into a bi--block of the same Type, and hence acts by permutations of
same--Type bi--blocks. Suppose a bi--block
$b$ is fixed by the permutation, so that $h|_b$ is an automorphism of the bi--block $b$. Since no bi--block $\neq II^\flat$ has a non trivial
automorphism acting freely on its nodes,
 $h$ should fix some node of $b$.
We conclude that all freely acting automorphisms
$g$ of the dashed category map bi--blocks into \emph{distinct}
same--Type bi--blocks. In particular, nodes $i$ and $g i$
belong to distinct bi--blocks for all $g\in \mathbb{G}$, $g\neq 1$. 

Let $X\in \mathsf{mod}\mathscr{T}$ be an
indecomposable,
so $q_\mathscr{T}(\mathbf{dim}\,X)=0,1$. One has
\begin{align}
2\,q_{\mathscr{T}}(\mathbf{dim}\,X)&=\sum_{\text{orbits of}\atop \text{0,I,II}}
\sum_{g\in\mathbb{G}}
y_{gb}^2+\frac{1}{2}\sum_{\text{orbits of}\atop \text{III,IV}}
\left\{\sum_{g\in\mathbb{G}}w_{g b,1}^2+
\sum_{g\in\mathbb{G}}w_{g b,2}^2\right\},\\
2\,q_{\mathscr{T}/\mathbb{G}}(\mathbf{dim}\,F_\lambda X)&=\sum_{\text{orbits of}\atop \text{0,I,II}}
\left(\sum_{g\in\mathbb{G}}
y_{gb}\right)^{\!\!2}+\frac{1}{2}\sum_{\text{orbits of}\atop \text{III,IV}}
\left\{\left(\sum_{g\in\mathbb{G}}w_{g b,1}\right)^{\!\!2}+
\left(\sum_{g\in\mathbb{G}}w_{g b,2}\right)^{\!\!2}\right\}
\end{align}
where $y_b$, (resp.\! $w_{b,1}$, $w_{b,2}$) is the
$\Z$--linear combinations of the $(\mathbf{dim}\,X)_i$ 
appearing in the polynomial $P_\mathfrak{b}$ of the $b$--th 
block (see eqns.\eqref{667nnhq}--\eqref{Qquad});
the essential point is that $y_b,\; w_{b,1},\;w_{b,w}\in\Z$ while
$w_{b,1}=w_{b,2}\mod2$.
If $\mathbf{dim}\,X$ is an imaginary root,
$y_b=w_{b,s}=0$ for all $b$, and so 
$\mathbf{dim}\,F_\lambda X$  is also an imaginary root. If $\mathbf{dim}\,X$ is a real root,
we have three possibilities: 
\begin{itemize}
\item[\textit{i)}] at one III/IV block
$b$ we have either $w_{b,1}=\pm 2$, $w_{b,2}=0$  or $w_{b,1}=0$, $w_{b,2}=\pm 2$,
and at all other blocks $y_{c}=w_{c,1}=w_{c,2}=0$;
\item[\textit{ii)}] for a type III/IV block $b$ 
and a 0/I/II block $c$
we have $w_{b,s}=\pm 1$ and $y_c=\pm 1$,  while $y_{d}=w_{d,1}=w_{d,2}=0$ for $d\neq b,c$;
\item[\textit{iii)}] for a pair of distinct I/II blocks\footnote{ The case of a pair of distinct Type 0 blocks is excluded since, at a Type $0$ block $b$,
$y_b\equiv (\mathbf{dim}\,F_\lambda X)_b\geq 0$. The same argument rules out pairs of Type $I^\flat$ or Type $II^\flat$ bi--blocks.}
$b$, $c$ we have $y_b=\pm1$ and $y_c=\pm1$,
all other variables being zero. 
\end{itemize}
In cases \textit{i)} and \textit{ii)}
$\mathbf{dim}\,F_\lambda X$ is automatically a real root of
$q_{\mathscr{T}/\mathbb{G}}(\cdot)$.
In case \textit{iii)} $\mathbf{dim}\,F_\lambda X$ is also a real root, \emph{unless} there exists
a $g\in\mathbb{G}$ such that $c=gb$,
while $y_b$ and $y_c$ have opposite signs.
In this last case $\mathbf{dim}\,F_\lambda X$ is
an imaginary root. The restriction
of $X$ to the two bi--blocks $b$ and $gb$ must have one of two possible forms: if $b$ is of Type I
\begin{align}
&X|_b\equiv\begin{gathered}\xymatrix{\C^m\ar[rr]&& \C^{m+1}}\end{gathered}
&&
X|_{gb}\equiv\begin{gathered}\xymatrix{\C^{k+1}\ar[rr]&& \C^k}
\end{gathered}\\
\intertext{the arrow being mono in the left figure and
epi in the right one, or, for Type II,}
&X|_b\equiv\begin{gathered}
\xymatrix{\C^\ell\ar@{..>}[drr]\\
\C^{\ell+m+1}\ar[u] && \C^m\ar[ll]}\end{gathered}
&&X|_{gb}\equiv\begin{gathered}
\xymatrix{\C^k\ar@{..>}[drr]\\
\C^{k+j}\ar[u] && \C^{j+1}\ar[ll]}\end{gathered}
\end{align}
the vertical arrows being epi, and the horizontal
arrow mono for the left figure and epi on the
kernel of the vertical arrow for the right one.
In both cases the module
\begin{equation}
F_\lambda X\big|_{\text{orbit}\atop
\text{of } b}=\bigoplus_{h\in\mathbb{G}} X\big|_{h b}
\end{equation}
 admits a one--parameter deformation which for Type I makes the arrow to be iso, and for Type II makes the horizontal arrow iso on the kernel of the vertical arrow. Then in both cases, whenever $\mathbf{dim}\,X$ is a real root while $\mathbf{dim}\,F_\lambda X$ is an imaginary root, we conclude that $F_\lambda X$ belongs to the projective closure of a $\mathbb{P}^1$ family of indecomposable modules\footnote{ $F_\lambda X$ is indecomposable. Indeed, let $h\in\mathbb{G}_X$. $h$ should fix both bi--blocks $b$ and $gb$, but an automorphism fixing a bi--block necessarily fixes a node,
while $\mathbb{G}_X$ acts freely on the nodes.
Hence $\mathbb{G}_X=(1)$, and the claim follows from \textbf{Fact 3}.}. This result generalizes eqn.\eqref{ttttq12} to $\cs[A_1]$ dashed categories which are not (nor can be reduced to) string
algebras.

\subsubsection{$\Pi(\scj)$ versus $\pi_1(\Sigma)$}
\label{s:versus}

Let $\Sigma$ be a surface with punctures
and marks on the boundaries, $Q$
the quiver of an ideal triangulation of $\Sigma$
\cite{triangulation1},
and $\mathscr{J}=\C Q/(\partial\cw)$ the corresponding Jacobian algebra/category\footnote{ We shall refer to $\C$--categories arising this way from an ideal triangulation of a surface with punctures and marks as \emph{a class $\cs[A_1]$ category.}}.
The authors of \cite{ABCJP} pose the question of the relation between the fundamental group of
the algebra, $\Pi(\scj)$, and the fundamental group of the corresponding surface $\pi_1(\Sigma)$. 
On the nose, there cannot be any simple relation:
$\pi_1(\Sigma)$ is a topological invariant of the surface $\Sigma$, while $\Pi(\scj)$ depends on the particular ideal triangulation as the following example shows. \medskip

\textbf{Example.}
The mutation class of
the punctured disk with $r$ marks on the boundary
contains the Dynkin quiver of type $D_r$ 
as well as the quiver consisting on a single oriented cycle of length $r$, $C_r$, with $\cw$ the cycle itself. 
The first Jacobian algebra, $\scj_1\equiv\C D_r$,
has a tree quiver, and hence is simply--connected
$\Pi(\scj_1)=0$, while the second one $\mathscr{J}_2=\C C_r/(\partial \cw)$ is well known to have
$\Pi(\scj_2)=\Z$.
\medskip

However, we are not really interested in $\Pi(\scj)$.
This universal covering group describes, \emph{via}
the Galois correspondence \eqref{classGalois},
\emph{all} possible coverings of $\scj$.
Most of the covering categories $\ca\to\scj$
described by the universal correspondence
do not correspond to any $\cn=2$ QFT.
A bounded $\C$--category $\ca$ should satisfy quite stringent conditions in order to correspond to a QFT:
to the very least, its quiver $Q$ should be $2$--acyclic and its ideal $I$ should arise from 
the differential of a non--degenerate superpotential
$\cw$ (in facts, we know that there are additional,  more restrictive, necessary conditions \cite{CNV}).
Morally speaking, the ``universal covering group''
we are really interested in, ``\;$\Pi(\scj)^\text{QFT}$\;'', is the group whose finite quotients 
are the Galois groups of the covers of $\scj$ by $\C$--categories
which \emph{do correspond} to $\cn=2$ QFTs.
We shall not attempt here to  construct rigorously a
would be `universal quantum--field--theoretical
Galois covering'. However, for the special case that $\scj$ is a class $\cs[A_1]$ category, we describe 
 all possible Galois covers $\mathscr{C}\to\scj$ of the class $\cs[A_1]$ category
$\scj$ by connected bounded $\C$--categories $\mathscr{C}$
which do correspond to \emph{complete}
$\cn=2$ QFTs.
\medskip

\textbf{Fact 4.} \textit{Let $F\colon\mathscr{J}^\prime\to\mathscr{J}$ be a Galois cover of connected bounded $\C$--categories, where $\mathscr{J}^\prime$, $\mathscr{J}$ are Jacobian categories of class $\cs[A_1]$ theories. 
In terms of Gaiotto UV surfaces $\Sigma^\prime_G$, $\Sigma_G$ and
quadratic differentials $\phi^\prime_2$, $\phi_2$ {\rm \cite{Gaiotto}}, $F$ is induced by a non--constant holomorphic map $\xi\colon \Sigma^\prime_G\to\Sigma_G$ whose only branch points are at
the \underline{irregular} punctures of $\phi_2$. One has
\begin{equation}\phi_2^\prime=\xi^*\phi_2.\end{equation}
Equivalently, $F$ is induced by an \emph{unbranched} cover of  
bordered surfaces $\mathsf{c}\colon\Sigma^\prime\to\Sigma$ {\rm\cite{CV11}}. }\medskip

\textbf{Fact 4} is proven in appendix \ref{proofpi}. We recall that the unramified covers
 $\mathsf{c}\colon \Sigma^\prime\to\Sigma$ are classified by the normal subgroups of $\pi_1(\Sigma^\circ)$, where $\Sigma^\circ$ is $\Sigma$ with the regular punctures ignored.
 
Let the quotient surface $\Sigma$ have genus $g$, $p$ punctures, and $b$ boundaries carrying $\{\ell_i\}_{i=1}^b$ marks. Let $\xi\colon \Sigma_G^\prime\to\Sigma_G$ be a covering of degree $d$
which is ramified only over the irregular punctures corresponding to the boundaries of $\Sigma$ \cite{CV11}.
The inverse image of the $i$--th irregular puncture
is a set of $s_i$ ramification points with ramification numbers
$m_{j,i}\geq 1$ ($j=1,\dots, s_i$, $i=1,\dots,b$) satisfying $\sum_j m_{j,i}=d$.
The covering surface $\Sigma^\prime$ has then
$b^\prime=\sum_{i}s_i$ boundary components
with $\ell^\prime_{j,i}=m_{j,i}\ell_i$ marks on them.
The topological invariants of the covering surface $\Sigma^\prime$ are
\begin{align}
&p^\prime= d\,p,\qquad\quad\qquad b^\prime=\sum_{i=1}^b s_i,
&&\ell_{j,i}^\prime=m_{j,i}\,\ell_i,\\ 
&2g^\prime-2=(2g-2)d+(b\,d- b^\prime),
&&\sum_{j=1}^{s_i}m_{ji}=d,\\
&n^\prime\equiv -3\,\chi(\Sigma^\prime)+\sum_{j,i}\ell^\prime_{j,i}=n\,d.
\end{align} 
Here $n$ ($n^\prime$) is the number of ideal arcs in any ideal triangulation of $\Sigma$ ($\Sigma^\prime$). It follows that, if $\mathsf{T}$ is an ideal triangulation of $\Sigma$, then $\mathsf{T}^\prime\equiv \mathsf{c}^{-1}(\mathsf{T})$ is an ideal triangulation of $\Sigma^\prime$.

By Hurwitz theory \cite{hurwitz}, the covering group $\mathsf{G}$ of $\mathsf{c}$ is a subgroup of the symmetric group $\mathfrak{S}_d$, acting \emph{transitively} on $(1,2,\dots, d)$, which is generated by the $b+2g$ elements of $\mathfrak{S}_d$
\begin{equation}\label{hur1}
\alpha_1,\dots, \alpha_b, \beta_1,\dots, \beta_g,\gamma_1,\dots,\gamma_g,
\end{equation}
which form a representation of $\pi_1(\Sigma^\circ)$,
i.e.\! satisfy the relation
\begin{equation}\label{hur2}
\alpha_1\cdots\alpha_b \beta_1\gamma_1\beta_1^{-1}\gamma_1^{-1}
\cdots \beta_g\gamma_g\beta^{-1}_g\gamma_g^{-1}=1,
\end{equation} 
and such that $\alpha_i$ belongs
to the conjugacy class of $\mathfrak{S}_d$ specified by the
partition of $d$
\begin{equation}\label{hur3}
m_{1, i}+m_{2, i}+\cdots +m_{s_i,i}=d.
\end{equation}
So $\mathsf{G}=\pi_1(\Sigma^\circ)/N$ for some
normal subgroup $N\triangleleft \pi_1(\Sigma^\circ)$. Conversely, any transitive subgroup $\mathsf{G}\subset \mathfrak{S}_d$ satisfying \eqref{hur1}--\eqref{hur3} corresponds to a cover $\mathsf{c}$.

By construction, $\mathsf{G}$ is an automorphism group of the ideal triangulation $\mathsf{T}^\prime$, hence of the covering Jacobian category $\mathscr{J}^\prime$. However, it is not necessarily true that
$\mathsf{G}$ acts freely on the objects of
$\mathscr{J}^\prime$, i.e.\! on  the arcs of $\mathsf{T}^\prime$, as the following example shows: 
\medskip

\textbf{Example 1.} Let $\Sigma$ be the pair of pants (a genus zero surface with three boundaries). 
By definition, a covering map $\xi\colon \Sigma_G^\prime\to \Sigma_G\equiv\mathbb{P}^1$,
which is branched only at the three irregular punctures, is given by
a Belyi function $\xi$ \cite{belyi}. A well known example of degree $d$ rational Belyi function is the Chebyshev polynomial of the first kind, $T_d(z)$. Identifying 
the symmetric group $\mathfrak{S}_d$ with the Weyl group $\mathsf{W}$ of $SU(d)$, we have $\mathsf{G}=\langle \sigma_e,\sigma_o\rangle\subset \mathsf{W}$, where 
$\sigma_e$ ($\sigma_o$) is the product of the
simple reflections at the even (odd) nodes of the bipartite Dynkin graph $A_{d-1}$. Clearly, $\mathsf{G}\cong
D_d$, the dihedral group of degree $d$, i.e.\!
the symmetry group of a regular $d$--gon.
$D_d$ does not act freely on the vertices of the $d$--gon. Correspondingly, $\mathsf{G}$ does not act freely on the arcs of $\mathsf{T}^\prime$.
\medskip

However, in this example we have a subgroup
$\mathbb{G}\equiv \langle \mathsf{Cox}\rangle\subset \mathsf{G}$, i.e.\! the cyclic group $\Z_d$ generated by the Coxeter element $\mathsf{Cox}\equiv \sigma_e\sigma_o\in\mathsf{W}$, which  does act
\emph{freely} and \emph{transitively} on the
$\mathsf{G}$--orbits. Then, as $\C$--categories,
$\mathscr{J}=\mathscr{J}^\prime/\mathbb{G}$,
and  it is the subgroup $\mathbb{G}$ which is 
the \emph{categorical} Galois group
(while $\mathsf{G}$ is the Galois group in the usual sense). Hence
\medskip

\textbf{Fact 5.} \textit{The covering of bordered surfaces $\mathsf{c}$ corresponds to a Galois $\mathbb{G}$--cover of class $\cs[A_1]$ categories
iff there is a subgroup $\mathbb{G}\subseteq \mathsf{G}\subseteq \mathfrak{S}_d$ which acts \emph{freely} and \emph{transitively}. In particular $d\equiv |\mathbb{G}|$.}
\medskip

It is easy to find examples with $\mathbb{G}=\mathsf{G}$ and Abelian (see section 3).
We present a simple example of \emph{non--Abelian} covering group $\mathsf{G}$ such that $\mathbb{G}\equiv \mathsf{G}$.
\medskip

\textbf{Example 2.} In the same context as in the
previous example, consider the Belyi rational function of degree 6 given by the expression of the
modular invariant $J(\tau)$ in terms of the
Legendre $\Gamma(2)$--modular function $\lambda(\tau)$
\begin{equation}\label{legendre}
J(\lambda)=\frac{4}{27}\,\frac{(1-\lambda+\lambda^2)^3}{\lambda^2 (1-\lambda)^2}.
\end{equation}
One has $\mathsf{G}\equiv PSL(2,\Z)/\Gamma(2)\cong \mathfrak{S}_3$ \cite{farkas}. 
Since $\mathsf{G}$ acts transitively as a subgroup of $\mathfrak{S}_6$, and, since $|\mathsf{G}|=6$,
$\mathsf{G}$ also acts freely, i.e.\! $\mathbb{G}=\mathsf{G}$. In physical terms, in the minimal case
(i.e.\! $\Sigma$ without punctures and a single mark per boundary)
eqn.\eqref{legendre} yields the Galois $\mathfrak{S}_3$--cover
of $SU(2)^3$ SYM with one half--trifundamental hypermultiplet by $SU(2)^{13}$ SYM
coupled to one half--trifundamental 
in each of the six representations of the form 
\begin{equation}
\Big(\overbrace{\mathbf{1},\dots,\mathbf{1}}^{\text{even $\#$ of $1$'s}},\mathbf{2},\mathbf{2},\mathbf{2},\mathbf{1},\dots,\mathbf{1}\Big)
\end{equation}
and a fundamental quark coupled to each \emph{odd--numbered} $SU(2)$ SYM system but the first and the 13--th ones which are each coupled to a $D_3$ AD system. 
If $\Sigma$ contains more punctures and marks,
eqn.\eqref{legendre} gives a corresponding $\mathfrak{S}_3$--cover of class $\cs[A_1]$ theories.

 \subsubsection{A mere computational tool: quiver covers}\label{quivercovers}
 
 In \S.\ref{s:versus} we discussed the
 Galois covers of Jacobian categories of $\cn=2$
 QFTs by other Jacobian categories of $\cn=2$ QFTs which lead to covering maps between their BPS spectra. However,
 the covering techniques 
 are useful to compute BPS spectra
independently of the fact that the covering
category is Jacobian or associated to a $\cn=2$ theory. From the computational point of view, what really matters is that the cover has a \emph{simple}
Representation Theory, not  its physical interpretation. 

An important special case is when
we have a class $\cs[A_1]$ with quiver $Q$
and another class $\cs[A_1]$ theory with quiver
$Q^\prime$, on which a group
$\mathbb{G}$ acts freely, such that $Q=Q^\prime/\mathbb{G}$.
The only reason why this is not a cover of
$\cs[A_1]$ categories is that, while the superpotential $\cw^\prime$ on $Q^\prime$ is
$\mathbb{G}$--invariant, its quotient is
not equal to $\cw$, and then
\begin{equation}
\C Q/(\partial \cw)\equiv \Big(\C Q^\prime/(\partial F^{\ast}\cw)\Big)\Big/\mathbb{G}  \neq \Big(\C Q^\prime/(\partial\cw^\prime)\Big)\Big/\mathbb{G}.
\end{equation}
From appendix \ref{proofpi}, 
 this happens when the corresponding
geometric cover of Gaiotto surfaces, $\xi\colon\Sigma_G^\prime\to\Sigma_G$, is \textit{branched over some \underline{regular} puncture of the quadratic differential
$\phi_2$. }

To compute the BPS spectrum of a class $\cs[A_1]$ model with regular punctures it is convenient to use the trick that we call \emph{gentling}
and review in appendix \ref{ap:gentling}. There the procedure is explained
from two different points of view: the 4d gauge theory one, and the surgery of bordered surfaces $\Sigma$ \cite{CV11}. Here we rephrase it in terms of Gaiotto's quadratic differentials $\phi_2$ on the UV curve $\Sigma_G$. The trick is to deform each
regular double pole of $\phi_2$ into a 
order three pole with an infinitesimal
coefficient
\begin{equation}
a_i\!\left(\frac{dz}{z-z_i}\right)^{\!\!2}+\text{less singular}\longrightarrow \left(a_i+\frac{\epsilon}{z-z_i}\right)\!\!\left(\frac{dz}{z-z_i}\right)^{\!\!2}+\text{less singular}
\end{equation}
One then computes the easier BPS spectrum of the $\C$--category associated with a surface with no regular
punctures. In the limit $\epsilon\to 0$ one almost gets back
the original BPS spectrum, except for $p$ vector multiplets
which, geometrically, correspond to minimal WKB geodesics going around each of the $p$ cubic punctures.

The effect of the gentling procedure
is to replace $\pi_1(\Sigma^\circ)$
with $\pi_1(\Sigma)$ where now we think of the punctures of $\Sigma$ as small holes.
Although this would not give us a covering of
BPS spectra in the strict sense, it would give us 
a computational tool which is practically equivalent to a \emph{bona fide} cover since the
discrepancy is just a known finite set
of BPS states which should be disregarded.  To distinguish these situations from the genuine Galois covers of QFTs, we shall call
them \emph{quiver covers}.

\subsubsection{Quotients of triangular factors
and invariant Lie subalgebras}

As discussed in \S.\ref{GIM-EALA}, to a
triangular algebra (or to a dashed category) $\sct$
we may associate a Slodowy GIM Lie algebra
$\mathfrak{L}_A$ through its Cartan matrix $A$. If $\mathbb{G}$ is an admissible group of automorphisms of $\sct$, it induces an automorphism group of the Dynkin bi--graph of 
$\mathfrak{L}_A$ and hence a group of outer automorphisms of the Lie algebra $\mathfrak{L}_A$.
To the quotient functor $F\colon \sct\to\sct/\mathbb{G}$ there corresponds a Lie algebra
embedding
\begin{equation}
\mathfrak{F}\colon \mathfrak{L}_A^\mathbb{G}\to \mathfrak{L}_A,\qquad\quad \mathfrak{L}_A^{\mathbb{G}}\equiv \mathfrak{L}_{A/\mathbb{G}},
\end{equation}
where $\mathfrak{L}_A^\mathbb{G}$ is the $\mathbb{G}$--invariant Lie subalgebra.
The Chevalley--Serre generators of $\mathfrak{L}_A^\mathbb{G}$
\begin{equation}
\big\{e_{\mathbb{G}i}, f_{\mathbb{G}i}, h_{\mathbb{G}i}\big\}
\end{equation} 
are labelled by the objects $\mathbb{G}i$ of $\mathscr{T}/\mathbb{G}$. Explicitly
\begin{equation}
e_{\mathbb{G}i}=\sum_{g\in\mathbb{G}}e_{gi},\qquad f_{\mathbb{G}i}=\sum_{g\in\mathbb{G}}f_{gi},
\qquad h_{\mathbb{G}i}=\sum_{g\in\mathbb{G}}h_{gi}.
\end{equation} 
It is easy to see that the Dynkin bi--graph
of $\mathfrak{L}_A^\mathbb{G}$ is the underlying bi--graph of the bi-quiver of $\mathscr{T}/\mathbb{G}$, i.e.\! $\mathfrak{L}_A^{\mathbb{G}}\equiv \mathfrak{L}_{A/\mathbb{G}}$. 

We note that when $\mathfrak{L}_A$ is
an extended affine Lie algebra of simply--laced type, and the quiver $\overline{Q}$ of the quotient category $\mathscr{T}/\mathbb{G}\equiv \C \overline{Q}/\overline{I}$ is $2$--acyclic,
$\mathfrak{L}^\mathbb{G}_A$ is also an
extended affine Lie algebra of simply--laced type.

\subsubsection{Pulled back central charge}\label{pullbackZ}
  Suppose we have a Galois cover $F\colon\ca\to\cb$ while on $\mathsf{mod}\,\cb$ we are given a central charge (stability function) i.e.\!
a group homomorphism\footnote{ As always
$K^0(\mathscr{A})$ stands for the Grothendieck group of the Abelian category $\mathscr{A}$. Given $X\in\mathscr{A}$ we write $[X]\in K^0(\mathscr{A})$ for its class.} $Z\colon K^0(\mathsf{mod}\,\cb)\to \C$ which maps the positive cone of
the Grothendieck group
 $K^0(\mathsf{mod}\,\cb)$ into the upper half--plane. By \textit{the pulled back central charge on 
 $\mathsf{mod}\,\ca$} we mean the
 group homomorphism $F^\lambda Z\colon K^0(\mathsf{mod}\,\ca)\to \C$ given by
 \begin{equation}
 F^\lambda Z\!\big([X]\big)= Z\!\big([F_\lambda X]\big).
 \end{equation}

\section{First examples and applications}

In this section we collect examples and some simple 
applications of the covering
techniques, as a warm up for more serious computations in later sections.\medskip

The simplest possible non--free $\cn=2$ systems are the Argyres--Douglas models. The corresponding 
Jacobian algebras are representation--finite, and the $\cn=2$ models all whose Jacobian algebras are representation--finite are Argyres--Douglas models \cite{CV11}.
We have the following result
\smallskip

\textbf{Fact} \cite{galsurvery}. \textit{Let $F\colon \ca\to\cb$ be a Galois cover of \underline{bounded} $\C$--linear categories with Galois group $\mathbb{G}$.
Then $\cb$ is representation--finite if and only if
\emph{(i)} $\mathbb{G}$ acts freely on the isoclasses of indecomposables and \emph{(ii)}
$\ca$ is representation--finite. In that case
$F_\lambda$ is surjective. }
\medskip

In other words, if an Argyres--Douglas theory has a (finite) covering $\cn=2$ theory, that theory should also be an Argyres--Douglas model. From \S.\ref{s:versus}, we know that the possible complete covers of
Argyres--Douglas models (of type $A,D$)
are controlled by the fundamental group of the
corresponding surface, $\pi_1(\Sigma_\mathrm{AD}^\circ)$. Since, $\Sigma^\circ_\mathrm{AD}$ is a disk (with marks), these Argyres--Douglas theories do not have any non--trivial\footnote{ A \emph{trivial} cover is one in which the covering theory consists of several
non--interacting copies of the quotient theory.
An example of trivial cover is the Galois cover of Argyres--Douglas theories
$\C D_2\to \C A_1$, i.e.\! two free hypers
doubly covering one free hyper. See appendix \ref{proofpi} for a list of such trivial exceptions.} QFT covers. (Of course, for type $D_p$ we have quiver covers in the sense of \S.\ref{quivercovers}; this is a very well known example of cover of $\C$--categories discussed at length, say, in ref.\!\cite{gal5},  or \cite{assem1} Chapter V).

The next simpler class of theories are the
$\cn=2$ gauge theories with gauge group $SU(2)$.
These models do have interesting Galois covers.

\subsection{$\Z_k$ covers in $SU(2)$ SQCD}

Let us start with our motivating example $N_f=2$ SQCD with $G_\mathrm{gauge}=SU(2)$ as a double cover of
pure SYM. The quivers of the two theories
are, respectively, the affine quivers of type $\widehat{A}(2,2)$ and $\widehat{A}(1,1)$ with
alternating orientation\footnote{ See \cite{CV11} for our nomenclature about quivers.
Since affine quivers are acyclic, 
the superpotential is zero.}
\begin{equation}\label{quileftright}
\begin{gathered}
\xymatrix{& \bullet_1\\
\circ_1\ar[ur]\ar[dr] && \circ_2\ar[ul]\ar[dl]\\
& \bullet_2}\\
\widehat{A}(2,2)
\end{gathered}\hskip 3cm
\begin{gathered}
\xymatrix{\bullet\\
\\
\circ\ar@<0.8ex>@/^2.5pc/[uu]\ar@<-0.8ex>@/_2.5pc/[uu]}\\ 
\widehat{A}(1,1)
\end{gathered}
\end{equation}
The quiver on the left has a freely acting $\Z_2$ automorphism
which  interchanges the
nodes of the same color (black/white).
From the figure, it is clear that its orbit category $\C \widehat{A}(2,2)/\Z_2$ is the category of the quiver on the right, $\C \widehat{A}(1,1)$.
More generally, let $p,q$ be two positive integers,
and choose $k\mid \gcd(p,q)$. We have
the $\Z_k$ Galois cover
\begin{equation}\label{kcoversu2}
\C\widehat{A}(p,q)\longrightarrow \C\widehat{A}(p,q)/\Z_k \equiv \C \widehat{A}(p/k,q/k)
\end{equation}
(the orientation of the arrows is chosen 
in the obvious way to be consistent with the quotient). In this case, the covering $\cn=2$ theory is $SU(2)$ SYM coupled to two Argyres--Douglas systems of types $D_p$ and $D_q$, while the quotient theory is $SU(2)$ SYM coupled to two AD systems\footnote{ In our conventions \cite{CV11}, $D_1$ is the empty theory, and $D_2$ a doublet of free quarks. } of types $D_{p/k}$ and $D_{q/k}$
\cite{CV11}.
 Choosing $(p/k,q/k)=(1,1),(2,1),$ and $(2,2)$
the quotient theory is, respectively, $SU(2)$
SQCD with $N_f=0,1,2$ fundamental quarks.

Hence $SU(2)$ SQCD coupled to at most two flavors (or, more generally, at most two Argyres--Douglas systems) admit $\Z_k$ Galois covers for all $k\in\mathbb{N}$. These are their only covers by complete $\cn=2$ QFT
(in fact, by any $\cn=2$ QFT)
since, for all these models, 
$\Sigma\equiv\Sigma^\circ=\C^\times$,
and $\pi_1(\Sigma)\equiv\pi_1(\Sigma^\circ)=\Z$.

For later reference, we describe the covering in terms of the corresponding 2d (2,2) theory \cite{CNV,CV11}. 4d $SU(2)$ SYM coupled to $D_p$ and $D_q$ AD systems corresponds to the (2,2)
Landau--Ginzburg model
\begin{equation}\label{exppexpq}
W(Z)=e^{p Z}+e^{-q Z},\qquad \text{with the identification }Z\sim Z+2\pi i.
\end{equation}
If $k\mid \gcd(p,q)$, the (2,2) model 
corresponding to the quotient theory
$SU(2)$ SYM with $D_{p/k}$ and $D_{q/k}$ AD
is the same Landau--Ginzburg model \eqref{exppexpq} but with the rescaled field identification
\begin{equation}\label{exppexpq2}
Z\sim Z+2\pi i/k,
\end{equation}
which gives a cover in the $tt^*$ sense \cite{ttstar,CV92}.

In terms of Lie algebras $\mathfrak{L}_A$, the $\Z_k$ cover \eqref{kcoversu2} becomes
the embedding of affine Kac--Moody algebras
\begin{equation}
A^{(1)}_{p/k+q/k-1} \equiv \big(A^{(1)}_{p+q-1}\big)^{\Z_k}\longrightarrow  A^{(1)}_{p+q-1}.
\end{equation}
\smallskip

$N_f=3$ SQCD (or more generally, $SU(2)$ SYM
with two fundamental quarks and a $D_p$ AD)
corresponds to the disk $D$ with two punctures and
2 marks (resp.\! $p$ marks) on the boundary.
We have $\pi_1(D^\circ)=(1)$, and hence its $\C$--category has no strict--sense cover by the $\C$--category of a $\cn=2$ QFT. Of course, 
since $\pi_1(D)=\langle x, y\rangle$, there are plenty of quiver covers in the sense of \S.\ref{quivercovers}. 
Likewise, $N_f=4$ SQCD has only covers in the quiver sense; they are
useful computational tools.

\subsubsection{$SU(2)$ SQCD BPS spectra
as Galois covers}\label{su2bpscov}

We look at the maps between the BPS spectra of $SU(2)$ SYM coupled to different pairs of AD systems which are 
induced by the $\Z_k$ covering functors
$F^\lambda$, $F_\lambda$ of \S.\ref{sec:coveringfunctors}.
The algebras associated to these models,
$\C\widehat{A}(p,q)$, are hereditary (hence
triangular with $q(\boldsymbol{x})$ a Kac--Moody Tits form),
string, and class $\cs[A_1]$, so all three
special techniques of \S\S.\,\ref{special cases}, \ref{coversofsa1} apply.

An acyclic affine quiver corresponds to an asymptotically--free $SU(2)$ gauge theory \cite{CV11}. The $SU(2)$
magnetic charge $m(X)$
of the BPS particle described by the module $X$
is equal to the Ringel defect of the module \cite{cattoy},
i.e.\! to the sum of the dimensions at the source nodes (the white ones) minus the sum of the dimensions at the sinks (the black nodes), cfr.\! eqn.\eqref{defect}.

The BPS spectrum of a complete theory contains only hypermultiplets (corresponding to rigid bricks of the algebra) and vector multiplets ($\mathbb{P}^1$ families of bricks with self--extensions).
In any chamber, an asymptotically--free $SU(2)$ gauge theory has at most \emph{one} BPS vector
multiplet, the $W$ boson; in other words,
the associated algebra $\mathscr{J}$ has a unique $\mathbb{P}^1$--family of non--rigid bricks $W_\mu$ ($\mu\in\mathbb{P}^1$). Being unique, this family is fixed by all
automorphisms of $\mathscr{J}$, i.e.\! $g\, W_\mu\cong W_\mu$
for all $g\in\Z_k$.
From \S\S.\,\ref{special cases}, \ref{coversofsa1}
it follows that the $W$ boson modules of the covering theory, $\widetilde{W}_\mu$,
are the pull back of the ones for the quotient theory,
$W_\mu$,
\begin{equation}
\widetilde{W}_{\mu^k}=F^\lambda W_\mu.
\end{equation}
Since $F_\lambda$ is adjoint to $F^\lambda$,
this implies that the push down functor $F^\lambda$
preserves the magnetic charge of the module
\begin{equation}
m(F_\lambda X)=m(X).
\end{equation}
Bricks with $m(X)\neq 0$ are rigid and
cannot be in the closure of the $W$ family (since the $W$ boson has zero magnetic charge).
By \S\S.\,\ref{special cases}, \ref{coversofsa1}
this implies that $F_\lambda$ sets a
$k$--to--1 correspondence between the dyons of
the covering theory and the dyons of the quotient theory. This is our $k$--fold cover between dyon towers; the example $N_f=2\to N_f=0$ SQCD  corresponds to $k=2$.

It remains to consider the rigid bricks of the covering theory
with $m(X)=0$ which correspond to the (finitely many) BPS states of the covering matter system \cite{cattoy}: in SQCD
they are just the quark states.
Let $X\in\mathsf{mod}\,\C\widehat{A}(p,q)$ be such a brick. By \S\S.\,\ref{special cases}, \ref{coversofsa1}, $F_\lambda X\in\mathsf{mod}\,\C\widehat{A}(p/k,q/k)$ is indecomposable, but not necessarily a brick nor rigid. If $m(X)=0$,
$F_\lambda X$ is a rigid brick of the quotient category,
if and only if $\mathbf{dim}\,F_\lambda X<\delta\equiv \mathbf{dim}\, W_\mu$ (for $k=2$
this is equivalent to $\mathbf{dim}\,F_\lambda X\neq\delta$). 

To the push down functor $F_\lambda$, restricted to the rigid bricks having vanishing magnetic charge,
we may give a different interpretation.
These bricks describe the BPS states of the covering matter $D_p\oplus D_q$ Argyres--Douglas system, 
and the Galois cover $F\colon \C \widehat{A}(p,q)\to \C\widehat{A}(p/k,q/k)$ induces a `covering'
of the respective matter systems
\begin{equation}
D_p\oplus D_q\to D_{p/k}\oplus D_{q/k}
\end{equation}
which is not a strict--sense cover of $\cs[A_1]$ categories (since, as shown at the beginning of this
section, there are none) but merely a quiver cover in the sense of  \S.\ref{quivercovers}. 
Then the would be push down $f_\lambda$
maps bricks of the covering matter category
into representations of the quotient matter
quiver which may or may not satisfy the
Jacobian relations $\partial\cw=0$. Only those which satisfy the relations are  actual BPS states of the quotient matter sector; they are precisely the ones with $\mathbf{dim}\,f_\lambda X<\delta$, in agreement with the previous result.

Since (cfr.\! \S\S.\,\ref{special cases}, \ref{coversofsa1}) $F_\lambda$ is onto
the rigid bricks of the quotient category,
the full BPS spectral correspondence is
\begin{equation}\label{ttable20}
\begin{tabular}{ccc@{\hspace{18pt}}c}\hline\hline
&covering model & quotient model & \\\hline
vector multiplet & $F^\lambda W_\mu$ & $W_\mu$ & 1--to--$k$\\
dyonic hypers & $X$ & $F_\lambda X$ & $k$--to--1\\
$m=0$ hypers & $X$ & $F_\lambda X$,\ 
if $\mathbf{dim}\,F_\lambda X<\delta$ & $k$--to--1
 \\\hline\hline
\end{tabular}
\end{equation}
More precisely, if the covering model is
equipped with the pull back $F^\lambda Z$ central charge, the stable particles of one column
of table \eqref{ttable20} map
in the corresponding stable particles of the other column. In the maximal chamber \cite{ACCERV1}
all states in the table are stable, and the correspondence is fully realized.

For instance, in the $N_f=2\to N_f=0$ cover, 
no rigid brick $X$ with $m(X)=0$ of the covering
theory satisfies the condition $\mathbf{dim}\,F_\lambda X<\delta$, and the BPS spectrum of the quotient theory contains only the $W$ boson and
the dyonic towers. Of course, this is the right result.

\subsection{$\Z_k$ covers of pure SYM with simply--laced gauge groups}\label{kkk12qn}

The above construction of $\Z_k$ Galois covers of pure $SU(2)$ SYM may be extended directly to
$\Z_k$ covers of pure SYM with any simply--laced gauge group $G$.
This is basically the idea of \cite{infty,CDZG}. Let us review their construction in the language of $tt^*$ covers for the corresponding 2d (2,2) model as in eqns.\eqref{exppexpq}--\eqref{exppexpq2}. The (2,2) model corresponding to SYM with gauge group $G$ is \cite{infty,tack}
\begin{equation}\label{lglgas}
W(Z,X,Y)=e^Z+e^{-Z}+W_G(X,Y), \qquad
Z\sim Z+2\pi i,
\end{equation}
where the polynomial $W_G(X,Y)$ is the $ADE$ minimal singularity corresponding to the simply--laced
Lie group $G$. The $tt^*$ $\Z_k$--Galois cover
is obtained by replacing $Z\mapsto k\,Z$
in eqn.\eqref{lglgas} while keeping $Z\sim Z+2\pi i$
\cite{infty,CDZG}.

\subsubsection{Quivers and their automorphisms}
In the mutation class of quivers with superpotentials for the $\Z_k$ cover of SYM with gauge group $G$ there are  two preferred `tensor product' quivers \cite{keller}
\begin{equation}
\text{triangle tensor product:}\ \widehat{A}(k,k)\boxtimes G,\qquad \text{square tensor product:}\ \widehat{A}(k,k)\,\square\, G,
\end{equation}
where $G$ stands for the Dynkin quiver of the gauge group with an alternating orientation
so that odd numbered nodes are sources and
even numbered ones are sinks.
The affine quiver $\widehat{A}(k,k)$ is also taken
in the alternating orientation. These orientations are chosen for convenience; all orientations are mutation equivalent, and hence describe the same physics. Both product quivers have a `projection' 
$\mathsf{p}$ on the second factor, i.e.\! on the Dynkin quiver $G$; the inverse image of a simple root of $G$, $\mathsf{p}^{-1}(\alpha_i)$, is a
full affine subquiver $\widehat{A}(k,k)$. For both forms of the quiver we label the nodes
as $\bullet_{a,i}$,
$\circ_{a,i}$ with $i\in G$ and $a\in\Z_k$ (the index $a$ being defined mod $k$).
Nodes $\bullet_{a,i}$ (resp.\! $\circ_{a,i}$)
are the sinks (sources) of the $i$--th affine subquiver $\widehat{A}(k,k)$.
The exchange matrix
of the square quiver is
\begin{equation}
B_{\bullet_{a,i},\bullet_{b,j}}=B_{\circ_{a,i},\circ_{b,j}}=0,\qquad B_{\circ_{a,i},\bullet_{b,j}}=\big(\delta_{a,b+1}-\delta_{ab}\big)\delta_{ij}+C_{ij}\,\delta_{ab}
\end{equation}
where $C$ is the Cartan matrix of $G$. The superpotential $\cw_\square$ of the $\widehat{A}(k,k)\,\square\,G$ quiver is the sum of all its oriented squares.
The exchange matrix of the triangle form is \cite{cattoy}
\begin{equation}
B_{\widehat{A}(k,k)\boxtimes G}=S_{\widehat{A}}\otimes S_G-S_{\widehat{A}}^t\otimes S_G^t
\end{equation}
where $B_{\widehat{A}}(k,k)$ (resp.\! $S_G$)
is the `Stokes' matrix\footnote{ In the math literature \cite{ringel,assem1} $S$ is called the inverse of the Cartan matrix of the algebra; we avoid that terminology since it is very confusing in the present context.}, i.e.\! the matrix such that the quiver exchange matrix and the
Tits matrix of the triangular algebra $\C\widehat{A}(k,k)$ (resp.\!\! $\C G$) are given by $B_{\widehat{A}}=S_{\widehat{A}}-S_{\widehat{A}}^t$ and $A_{\widehat{A}}=S_{\widehat{A}}+S_{\widehat{A}}^t$ (resp.\!\!
$B_{G}=S_{G}-S_{G}^t$ and $A_{G}=S_{G}+S_{G}^t$).
 The superpotential $\cw_\triangle$ of the $\widehat{A}(k,k)\boxtimes G$ quiver is a certain sum over its oriented triangles.
 $\cn=2$ super--Yang--Mills with simply laced gauge group $G$ corresponds to the case $k=1$ of the above
  construction \cite{CNV,ACCERV2,cattoy}.
  
Clearly both forms of the quiver,  $\widehat{A}(k,k)\boxtimes G$ and $\widehat{A}(k,k)\,\square\,G$,
have a $\Z_k$--automorphism group, inherited from the
first factor in the tensor product, $\widehat{A}(k,k)$,
which acts freely on the objects
\begin{equation}
\Z_k\ni s\colon (\bullet_{a,i},\circ_{a,i})\longmapsto 
(\bullet_{a+s,i},\circ_{a+s,i}).
\end{equation}
Since this $\Z_k$ action leaves invariant the superpotentials, $\cw_\triangle$ and $\cw_\square$, these quiver automorphisms extend to automorphisms of the Jacobian categories
\begin{equation}
\mathscr{J}_G^\triangle(k,k)=\C\widehat{A}(k,k)\boxtimes G\big/(\partial\cw_\triangle),\qquad 
\mathscr{J}_G^\square(k)=\C\widehat{A}(k,k)\,\square\, G\big/(\partial\cw_\square).
\end{equation}
 Hence, for all $\ell\mid k$ we may form 
 Galois covers
 \begin{align}
 F_\triangle\colon \mathscr{J}_G^\triangle(k,k)&\longrightarrow \mathscr{J}_G^\triangle(k,k)\big/\Z_\ell\cong \mathscr{J}_G^\triangle(k/\ell,k/\ell)\\
 F_\square\colon \mathscr{J}_G^\square(k)&\longrightarrow \mathscr{J}_G^\square(k)\big/\Z_\ell\cong \mathscr{J}_G^\square(k/\ell)
 \end{align}
 whose quotient categories have the same `tensor product' form with $k$ replaced by $k/\ell$, as it is
 obvious from the geometric description of the cover around eqn.\eqref{lglgas}. Then both the cover and quotient categories correspond to well defined $\cn=2$ QFTs.
 
 The triangle construction may be further generalized to
 tensor products of the form $\widehat{A}(p,q)\boxtimes G$ with $p\neq q$. Again, if
 $\ell\mid\gcd(p,q)$, we have a Galois covering
 \begin{equation}
 F_\triangle\colon \mathscr{J}_G^\triangle(p,q)\longrightarrow \mathscr{J}_G^\triangle(p,q)\big/\Z_\ell\cong \mathscr{J}_G^\triangle(p/\ell,q/\ell).
 \end{equation}
 
  \begin{figure}
 \begin{equation*}
 \begin{xy} 0;<1pt,0pt>:<0pt,-1pt>:: 
(59,68) *+{\circ_{1,2}} ="0",
(40,48) *+{\circ_{2,2}} ="1",
(0,68) *+{\bullet_{1,2}} ="2",
(99,48) *+{\bullet_{2,2}} ="3",
(59,20) *+{\bullet_{1,1}} ="4",
(0,20) *+{\circ_{1,1}} ="5",
(99,0) *+{\circ_{2,1}} ="6",
(40,0) *+{\bullet_{2,1}} ="7",
"0", {\ar"2"},
"0", {\ar"3"},
"4", {\ar"0"},
"1", {\ar"2"},
"1", {\ar"3"},
"7", {\ar"1"},
"2", {\ar"5"},
"3", {\ar"6"},
"5", {\ar"4"},
"6", {\ar"4"},
"5", {\ar"7"},
"6", {\ar"7"},
\end{xy}
 \end{equation*}
 \caption{\label{klein}The quiver $\widehat{A}(2,2)\,\square\,A_2$: a cube with oriented squares on the four lateral faces.}
 \end{figure}
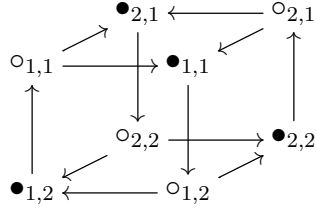
 
The square category $\mathscr{J}_G^\square(k)$ may have a larger group of automorphisms acting freely on the objects. The simplest instance is
$\widehat{A}(2,2)\,\square\,A_2$ whose quiver
is shown in figure \ref{klein}. The freely acting automorphisms of this category form
the Klein Vierergruppe $V_4\simeq \Z_2\times \Z_2$
\begin{equation}
(\bullet_{1,i},\circ_{1,i})\leftrightarrow
(\bullet_{2,i},\circ_{2,i})\quad \text{and}\quad
(\bullet_{a,1},\circ_{a,1})\leftrightarrow
(\bullet_{a,2},\circ_{a,2}).
\end{equation}
The quotient of $\mathscr{J}_{A_2}^\square(2)$ with respect to the full $V_4$ group has a non--$2$--acyclic quiver. So, while such quotients by enhanced automorphism groups may be quite useful to actually compute the BPS spectrum at special points in parameter space which preserve such enhanced symmetry, they do not describe
Galois covers of BPS spectra of $\cn=2$ QFTs.

 \subsubsection{Covers of pure SYM from Minahan--Nemeshanski SCFT}

 As explained in detail in \cite{infty,CDZG},
 the category $\mathscr{J}^\triangle_G(p,q)$
 corresponds to $\cn=2$ SYM with group $G$
 which gauges
 the diagonal flavor (sub)group $G$ of
 a pair of $\cn=2$ SCFTs of types $D_p(G)$
 and $D_q(G)$. In the notations of \cite{infty,CDZG},
 $D_1(G)$ is the trivial SCFT. When the `matter' sector consists of two copies of the \emph{same} SCFT $D_p(G)$,
it is more convenient to use the mutation 
equivalent square category
 $\mathscr{J}^\square_G(p)$.

 If $k\mid\gcd(p,q)$, the Galois cover 
 \begin{equation}\label{ttthesecorv}
 F\colon \mathscr{J}^\triangle_G(p,q)\to
 \mathscr{J}^\triangle_G(p/k,q/k)
 \end{equation}
  then corresponds to replacing the `matter' sector
 $D_p(G)\oplus D_p(G)$ with 
 $D_{p/k}(G)\oplus D_{q/k}(G)$.

   \begin{figure}
\begin{gather*}
\xymatrix{\fbox{\text{ $\phantom{\Bigg|}E_6\phantom{\Bigg|}$ MN} }\ar@{-}[rr] 
&& *++[o][F-]{ \phantom{\Bigg|}SO(8)\phantom{\Bigg|} }\ar@{-}[rr] &&\fbox{\text{ $\phantom{\Bigg|}E_6\phantom{\Bigg|}$ MN} }}\\
\text{$\cn=2$ theory corresponding to the $\C$--category }\mathscr{J}^\square_{SO(8)}(2)
\end{gather*}
\begin{gather*}
\xymatrix{*++[o][F-]{ \phantom{\Bigg|}SU(2))\phantom{\Bigg|} }\ar@{-}[r] &\fbox{\text{ $\phantom{\Bigg|}E_7\phantom{\Bigg|}$ MN} }\ar@{-}[r] 
& *++[o][F-]{ \phantom{\Bigg|}SO(10)\phantom{\Bigg|} }\ar@{-}[r] &\fbox{\text{ $\phantom{\Bigg|}E_7\phantom{\Bigg|}$ MN} }
\ar@{-}[r]& *++[o][F-]{ \phantom{\Bigg|}SU(2)\phantom{\Bigg|} }}\\
\text{$\cn=2$ theory corresponding to the $\C$--category }\mathscr{J}^\square_{SO(10)}(2)\end{gather*}
\begin{gather*}
\xymatrix{*++[o][F-]{ \phantom{\Bigg|}SU(3))\phantom{\Bigg|} }\ar@{-}[r] &\fbox{\text{ $\phantom{\Bigg|}E_8\phantom{\Bigg|}$ MN} }\ar@{-}[r] 
& *++[o][F-]{ \phantom{\Bigg|}\;\;E_6\;\;\phantom{\Bigg|} }\ar@{-}[r] &\fbox{\text{ $\phantom{\Bigg|}E_8\phantom{\Bigg|}$ MN} }
\ar@{-}[r]& *++[o][F-]{ \phantom{\Bigg|}SU(3)\phantom{\Bigg|} }}\\
\text{$\cn=2$ theory corresponding to the $\C$--category }\mathscr{J}^\square_{E_6}(2)\end{gather*}
\caption{\label{MNcorX}The $\cn=2$ theories whose BPS spectra double covers that of pure $\cn=2$ SYM with gauge groups $G=SO(8)$, $SO(10)$, and $E_6$. The circles stands for $\cn=2$ SYM sectors, the rectangles for Minahan--Nemeshansky $E_r$ SCFTs.}
\end{figure}
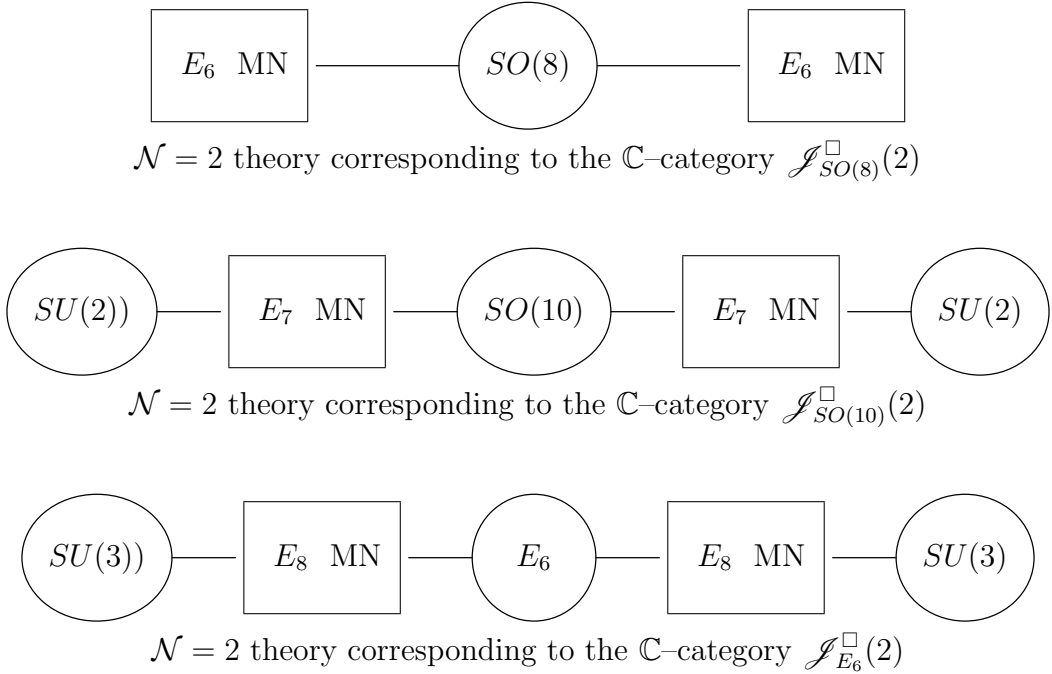

 In particular, if $p=q=k$, we get a $\Z_k$--cover of the category corresponding to \emph{pure} SYM with gauge group $G$, by the same gauge theory coupled to two copies
 of $D_k(G)$. If $G=SU(2)$ and $k=2$, this reduces to the GMN double cover of $SU(2)$ SYM by the same theory coupled to two fundamental quarks. So the covers \eqref{ttthesecorv} are direct generalizations of our motivating example.

 A more sophisticated example is obtained by replacing $SU(2)$ SYM by $SO(8)$ SYM 
and the two doublets of free quarks, $D_2(SU(2))$,
by their $SO(8)$ counterpart $D_2(SO(8))$,
which is \cite{infty,CDZG} the 
$E_6$
 Minahan--Nemeshanski model \cite{MN1}.
Thus the $\cn=2$ theory obtained by gauging 
the $SO(8)_\mathrm{diag}\subset E_6\times E_6$
 flavor symmetry of two copies of  the 
 $E_6$
 Minahan--Nemeshanski SCFT
corresponds to the category $\mathscr{J}^\square_{SO(8)}(2)$, which is a $\Z_2$ Galois cover of the pure $SO(8)$ SYM category
$\mathscr{J}^\square_{SO(8)}(1)$.

Likewise, for $p=q=k=2$ and gauge groups
$G=SO(10)$ and $G=E_6$ we get `matter'
systems involving the Minahan--Nemeshanski theories with flavor symmetry, respectively,
$E_7$ and $E_8$ \cite{MN1,MN2} together with some further SYM sectors, see figure \ref{MNcorX}.
Analogously, the double cover of the
$SU(3)$ SYM $\C$--category, $\mathscr{J}^\square_{SU(3)}(2)$, corresponds to  SYM
gauging the diagonal flavor symmetry of two copies of the Argyres--Douglas system of type $D_4$
\cite{infty,CDZG}.

\subsubsection{Galois covers between $SU$ and $USp$/$SO$ linear quiver theories}

\begin{figure}
\begin{scriptsize}
\begin{equation*}
\xymatrix{*++[o][F-]{\phantom{\Bigg|}SU(m)\phantom{\Bigg|}}\ar@{-}[r] &
*++[o][F-]{\phantom{\Bigg|}SU(2m)\phantom{\Bigg|}}\ar@{-}[r] &
*++[o][F-]{\phantom{\Bigg|}SU(3m)\phantom{\Bigg|}}\ar@{-}[r]&\cdots\cdots\cdots\ar@{-}[r]&
*++[o][F-]{\phantom{\Bigg|}SU(N-m)\phantom{\Bigg|}}\ar@{-}[r] & *++[o][F=]{\phantom{\Bigg|}SU(N)\phantom{\Bigg|}}\\
*++[o][F-]{\phantom{\Bigg|}SU(n)\phantom{\Bigg|}}\ar@{-}[r] &
*++[o][F-]{\phantom{\Bigg|}SU(2n)\phantom{\Bigg|}}\ar@{-}[r] &
*++[o][F-]{\phantom{\Bigg|}SU(3n)\phantom{\Bigg|}}\ar@{-}[r] &\cdots\cdots\cdots\ar@{-}[r]
 &
*++[o][F-]{\phantom{\Bigg|}SU(N-n)\phantom{\Bigg|}}\ar@{-}@/_2pc/[ru]&}
\end{equation*}
\vskip 18pt
\caption{\label{SUlin}The $SU$ linear quiver gauge theories
which correspond to bounded $\C$--categories of the
$\mathscr{J}^\triangle_{SU(N)}(p,q)$ class. $m,n$ are divisors of $N$, and $p=N/m$, $q=N/n$. Circles denote $\cn=2$ SYM sectors and links
bifundamental hypermultiplets. }
\begin{equation*}
\xymatrix{*++[o][F-]{\phantom{\Bigg|}SU(mk)\phantom{\Bigg|}}\ar@{-}[r] &
*++[o][F-]{\phantom{\Bigg|}SU(2mk)\phantom{\Bigg|}}\ar@{-}[r] &
*++[o][F-]{\phantom{\Bigg|}SU(3mk)\phantom{\Bigg|}}\ar@{-}[r]&\cdots\cdots\cdots\ar@{-}[r]&
*++[o][F-]{\phantom{\Bigg|}SU(N-mk)\phantom{\Bigg|}}\ar@{-}[r] & *++[o][F=]{\phantom{\Bigg|}SU(N)\phantom{\Bigg|}}\\
*++[o][F-]{\phantom{\Bigg|}SU(nk)\phantom{\Bigg|}}\ar@{-}[r] &
*++[o][F-]{\phantom{\Bigg|}SU(2nk)\phantom{\Bigg|}}\ar@{-}[r] &
*++[o][F-]{\phantom{\Bigg|}SU(3nk)\phantom{\Bigg|}}\ar@{-}[r] &\cdots\cdots\cdots\ar@{-}[r]
 &
*++[o][F-]{\phantom{\Bigg|}SU(N-nk)\phantom{\Bigg|}}\ar@{-}@/_2pc/[ru]&}
\end{equation*}
\end{scriptsize}
\caption{The $\Z_k$ Galois  quotient of the quiver gauge theory in figure \ref{SUlin}.\label{linquot}  }
\end{figure}

The linear quiver $\cn=2$ gauge theories in figure
\ref{SUlin},
 where $m$, $n$, are divisors of $N$,
have BPS quivers in the mutation class
of $\widehat{A}(N/m,N/n)\boxtimes SU(N)$.
The beta--functions of all YM couplings vanish,
but for the one corresponding to the
$SU(N)$ group (the one in the double--line
circle) which is just asymptotically free.
Indeed the two `matter' sectors, $D_{N/m}(SU(N))$
and $D_{N/n}(SU(N))$, correspond to the 
two disconnected quiver theories one obtains by
deleting the asymptotically--free double--lined gauge node \cite{CDZG}.

In the same fashon, if $m,n$ are divisors of $N-1$,
the matter systems $D_{(N-1)/m}(SO(2N))$,
$D_{(N-1)/n}(SO(2N))$ correspond to
$USp/SO$ linear quivers \cite{tack2}, see \cite{CDZG}.
These quivers have different forms,
depending on the parity of the two integers
$(N-1)/m$
and $(N-1)/n$, see ref.\!\cite{CDZG}
for details. 

If there is an integer
$k$ such that $mk, nk\mid N$, the $\C$--category $\mathscr{J}^\triangle_{SU(N)}(N/m,N/n)$
associated to the $\cn=2$ theory in figure \ref{SUlin}
has a $\Z_k$ Galois quotient which corresponds to a linear quiver model of the same kind, see figure \ref{linquot}.
The $USp/SO$ linear quivers have similar $\Z_k$
Galois covers, provided $mk$ and $nk$
both divide $(N-1)$.

It is remarkable that
pairs of linear quiver gauge theories --- as the pair in figures \ref{SUlin} and \ref{linquot} --- which look
very different, have BPS spectra which (in $\Z_k$ symmetric chambers) are
simply related \emph{via} the $\Z_k$ covering functors $F_\lambda$
and $F^\lambda$.

\subsubsection{Galois covers of BPS spectra of SYM} \label{rrrrr6519}

The Galois cover of $\C$--categories
 \begin{equation}
 \mathscr{J}^\triangle_G(p,q)\to
\mathscr{J}^\triangle_G(p/k,q/k)
\end{equation} 
induces covering functors $F^\lambda$, $F_\lambda$ which relate the BPS spectra
of two $\cn=2$ theories with the same gauge group $G$ but different `matter' content, at
$\Z_k$ symmetric points in the parameter space of the covering theory.

\paragraph{Finite chambers.} Pure SYM with simply--laced gauge group $G$
has a strongly--coupled finite chamber
\cite{ACCERV2} whose spectrum consists
of two hypermultiplet dyons per positive root of $G$.
In facts, under the identification of the
lattice of conserved charges for pure SYM
with $\Gamma_{A^{(1)}_1}\otimes \Gamma_G$, where by $\Gamma_G$ (resp.\! $\Gamma_{A^{(1)}_1}$) we mean the root lattice of the Lie algebra
$G$ (resp.\! $A^{(1)}_1$), the charge vectors
of the stable dyons  at strong YM coupling
 take the form
\begin{equation}
\alpha_a\otimes \beta\in \Gamma_{A^{(1)}_1}\otimes \Gamma_G\qquad\quad\begin{aligned}&\alpha_a\
\text{is a simple root of }A^{(1)}_1\\ 
&\beta\ \text{a positive root of }G.
\end{aligned}
\end{equation}
More generally, the strong coupling BPS spectrum of $G$ SYM coupled to $D_p(G)$ and $D_q(G)$
contains hypermultiplets whose charge vectors make $(p+q)$ copies of the positive roots
of $G$
\begin{equation}
\alpha_a\otimes \beta\in \Gamma_{A^{(1)}_{p+q-1}}\otimes \Gamma_G\qquad\quad\begin{aligned}&\alpha_a\
\text{is a simple root of }A^{(1)}_{p+q-1}\\ 
&\beta\ \text{a positive root of }G.
\end{aligned}
\end{equation} 
Let $k\mid\gcd(p,q)$ and consider the 
Galois group $\Z_k$ acting on $\mathscr{J}^\triangle(p,q)$. The Galois group acts freely
on the generators $\alpha_a\otimes \beta_i$ of the lattice
$\Gamma_{A^{(1)}_{p+q-1}}\otimes \Gamma_G$
so that the push down functor 
$F_\lambda\colon\mathsf{mod}\mathscr{J}^\triangle_G(p,q)\to\mathsf{mod}\mathscr{J}^\triangle_G(p/k,q/k)$ send stable bricks into stable bricks, and gives a $k$--to--1
correspondence of the strong coupling spectra of the two theories. This result may be seen as a direct consequence of \textbf{Fact 3}.

\paragraph{The `dual' weak coupling chamber.}
The models with a square product quiver $G^\prime\,\square\,G$  have a `dual' (in the sense of \cite{CNV}) weak coupling chamber which has
 a very simple BPS spectrum. This remains true
 even if the Dynkin graph $G^\prime$ is affine
 as in the present class of theories.
To get this `dual' chamber for SYM coupled to
$D_p(G)$, $D_q(G)$  AD one has to fine--tune the  parameters to exceptional alignments in the $Z$--plane; hence the physical relevance 
of the resulting spectrum may be questioned. Mathematically, however, it makes sense, and the BPS spectrum consists of hypers and 
 vector multiplets only with charge vectors 
\begin{equation}
\alpha\otimes \beta_i\in \Gamma_{A^{(1)}_{p+q-1}}\otimes \Gamma_G\qquad\begin{aligned}&\alpha\ \text{is }\begin{cases}
\text{a  real Schur root of }A^{(1)}_{p+q-1}\ \text{for hypers}\\ 
\text{imaginary Schur root of }A^{(1)}_{p+q-1}\ \text{for vectors}
\end{cases}\\
&\beta_i\ \text{is a simple root of }G.
\end{aligned}
\end{equation} 
Then BPS spectrum is $r(G)$ copies of 
the spectrum of $SU(2)$ SYM with $D_p\oplus D_q$ matter at weak coupling, one copy per simple root of $G$. The spectral cover given by $F_\lambda$
works exactly as in \S.\,\ref{su2bpscov} copy by copy.
Again the idea that the cover of the BPS spectrum is realized chamber by chamber is confirmed.

\paragraph{More general chambers.} The covering of BPS spectra works smoothly in all chambers. The details of the spectra of $\Z_k$ covers of SYM
will be presented elsewhere.

\subsection{Abelian Galois covers of class $\cs[A_1]$ models}

\paragraph{Genus $>1$ curves.} Suppose we have a Gaiotto UV curve $\Sigma_G$ with genus $g>1$ and $p$ regular punctures such that
\begin{equation}
k\mid (g-1),\qquad k\mid p,
\end{equation}
 which has a freely acting $\Z_k$ automorphism group
which permutes the punctures. The
quotient $\Sigma_G^\prime\equiv \Sigma_G/\Z_k$ is a Gaiotto surface
with $(g^\prime-1)=(g-1)/k$ and $p^\prime=p/k$.
The cover $\Sigma_G\to \Sigma_G^\prime$
is \emph{unbranched} (i.e.\! the corresponding $\cn=2$ gauge theories are coupled only to half--trifundamentals and bi--fundamentals, but
no AD system).   

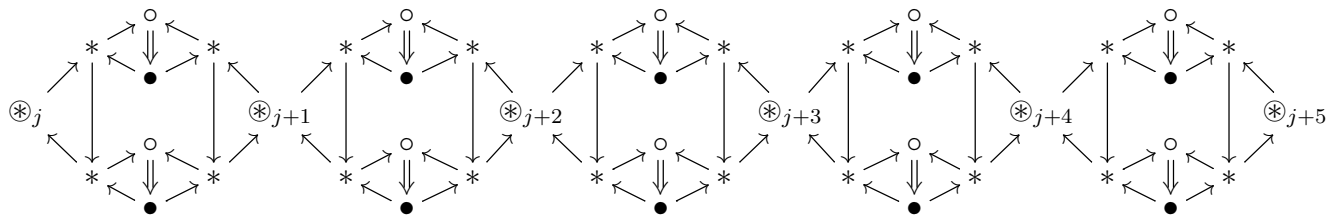
\begin{figure}
\begin{equation*}
\begin{xy} 0;<1pt,0pt>:<0pt,-1pt>:: 
(240,0) *+{\circ} ="0",
(240,24) *+{\bullet} ="1",
(240,49) *+{\circ} ="2",
(240,73) *+{\bullet} ="3",
(216,61) *+{\ast} ="4",
(264,61) *+{\ast} ="5",
(216,12) *+{\ast} ="6",
(264,12) *+{\ast} ="7",
(289,37) *+{\circledast_{j+3}} ="8",
(191,37) *+{\circledast_{j+2}} ="9",
(311,12) *+{\ast} ="10",
(311,61) *+{\ast} ="11",
(336,0) *+{\circ} ="12",
(336,24) *+{\bullet} ="13",
(336,49) *+{\circ} ="14",
(336,73) *+{\bullet} ="15",
(359,61) *+{\ast} ="16",
(359,12) *+{\ast} ="17",
(384,37) *+{\circledast_{j+4}} ="18",
(409,12) *+{\ast} ="19",
(409,61) *+{\ast} ="20",
(433,0) *+{\circ} ="21",
(433,24) *+{\bullet} ="22",
(433,49) *+{\circ} ="23",
(433,73) *+{\bullet} ="24",
(455,12) *+{\ast} ="25",
(455,61) *+{\ast} ="26",
(480,37) *+{\circledast_{j+5}} ="27",
(169,12) *+{\ast} ="28",
(169,61) *+{\ast} ="29",
(144,49) *+{\circ} ="30",
(144,73) *+{\bullet} ="31",
(144,24) *+{\bullet} ="32",
(144,0) *+{\circ} ="33",
(121,12) *+{\ast} ="34",
(121,61) *+{\ast} ="35",
(96,37) *+{\circledast_{j+1}} ="36",
(71,12) *+{\ast} ="37",
(71,61) *+{\ast} ="38",
(47,0) *+{\circ} ="39",
(47,24) *+{\bullet} ="40",
(47,49) *+{\circ} ="41",
(47,73) *+{\bullet} ="42",
(25,12) *+{\ast} ="43",
(25,61) *+{\ast} ="44",
(0,37) *+{\circledast_j} ="45",
"0", {\ar@2"1"},
"6", {\ar"0"},
"7", {\ar"0"},
"1", {\ar"6"},
"1", {\ar"7"},
"2", {\ar@2"3"},
"4", {\ar"2"},
"5", {\ar"2"},
"3", {\ar"4"},
"3", {\ar"5"},
"6", {\ar"4"},
"4", {\ar"9"},
"7", {\ar"5"},
"5", {\ar"8"},
"9", {\ar"6"},
"8", {\ar"7"},
"8", {\ar"10"},
"11", {\ar"8"},
"9", {\ar"28"},
"29", {\ar"9"},
"10", {\ar"11"},
"10", {\ar"12"},
"13", {\ar"10"},
"11", {\ar"14"},
"15", {\ar"11"},
"12", {\ar@2"13"},
"17", {\ar"12"},
"13", {\ar"17"},
"14", {\ar@2"15"},
"16", {\ar"14"},
"15", {\ar"16"},
"17", {\ar"16"},
"16", {\ar"18"},
"18", {\ar"17"},
"18", {\ar"19"},
"20", {\ar"18"},
"19", {\ar"20"},
"19", {\ar"21"},
"20", {\ar"23"},
"24", {\ar"20"},
"22", {\ar"19"},
"22", {\ar"25"},
"21", {\ar@2"22"},
"25", {\ar"21"},
"23", {\ar@2"24"},
"26", {\ar"23"},
"24", {\ar"26"},
"25", {\ar"26"},
"27", {\ar"25"},
"26", {\ar"27"},
"28", {\ar"29"},
"32", {\ar"28"},
"28", {\ar"33"},
"29", {\ar"30"},
"31", {\ar"29"},
"30", {\ar@2"31"},
"35", {\ar"30"},
"31", {\ar"35"},
"33", {\ar@2"32"},
"32", {\ar"34"},
"34", {\ar"33"},
"34", {\ar"35"},
"36", {\ar"34"},
"35", {\ar"36"},
"36", {\ar"37"},
"38", {\ar"36"},
"37", {\ar"38"},
"37", {\ar"39"},
"40", {\ar"37"},
"38", {\ar"41"},
"42", {\ar"38"},
"39", {\ar@2"40"},
"43", {\ar"39"},
"40", {\ar"43"},
"41", {\ar@2"42"},
"44", {\ar"41"},
"42", {\ar"44"},
"43", {\ar"44"},
"45", {\ar"43"},
"44", {\ar"45"},
\end{xy}
\end{equation*}
\caption{\label{ggreater1} A piece of a triangulation quiver (with $\cw$) for a surface $\Sigma$ of genus $g=k+1$ with $k$ regular punctures; the pattern repeats for $j\in\Z$ and the nodes are periodically identified mod $k$,
$\circledast_j\sim\circledast_{j+k}$.
For all
$m\mid k$, it is a \emph{unbranched} $\Z_m$ cover of the quiver (with $\cw$) of a surface of genus $g^\prime=k/m+1$ and $k/m$ punctures obtained by
$\circledast_j\sim\circledast_{j+k/m}$.
}
\end{figure}

To the geometric cover $\Sigma_G\to\Sigma^\prime_G$ there correspond a Galois cover of
triangulation categories. The $\Z_k$
symmetric quiver for the minimal number of punctures ($p=k$) is represented in figure
\ref{ggreater1}. To add further punctures
just replace in figure \ref{ggreater1}
some $\circledast$ nodes, or some pair of triangles forming a Kronecker subquiver,
by the quiver of a multi--punctured cylinder
\begin{equation}
\begin{gathered}\xymatrix{
& \circ\ar@<0.4ex>[dd]\ar@<-0.4ex>[dd] & & \circ\ar@<0.4ex>[dd]
\ar@<-0.4ex>[dd] &&& \circ\ar@<0.4ex>[dd]\ar@<-0.4ex>[dd] &\\
\ast \ar[ru] & &\ast\ar[ul]\ar[ur] & &\ast \ar[ul]\ar@{..}[r]|{\cdots\cdots\cdots}&\ast\ar[ur]& &\ast\ar[ul]\\
& \bullet\ar[ul]\ar[ur] & &\bullet\ar[ul]\ar[ur]  &&&\bullet_{p}\ar[ur]\ar[ul]&}\end{gathered}
\end{equation}
The generalization to irregular punctures may then be obtained by the gentling trick (appendix \ref{ap:gentling}).

In conclusion, all $g>1$ class $\cs[A_1]$
categories with $p\geq g-1$ regular punctures 
 have natural $\Z_k$ Galois covers, for
$k\mid \gcd(g-1,p)$,
which may be used to relate the BPS spectra of different theories (at special points in their parameter spaces). 

\paragraph{Genus one curves.} The Galois cover between genus one $\cs[A_1]$ theories --- corresponding to \emph{isogenies} of elliptic curves --- is discussed in section \ref{genusone}, where
we describe their BPS spectra in detail. We refer to that section.

\paragraph{Genus zero curves.} In the case of genus zero, according to \S.\ref{s:versus},
to have a \emph{strict sense} Galois cover we need at least two irregular punctures to serve as branching points of the cover. Then suppose 
(for simplicity) that we have precisely two irregular
punctures, that is, $\Sigma$ has two boundary components, respectively with $r_1$ and $r_2$ 
marks, and $p$ punctures. The $\Z_k$--cover
$\Sigma^\prime$ has also two boundary components, with $kr_1$ and $kr_2$ marks, and
$pk$ punctures.  

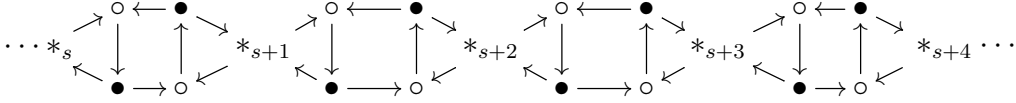
\begin{figure}
\begin{equation*}
\begin{xy} 0;<1pt,0pt>:<0pt,-1pt>:: 
(30,0) *+{\circ} ="0",
(30,30) *+{\bullet} ="1",
(54,0) *+{\bullet} ="2",
(54,30) *+{\circ} ="3",
(85,15) *+{\ast_{s+1}} ="4",
(111,0) *+{\circ} ="5",
(111,30) *+{\bullet} ="6",
(143,0) *+{\bullet} ="7",
(143,30) *+{\circ} ="8",
(171,15) *+{\ast_{s+2}} ="9",
(198,0) *+{\circ} ="10",
(198,30) *+{\bullet} ="11",
(230,0) *+{\bullet} ="12",
(230,30) *+{\circ} ="13",
(257,15) *+{\ast_{s+3}} ="14",
(288,30) *+{\bullet} ="15",
(288,0) *+{\circ} ="16",
(311,0) *+{\bullet} ="17",
(311,30) *+{\circ} ="18",
(0,15) *+{\cdots \ast_s} ="19",
(341,15) *+{\phantom{mn}\ast_{s+4}\cdots} ="20",
"0", {\ar"1"},
"2", {\ar"0"},
"19", {\ar"0"},
"1", {\ar"3"},
"1", {\ar"19"},
"3", {\ar"2"},
"2", {\ar"4"},
"4", {\ar"3"},
"4", {\ar"5"},
"6", {\ar"4"},
"5", {\ar"6"},
"7", {\ar"5"},
"6", {\ar"8"},
"8", {\ar"7"},
"7", {\ar"9"},
"9", {\ar"8"},
"9", {\ar"10"},
"11", {\ar"9"},
"10", {\ar"11"},
"12", {\ar"10"},
"11", {\ar"13"},
"13", {\ar"12"},
"12", {\ar"14"},
"14", {\ar"13"},
"15", {\ar"14"},
"14", {\ar"16"},
"16", {\ar"15"},
"15", {\ar"18"},
"17", {\ar"16"},
"18", {\ar"17"},
"17", {\ar"20"},
"20", {\ar"18"},
\end{xy}
\end{equation*}
\caption{The $\Z_k$--symmetric triangulation quiver for the genus zero surface with two boundaries, each with $k$ marks, and $k$ regular punctures. The subscript $s$ is periodically identified mod $k$.
The superpotential $\cw$ is the sum of the oriented triangles and the oriented squares.\label{zerogenus}}
\end{figure}

As an illustration, the covering quiver with freely acting $\Z_k$--symmetry corresponding to $r_1=r_2=p=1$ is represented in figure \ref{zerogenus}. In this example, the covering theory is (in some regime) a generalized linear quiver with gauge group $SU(2)^{k+1}$, $k$ bifundamental quarks, and two AD systems of type $D_k$; the quotient model is just $SU(2)^2$ SYM coupled to a bifundamental.
The covering BPS spectrum of this model (in a finite chamber) is described explicitly in \S.\ref{091cb}.  

\section{Non-Abelian $\mathbb{G}$: Modular curves, \emph{dessins d'enfants}}

All examples of spectral Galois cover in the previous section had an \emph{Abelian} Galois group
$\mathbb{G}$. It is natural to look for examples 
where the Galois group $\mathbb{G}$ is genuinely non--Abelian.
We already described a class of $\cs[A_1]$ models
which are $\mathbb{G}\equiv\mathfrak{S}_3$ covers of
$\cn=2$ QFTs, namely the last example in \S.\ref{s:versus}. In that case $\mathbb{G}$ was realized as a degree 6 representation of the modular group $PSL(2,\Z)$.  In this section we consider in more detail the coverings between Jacobian categories whose Galois group
$\mathbb{G}$ arises from a finite--dimensional representation of $PSL(2,\Z)$. We stress that the following construction may be applied to class $\cs[G]$ theories for all $G=ADE$. However, to be very explicit, in this section we limit ourselves
to class $\cs[A_1]$ categories.
The BPS spectra of the corresponding class
$\cs[A_1]$ theories have magical properties.

\subsection{Principal modular curves}
The last example of \S.\ref{s:versus} was based on the modular cover\footnote{ Here $\overline{\mathbb{H}}=\mathbb{H}\cup \mathbb{Q}\cup i\infty$, with $\mathbb{H}$ the upper half plane.}
\begin{equation}
\mathbb{P}^1\cong X(2)\equiv \overline{\mathbb{H}}\big/\Gamma(2)\xrightarrow{\ \xi\ }\overline{\mathbb{H}}\big/PSL(2,\Z)\equiv X(1)\cong\mathbb{P}^1.
\end{equation} 
Here $X(N)$ is the (compactified) modular curve $\overline{\mathbb{H}}/\Gamma(N)$ where
$\Gamma(N)$ is the \emph{level $N$ principal congruence subgroup} of the modular group $SL(2,\Z)$ \cite{elliptic,farkas}:
\begin{equation}
\Gamma(N)=\left\{\begin{pmatrix} a & b\\ c & d\end{pmatrix}\in SL(2,\Z)\; \colon\; \begin{pmatrix} a & b\\ c & d\end{pmatrix}=\begin{pmatrix} 1 & 0\\ 0 & 1\end{pmatrix}\mod N\right\}.
\end{equation}
Going through the argument of \S.\ref{s:versus}, we realize that we may replace $\Gamma(2)$ by any
congruence subgroup $\Gamma$ of $SL(2,\Z)$ provided it is \emph{normal} and \emph{torsionless}. Both conditions hold for 
the principal congruence subgroups $\Gamma(N)$
for all $N\geq 2$, and we focus on this class of modular subgroups. The cover $X(N)\xrightarrow{\ \xi_N\ } X(1)$ has degree \cite{elliptic}
\begin{equation}
d(N)=
\begin{cases}\frac{1}{2}N^3 \prod_{p\mid N}(1-1/p^2) & \text{if }N>2,\\
6 & \text{if }N=2.
\end{cases}
\end{equation}
Note that $6\mid d(N)$ and $N\mid d(N)$ for all $N\geq 2$. The genus $g(N)$ of the Riemann surface $X(N)$ is zero for $N=1,2$
while for $N>2$ 
\begin{equation}
g(N)=1+\frac{(N-6)\,d(N)}{12N},
\end{equation}
from which we see that $g(N)=0$ iff $N\leq 5$
and $g(N)=1$ iff $N=6$.
The three Hurwitz partitions 
of $X(N)\xrightarrow{\ \xi_N\ } X(1)$ are
\begin{equation}
d(N)= N+N+N+\cdots+N, \quad d(N)=2+2+2+\cdots+2,\quad d(N)=3+3+3+\cdots +3.
\end{equation}
 
Suppose we have a class $\cs[A_1]$ theory defined by a quadratic differential $\phi_2$ on $X(1)\equiv \mathbb{P}^1$ with three irregular poles
at $\tau=i\infty$, $i$, and $e^{\pi i/3}$ of respective order
$\ell_1+2$, $\ell_2+2$ and $\ell_3+2$,
as well as $k$ ordinary double poles at generic positions. The pulled back quadratic
differential $\xi_N^*\phi_2$ defines a Gaiotto
theory with UV surface $X(N)$ of genus $g(N)$,
$d(N)\,k$ regular punctures, 
and 
\begin{equation}
b(N)=\frac{5N+6}{6N}\,d(N)
\end{equation}
irregular punctures, i.e.\! boundary components for the
associated bordered surface $\Sigma$. For $N>2$,
$d(N)/N$ boundary components have $N\ell_1$ marks,
$d(N)/2$ of them have $2\ell_2$ marks,
and $d(N)/3$ have $3\ell_3$ marks. 
The number of ideal arcs in the covering
triangulation is then 
\begin{equation}
n(N)=\Big(3+3k+\sum_i\ell_i\Big)d(N),
\end{equation} as it should.
The categorical Galois group is
\begin{equation}
\mathbb{G}=PSL(2,\Z/N\Z).
\end{equation} 
Of course, $|\mathbb{G}|=d(N)$.
In some corner of its parameter space, the resulting $\cn=2$ QFT is a gauge theory with gauge group
\begin{equation}
G_\mathrm{gauge}=SU(2)^{m(N)},\qquad \text{where } m(N)=\frac{23}{12}\,d(N)+\frac{d(N)}{2N}+k\,d(N),
\end{equation}
coupled (in a subtle way) to half--trifundamentals, bi--fundamentals,
fundamentals, and Argyres--Douglas systems of types $D_3$ and $D_N$.

For instance, if we take the Galois group to be the
icosahedral group $\mathfrak{A}_5\equiv PSL(2,\mathbb{F}_5)$, corresponding to $N=5$, and no regular punctures, $k=0$, we get a $SU(2)^{126}$ gauge theory.

A special case is $N=7$; the order of the
symmetry group of the covering surface, $d(7)$,
saturates the Hurwitz general bound on the order $h$ of the  automorphism group of a Riemann surface
of given
genus $g$ 
\begin{equation}
h\leq 84(g-1).
\end{equation} 
In this case the 
Galois group is the Hurwitz group \cite{coxeterz}. The minimal theory
with this symmetry has gauge group $SU(2)^{334}$.

\subsection{The modular quivers $Q(N)$ and their flat cousins $Q(N)^\mathrm{fl}$}\label{ffflatcou}

For simplicity, in this subsection we assume $k=0$ and $\ell_1=\ell_2=\ell_3=1$. The extension to the general case is, in principle, straightforward.

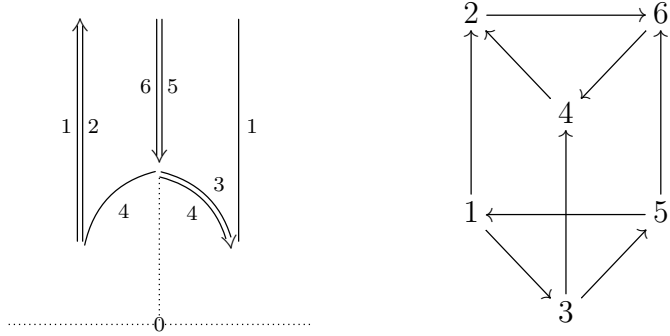
\begin{figure}
\begin{equation*}
\begin{gathered}
\xymatrix{&&\ar@{=>}[dd]_6^5&\\
&&\\
&&\ar@<-0.7ex>@{=>}@/^0.6pc/[dr]_4^3\\
&\ar@{=>}[uuu]^1_2\ar@<-0.4ex>@{-}@/^0.6pc/[ur]_4&&\ar@{-}[uuu]_1\\
\ar@{..}[rrrr]|0 &&\ar@{..}[uu]&&}
\end{gathered}\qquad\qquad
\begin{gathered}
\xymatrix{2\ar[rr] && 6\ar[dl]\\
& 4\ar[lu]\\
1\ar[uu]\ar[dr] && 5\ar[uu]\ar[ll]\\
& 3\ar[uu]\ar[ur]}
\end{gathered}
\end{equation*}
\caption{\label{letrr}\textsc{Left:} the $\Z_3$--symmetric ideal triangulation $T_\cf$ of a pair of pants with one mark on each of the three boundaries as uniformized by the fundamental domain $\cf$ of the modular group (schematic). Ideal arcs with the same number are identified by the action of $PSL(2,\Z)$. The arrows point toward the common vertex of two ideal sides of a triangle whose third side is a boundary arc. \textsc{Right:} the triangulation quiver $Q(1)$ of the symmetric ideal triangulation: nodes of $Q(1)$ are numbered as the corresponding arcs in the triangulation. }
\end{figure}

The fundamental domain of $\Gamma(N)$
has the form 
\begin{equation}\label{fundomain}
\cf(N)=\bigcup_{g\in \mathbb{G}}g(\cf)
\end{equation}
 where $\cf$ is the usual fundamental domain of the modular group $PSL(2,\Z)$. The $\mathbb{G}$--invariant ideal triangulations of the covering surface $\Sigma^\prime$ are constructed by pulling back the ideal
triangulations for the pair of pants with one mark on each boundary. Identifying the pair of pants with 
the fundamental domain $\cf$ with the cusp and elliptic points removed, a convenient (in fact, the most symmetric) ideal triangulation of $\cf$,
is shown in figure \ref{letrr}; we denote it as $T_\cf$.
Then a nice $\mathbb{G}$--invariant triangulation $T(N)$ of $\Sigma^\prime$ is obtained by gluing together
the images $g(T_\cf)$ for $g\in\mathbb{G}$, cfr.\! eqn.\eqref{fundomain}.

From the figure it is obvious how to construct the triangulation quiver $Q(N)$ for the pulled back triangulation $T(N)$. Draw 
the \emph{underlying} ideal triangulation $T(N)^\mathrm{fl}$ of the fundamental domain
$\cf(N)$, i.e.\! the pull back of the triangulation $T^\mathrm{fl}_\cf$ which underlies $T_\cf$: by this we mean
the triangulation obtained by replacing marked boundaries with ordinary punctures. $T(N)^\mathrm{fl}$ is just a convenient $\mathbb{G}$--invariant ideal triangulation of a genus $g(N)$ closed surface
$\Sigma(N)^\mathrm{fl}$ with $3\,d(N)$ punctures. Then replace the ideal arcs of $T(N)^\mathrm{fl}$ with \emph{pairs} of arcs (as in the left part of fig.\,\ref{letrr}) and orient them  in such a way that the boundaries of all triangles of $T(N)^\mathrm{fl}$ form oriented cycles. Call \emph{upper} (resp.\!
\emph{lower}) a triangle of $T(N)^\mathrm{fl}$ if its boundary is counter--clockwise (resp.\! clockwise) oriented.
The block decomposition of the quiver $Q(N)^\mathrm{fl}$ of the underlying ideal triangulation $T(N)^\mathrm{fl}$ consists of $d(N)$ Type II blocks associated to upper triangles and 
$d(N)$ Type II blocks for the lower triangles.
In $Q(N)^\mathrm{fl}$ an upper (lower) Type II block may be glued only to lower (upper) Type II blocks, and all their white nodes are glued in pairs in $Q(N)^\mathrm{fl}$ (since $\Sigma(N)^\mathrm{fl}$ is closed). We get the desired quiver $Q(N)$ from $Q(N)^\mathrm{fl}$ by ``separation of the glued white nodes in the vertical direction'', that is,
at each node of $Q(N)^\mathrm{fl}$ we undo the gluing, insert a Type I block going from the white node of the lower Type II block to the white node of the upper Type II block, and glue back. For instance, if $Q(1)^\mathrm{fl}$
is the quiver of the sphere with 3 punctures, obtained by gluing two oppositely oriented Type II blocks, the separation in the vertical direction produces $Q(1)$ which is the well--known prism quiver in figure \ref{letrr}.
We say that $Q(N)^\mathrm{fl}$ is obtained by \emph{flattening}
$Q(N)$.

In principle, to compute the flattened out quiver $Q(N)^\mathrm{fl}$, one has to pull back the triangulation
$T_\cf^\mathrm{fl}$ to the fundamental domain
of $\Gamma(N)$ by gluing the various copies through the decomposition \eqref{fundomain}. In practice, we may avoid going through that, and get $Q(N)^\mathrm{fl}$ by elementary  symmetry considerations. 

\begin{figure}
\centering
\includegraphics[width=1\textwidth]{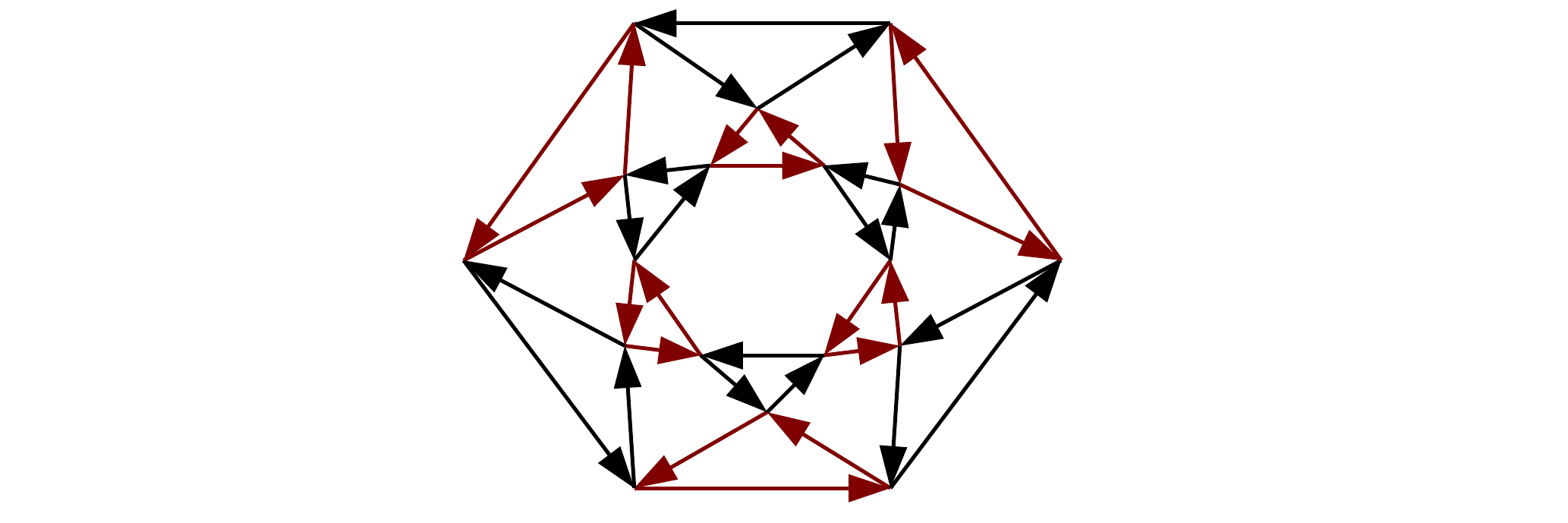}  
\caption{The flattened out quiver $Q(2)^\mathrm{fl}$ for 
the modular $\mathfrak{S}_3$--cover of the pair of pants with one mark per boundary. Upper (lower) triangles are drawn in red (black). }
 \label{quiverdia}
\end{figure}

To illustrate the procedure, we consider again our original example $N=2$. Since the
Hurwitz partitions for the cover $J=J(\lambda)$ of eqn.\eqref{legendre}
are $(2,2,2)$, $(2,2,2)$, and $(3,3)$, the 
underlying triangulation $T(2)^\mathrm{fl}$ is based 
on six punctures of valency 4 and two punctures
of valency 6, while $T(2)^\mathrm{fl}$ contains $6\times 2=12$ (ordinary) triangles.
The underlying graph $L(2)^\mathrm{fl}$ of the flattened out quiver $Q(2)^\mathrm{fl}$ is then given by the edges of a tiling of the sphere $S^2$ by 12 triangles,
6 squares, and 2 hexagons. $Q(2)^\mathrm{fl}$ is then obtained by making all the 20 polygons of $L(2)^\mathrm{fl}$ into oriented cycles.
This determines uniquely $Q(2)^\mathrm{fl}$ to be the diamond quiver in figure \ref{quiverdia}. In the figure we have distinguished the arrows belonging to upper/lower Type II blocks by their color \emph{red} resp.\! \emph{black}. The automorphism group of $Q(2)^\mathrm{fl}$  is the dihedral group 
$D_{12}$ (which is an automorphism of the full Jacobian category of the sphere with 8 regular punctures); however only the subgroup $D_6\cong \mathfrak{S}_3\subset D_{12}$ preserves the coloring of the arrows, and hence is an automorphism of the
vertically separated quiver $Q(2)$ (and category)
of the Galois cover.
$\mathfrak{S}_3$ acts freely on both the nodes and the Type II blocks. Hence we recover the correct Galois group $\mathfrak{S}_3$ of the QFT described at the end of \S.\ref{s:versus}. 
The quiver $Q(2)$ is obtained by separating vertically the red and black blocks in figure
\ref{quiverdia}. 

In the general case, the flattened quiver $Q(N)^\mathrm{fl}$ is obtained by cyclically orienting
the boundaries of a polygonal tiling of
a surface $\Sigma(N)$ of genus $g(N)$, invariant under its automorphism group $PSL(2,\Z/N\Z)$,  consisting of the following polygonal tiles:
\begin{equation*}
\begin{aligned}
&\bullet\quad 2\,d(N)\ \ \text{triangles} &&\phantom{mmmmm}
&&\bullet\quad \frac{d(N)}{2}\ \ 4\text{--gons}\\
&\bullet\quad \frac{d(N)}{3}\ \ \text{hexagons} &&\phantom{mmmmm}
&&\bullet\quad \frac{d(N)}{N}\ \ 2N\text{--gons.}
\end{aligned}
\end{equation*}
This tiling has $V= 3\,d(N)$ vertices,
$E=6\,d(N)$ edges, and $F=(17\,N+6)d(N)/6 N$
faces. The Euler characteristic of this tiling
is 
\begin{equation}
V-E+F=(6-N)d(N)/6\,N\equiv 2-2\,g(N),
\end{equation}
as it should. Since $\Sigma(N)\equiv X(N)$ is orientable,
there exists an orientation of the edges of the tiling so that all polygonal faces have cyclically oriented
boundaries. The tiling with this orientation is the
flattened quiver $Q(N)^\mathrm{fl}$. By vertical separation of upper/lower Type II blocks, one gets
the desired $\mathbb{G}$--invariant quiver $Q(N)$.

\subsection{Simple generalizations}\label{minexample}

Let $N$, $M$ be \emph{coprime}
integers $\geq 2$. 
The modular cover $X(N)\xrightarrow{\ \xi_{N}\ } X(1)$
may be generalized to 
\begin{equation}
X(NM)\xrightarrow{\ \xi_{N,M}\ } X(M).
\end{equation}
$\xi_{N,M}$ is a cover of a genus $g(M)$ surface $X(M)$ by a genus $g(NM)$ one $X(NM)$ with degree
\begin{equation}
d(N,M)\equiv \frac{d(NM)}{d(M)}=\begin{cases} 2\,d(N) &\text{if }N,M\geq 3\\
d(N) &\text{if }M= 2\ \text{or }N=2,
\end{cases}
\end{equation} 
branched over the $d(M)/M$ cusps
with equal Hurwitz partitions
\begin{equation}
d(N,M)= N+N+\cdots +N.
\end{equation}
The Galois group of the cover is
\begin{equation}
\mathbb{G}=SL(2,\Z/N\Z).
\end{equation}

Assuming the $i$--th boundary of the base
bordered surface $\Sigma(M)$ have
$\ell_i$ marks ($i=1,\dots, d(M)/M$),
and adding to it $k$ regular punctures,
we get a $\mathbb{G}$--cover QFT with
gauge group
\begin{equation}
SU(2)^{m(N,M)}\quad\text{where }\ 
m(N,M)=\frac{NM+2}{4 MN}\,d(NM)+k\,\frac{d(NM)}{d(M)}.
\end{equation}

\paragraph{Example.} The minimal $M$ for which the quotient surface $X(M)$ is not a sphere is $M=6$. The smallest possible coprime integer is
$N=5$.
This minimal non--spherical example corresponds to a Gaiotto theory defined by a torus
with at least $d(6)/6=12$ irregular punctures
(plus, possibly, regular ones) covered by a
Gaiotto theory on a surface of genus
\begin{equation}
g(30)=577
\end{equation}
having at least
\begin{equation}
\frac{d(30)}{30}=288
\end{equation}
 irregular punctures. The \emph{minimal}
 model in this class has gauge group
 \begin{equation}
 G_\mathrm{gauge}=SU(2)^{2304},
 \end{equation}
 and thousands of different matter sectors.
It is remarkable that we may explicitly compute the BPS spectra of gauge theories of such
enormous complexity\,!

\subsection{Generalization: regular Grothendieck's
\emph{dessins d'enfants}}

The above constructions may be
generalized in various ways. As a payoff we get a more intrinsic geometric interpretation
of the `separation in the vertical direction' method to construct 
the relevant quivers. 
As in \S.\ref{s:versus} we consider covers of the
sphere branched over three points.
It is customary to set the three branch points in $\mathbb{P}^1$ at $\infty$, $0$, and $1$
(before we set them at $i\infty$, $i$, and $e^{\pi i/3}$
mod $PSL(2,\Z)$). By definition
a covering $\xi\colon\Sigma_G\to \mathbb{P}^1$ branched over $\infty,1,0$ is given by a normalized Belyi function $\xi$
\cite{belyi,itzk}. It is a fundamental theorem 
of diophantine geometry that a smooth algebraic curve $\Sigma_G$ is defined over the algebraic closure $\mathbb{Q}^\mathrm{al}$ of $\mathbb{Q}$
if and only if there exists a Belyi function $\xi$ on $\Sigma_G$ \cite{belyi,itzk}. Moreover, in this case
there is a subgroup $\Gamma$ of some triangle
group\footnote{ We always assume $\Delta(p,q,r)$ to be \emph{minimal} i.e.\! $p$, $q$, $r$ are the least common multiplexes of all ramifications orders of $\xi$ above $0$, $1$, $\infty$, respectively.} 
$\Delta(p,q,r)$ such that $\Sigma_G\cong \mathbb{H}/\Gamma$ and \cite{itzk,wol2} 
\begin{align}
&\xi\colon \mathbb{H}/\Gamma\to \mathbb{H}/\Delta(p,q,r)\equiv \mathbb{P}^1\\
&\Delta(p,q,r)\equiv\Big\langle \sigma_0, \sigma_1,\sigma_\infty\; \Big|\; \sigma_0^p=\sigma_1^q=\sigma_\infty^r=\sigma_0\sigma_1\sigma_\infty=1\Big\rangle.
\end{align}

Belyi functions are encoded in Grothendieck's \emph{dessins d'enfants} \cite{itzk,groth,wol}.
We define\footnote{ In the literature there are slightly different definitions. We use the one in ref.\cite{wol}.} a \emph{dessin d'enfant} to be a  bi--partite graph $D$ on a compact oriented $2$--manifold $\Sigma_G$ such that $\Sigma_G\setminus D$ is a disjoint union of simply connected
sets. The vertices of $D$ are colored alternatively in black and white. If $\xi\colon \Sigma_G\to \mathbb{P}^1$ is a Belyi function, the corresponding \emph{dessin} is given
by $\xi^{-1}([0,1])$, which is a connected graph.
\begin{itemize}
\item[a)] the preimages of $0$ and $1$ are, respectively, the white and black vertices of $D$.
The valency of a vertex is equal to the ramification order of $\xi$ at these points;
\item[b)] each component of $\Sigma_G\setminus D$ is a cell containing precisely one pole of $\xi$. The cell is bounded by $2m$ edges, where $m$ is the order of the pole (ramification index over $\infty$).
\end{itemize}
The inverse correspondence \emph{dessin} $\rightarrow$ Belyi function is given by the Riemann existence theorem \cite{zvon}. 

A \emph{dessin}
$D\subset \Sigma_G$ defines a \emph{canonical
triangulation} $T_\mathrm{can}(D)$ of  $\Sigma_G$ whose
edges are $\xi^{-1}(]0,1[)\cup \xi^{-1}(]1,\infty[)\cup
\xi^{-1}(]\infty,0[)$. One constructs $T_\mathrm{can}(D)$ by
introducing a puncture in the center of each cell of
$\Sigma_G\setminus D$ and connecting it to the vertices of $D$. $T_\mathrm{can}(D)$ contains two kinds of triangles, \emph{upper} and \emph{lower},
according whether their vertices, ordered in the counter--clockwise direction, are labelled $0$, $1$, $\infty$ (resp.\!
$0,\infty, 1$); $\xi$ maps upper/lower triangles into the upper/lower half--plane \cite{magot}. 
\medskip

\textbf{Definition.} $D\subset \Sigma_G$ a
\emph{dessin of enfant}. Its canonical quiver
$Q(D)$ is obtained from the triangulation quiver
of $T_\mathrm{can}(D)$ by vertical separation of the
upper/lower triangles (cfr.\! previous subsection).
It describes the class $\cs[A_1]$ theory obtained by pulling back $SU(2)^3$ with half a trifundamental by the Belyi function $\xi$ associated to $D$. 
\medskip

Clearly, the modular quiver $Q(N)$ we obtained
previously by direct considerations are just the canonical ones for the \emph{dessins}
associated to the modular covers.

The automorphism group $\mathsf{G}$
of the \emph{dessin d'enfant} $D$
acts as an automorphism of the Jacobian category of the covering class $\cs[A_1]$ theory. 
As we saw in \textbf{example 1} of \S.\ref{s:versus},
in general $\mathsf{G}$ does not act freely on the nodes of $Q(D)$, i.e.\! on the edges of $D$. 
By definition, $\mathsf{G}$ acts transitively on the edges iff $D$ is a regular \emph{dessin} \cite{wol,wol2}; for genus $>1$, this happens precisely when $\Gamma\equiv $ is normal in $\Delta(p,q,r)$ (then it is also torsionless). In this special case \cite{wol2}, $\Sigma_G$ is a \emph{quasi--platonic surface} i.e.\! a surface with many automorphisms \cite{wol,wol2} and $\mathbb{G}=\Delta(p,q,r)/N$
is the group of automorphisms of the regular \emph{dessin}, and also a group of automorphisms of the covering surface $\Sigma_G$, isomorphic to the monodromy group and the covering group of the normal covering Belyi function 
\begin{equation}\label{eeeeebbz}
\xi\colon \Sigma_G\to \mathbb{H}/\Delta(p,q,r)\equiv \mathbb{P}^1.
\end{equation} 
By definition, in the regular case, the integers
$p$, $q$, and $r$ are divisors of the degree $d$
of $\xi$.

Suppose the quotient theory is
$SU(2)^3$ SYM coupled to half a trifundamental
and three Argyres--Douglas systems of
type $D_{\ell_1}$, $D_{\ell_2}$, and $D_{\ell_3}$.
Then its $\Delta(p,q,r)/N$--cover given by
a degree $d$ Belyi function $\xi$, corresponding to a regular \emph{dessin}
\eqref{eeeeebbz}, is a gauge theory with gauge group
\begin{equation}\label{tttttmmz}
SU(2)^m,\quad\text{where }m=\left(3+\frac{1}{p}+\frac{1}{q}+\frac{1}{r}\right)\frac{d}{2},
\end{equation}
coupled to $d$ half--trifundamentals together with
$d/p$ copies of $D_{p\ell_1}$ AD,
$d/q$ copies of $D_{q\ell_2}$ AD,
and $d/r$ copies of $D_{r\ell_3}$ AD.

\paragraph{Remark.} The above discussion was focused on the special case of coverings branched over \emph{three} irregular punctures (the case of 
two irregular punctures being Abelian and covered in the previous section). For $n>3$ branching irregular punctures one needs to use a multi--color generalization of the \emph{dessins d'infants}, see ref.\!\cite{petronio}.

\subsection{Finite chambers of \emph{dessin d'enfant} cover $\cn=2$ QFTs}

The BPS spectra of the $\cn=2$ QFTs
defined by modular covers $X(N)\to X(1)$
of \S.\ref{ffflatcou}
may
be obtained, at $\mathbb{G}$--invariant points,  by pulling back the BPS spectrum of
$SU(2)^3$ SYM coupled to half a tri--fundamental in the corresponding chamber. The BPS spectrum of the last model is well known in many chambers, and may be easily determined for all choices of the central charge $Z$ by the string/band construction
(see appendix \ref{stringbands}).
We begin by describing the BPS spectrum in a class of
finite chambers.

We note that the modular quivers $Q(N)$
decompose in the two complementary full subquivers
$Q(N)_\text{up}$ and $Q(N)_\text{down}$
over, respectively, the nodes of the upper and lower 
Type II blocks. All arrows in $Q(N)$
which connect the two complementary subquivers are oriented from $Q(N)_\text{down}$ to $Q(N)_\text{up}$
\begin{equation}\label{pppqw12}
\xymatrix{\text{\fbox{\ $\phantom{\Bigg|}Q(N)_{\text{down}}$\ }}\ar@<3ex>[rr] 
\ar@<1.5ex>[rr] 
\ar[rr]
\ar@<-1.5ex>[rr]
\ar@<-3ex>[rr] 
&&
\text{\fbox{\ $\phantom{\Bigg|}Q(N)_{\text{up}}\ $\ 
}}}
\end{equation}
Both subquivers, $Q(N)_\text{down}$ and 
$Q(N)_\text{up}$ consist of the \textit{disconnected union} of $d(N)$ copies
of the cyclic form $C_3$ of the $A_3$ quiver 
(with its superpotential). Correspondingly
the charge lattice $\Gamma$ of the covering theory may be identified with
\begin{equation}\label{zzzzmm9}
\Gamma\cong\bigoplus_{s=1}^{2\,d(N)}\Gamma_{C_3},\qquad\quad \Gamma_{C_3}\cong \Gamma_{A_3}\ \text{(the root lattice of $A_3$)}.
\end{equation}
We choose the central charge so that\footnote{ Note that this an open condition; our conclusions will 
apply to \emph{open} BPS chambers. The subtleties of $\mathbb{G}$--points being on (the intersection point of several) walls of marginal stability, do not apply to the present situation.}
\begin{equation}
 \text{$\arg Z(e_i)<\arg Z(e_j)$ for all $e_i\in Q(N)_{\text{down}}$ and $e_j\in Q(N)_{\text{up}}$.}
 \end{equation}
 Then a module $X$ is unstable unless
 $\mathrm{supp}\,X\subset Q(N)_{\text{down}}$
 or $\mathrm{supp}\,X\subset Q(N)_{\text{up}}$.
 In fact, a necessary condition for stability is
 \begin{equation}
 \mathrm{supp}\, X\subset \Big\{\text{a $C_3$ connected component of $Q(N)_{\text{down}}$
 or $Q(N)_{\text{up}}$}\Big\}.
 \end{equation}
Thus, with respect to the identification 
\eqref{zzzzmm9}, the stable modules $X$ have
charge vectors of the form
\begin{equation}
\begin{aligned}
&\mathbf{dim}\,X=0\oplus 0\oplus \cdots \oplus 0 \oplus \boldsymbol{x} \oplus 0\oplus\cdots \oplus 0\in \bigoplus_{s=1}^{2\,d(N)}\Gamma_{C_3},\\
&\boldsymbol{x}\in \Gamma_{C_3}\ \text{is a charge vector of a stable module for $A_3$ Argyres--Douglas.}
\end{aligned}
\end{equation}
Hence the BPS spectrum in these chambers 
consists of $2\,d(N)$ copies of the spectrum of 
the Argyres--Douglas of type $A_3$. The
various copies may correspond to spectra in different
chambers of Argyres--Douglas, but at a
$\mathbb{G}$--symmetric point all
upper (lower) copies belong to the same AD chamber.

Of course, this is just the pull back of the usual 
finite chambers for the prism quiver $Q(1)$ in the right
hand side of figure \ref{letrr}. The consistency
of this pulled back spectrum sets severe constraints on the structure of the quiver $Q(N)$, which essentially fix its structure to be the one
we described in \S.\ref{ffflatcou}.

Clearly, the same discussion holds, \emph{mutatis mutandis,} 
 for general
\emph{dessin d'enfant} covering of the $SU(2)^3$ theory with half a trifundamental, and in particular for the
quasi--platonic coverings. Indeed, all canonical quivers $Q(D)$ decompose into two complementary subquivers, $Q_\mathrm{down}$
and $Q_\mathrm{up}$ which are disconnected unions of Type II blocks, and which are connected by arrows in the same direction as in \eqref{pppqw12}. Hence\medskip

\textbf{Fact.}
\textit{All \emph{dessin d'enfant} class $\cs[A_1]$
$\cn=2$ model  (see eqn.\eqref{tttttmmz} for its description as a gauge theory) has a chamber whose BPS spectrum consists of $2\,d$ copies
 of the $A_3$ Argyres--Douglas spectrum, where
 $d$ is the degree of the associated Belyi function
 $\xi$.}

\subsection{\emph{Dessins d'enfants} models: general BPS chambers}

The BPS spectrum of a regular \emph{dessin d'enfant}
$\cn=2$ model with $\ell_1=\ell_2=\ell_3=1$ may be determined in a rather explicit way in \emph{any} BPS chamber. Indeed, the Jacobian algebra
$\mathscr{J}(D)$ of the canonical quiver of
a \emph{dessin} is a string algebra, and its
indecomposable modules may be constructed in terms of strings and bands.
The special techniques introduced in  
\S\S.\,\ref{special cases}, \ref{coversofsa1}
apply to all these theories, leading to a rather explicit description of their BPS spectra in infinite as well finite chambers, both \emph{per se} as well as
covers of the well understood $SU(2)^3$ with a half trifundamental theory. 
 
\section{Galois covers and the quantum monodromy $\mathbb{M}(q)$}\label{zzzuuq}

One purpose of BPS spectral covers
is to prove/verify the WCF, which is essentially the same as computing the
action of (the classical limit of)  the 
quantum monodromy $\mathbb{M}(q)$
on the quantum torus algebra $\mathbb{T}_Q(q)$
\cite{CNV}. 

We recall the basics of the formalism of ref.\!\cite{CNV}. Given a $2$--acyclic quiver $Q$
with exchange matrix $B$ and (reduced, non--degenerate) superpotential $\cw$,
the associated quantum torus algebra
$\mathbb{T}_Q(q)$ is defined as follows \cite{CNV,gon1,gon2,kell2}:
to each node $i$ of $Q$ (i.e.\! to each object of
the Jacobian category $\scj\equiv \C Q/(\partial\cw)$) we associate an invertible quantum operator $Y_i$, and set
\begin{equation}
\mathbb{T}_Q(q)=\C\langle\;Y_i,Y_j^{-1} \;\rangle\big/\mathbb{I}(q),
\end{equation}
where $\mathbb{I}(q)$ is the ideal generated by the commutation relations
(here $q\equiv e^{2\pi i\tau}\in\C^\times$) 
\begin{equation}
Y_i\,Y_j=q^{B_{ij}}\;Y_j\,Y_i.
\end{equation} 
Replacing the quiver $Q$ by its mutation at the $k$--th node,
$\mu_k(Q)$,  induces a correspondence\footnote{ Note that $\mathbb{T}_Q(q)$ is, up to unitary equivalence, a central extension of the algebra
of quantum operators $e^{i x_a}$, $e^{2\pi i\tau p_a}$, where $x_a$, $p_a$ are canonical operators in $L^2(\R^{\mathrm{rank}\,B/2})$, satisfying $[x_a,p_b]=i\delta_{ab}$. Since $\mathrm{rank}\,B$ is a mutation invariant \cite{triangulation1}, all algebras
$\mathbb{T}_{Q^\prime}(q)$, with $Q^\prime$ in the mutation class of $Q$, are abstractly the `same' algebra, and hence the correspondence 
$\mathbb{T}_Q(q)\to \mathbb{T}_{Q^\prime}(q)$ 
makes sense.}
$\mathbb{T}_Q(q)\to \mathbb{T}_{\mu_k(Q)}(q)$ of quantum torus algebras given explicitly by the
Fock--Goncharov mutation operation\footnote{ We recall that the $Y$'s before and after the modification belong to \emph{distinct} quantum torus algebras; to avoid cluttering we shall not make the distinction manifest in the notation, we hope this will not cause confusion.} $\phi_k$
\cite{CNV,ACCERV2,gon1,gon2,kell2}
\begin{equation} \label{xxxyyy}
\phi_k(Y_i)= \begin{cases}
Y_k^{-1} & \text{if } i=k\\
Y_i & \text{if there are no arrows $i\to k$ in $\mu_k(Q)$}\\
q^{-m^2/2}\, Y_i Y_k^m &
\text{if there are $m\geq 1$ arrows $i\to k$ in $\mu_k(Q)$.}
\end{cases}
\end{equation}

One extends the definition of the elements
of the torus algebra
to all charge vectors $\gamma\equiv \sum_in_i e_i\in\Gamma$ by setting $Y_{e_i}=Y_i$ and defining
\begin{equation}
Y_{\gamma_1+\gamma_2}= q^{-B(\gamma_1,\gamma_2)/2}\;
Y_{\gamma_1}\,Y_{\gamma_2},\qquad \text{where} \quad B(\gamma_1,\gamma_2)\equiv
B_{ij}\,n_{1,i}\,n_{2,j}.\end{equation} 
Then the quantum monodromy $\mathbb{M}(q)$ is
the ordered product \cite{CNV}
 \begin{equation}\label{wqrtz}
 \mathbb{M}(q)= \prod^\curvearrowleft_{(\gamma,s)\in\text{BPS}} \Psi(q^s\,Y_\gamma;q)^{(-1)^{2s}}
 \end{equation}
 where the product is over all the (Clifford vacua of the) stable BPS multiplets with charge $\gamma\in\Gamma$ and spin $s$; $\Psi(y;q)$
 is the following version of the compact quantum dilogarithm \cite{fadd}
\begin{gather}
\Psi(y;q)=\prod_{j=0}^\infty (1-q^{j+1/2}y)^{-1},\intertext{with functional equation }
\Psi(q^{\pm1}y;q)=(1-q^{\pm 1}y)^{\pm1/2}\,\Psi(y;q),
\end{gather}
and $\prod\limits^\curvearrowleft$ stands for the product ordered in increasing
 (cyclic) BPS phase $\arg Z(\gamma)$.
 The \emph{refined} WCF \cite{CNV,wallcrossingtopstrings,guk1,guk2} 
  is the statement
that the conjugacy class of $\mathbb{M}(q)$
is independent of the chamber used to define
it. The ordinary WCF is the classical limit, $q\to1$,
of the statement.

We stress a \emph{caveat} which one should always keep in mind. The form \eqref{wqrtz} of the quantum monodromy as a product of quantum dilogarithms
holds in the \emph{generic} case, that is, under the assumption that all BPS states which have
the same phase are mutually local
\begin{equation}\label{gencon}
\textbf{(genericity condition):}\ \gamma_1,\gamma_2\in\mathrm{BPS}\ \text{and }\arg Z(\gamma_1)=
\arg Z(\gamma_2)\ \Rightarrow\ \langle \gamma_1,\gamma_2\rangle_D=0.
\end{equation}
Otherwise one should work with the (quantum)
KS formula in its more general form\footnote{ For \emph{non--generic} phase configurations,
the BPS counting indices $\Omega(\gamma)$
are not even expected to be integers, cfr.\!
\textbf{Conjecture 6} in ref.\cite{KS1}.} \cite{KS1}.
Although mathematically the stability conditions and the quantum monodromy make perfect sense at such non--generic points, and all general theorems (properly formulated) still hold, \emph{physically} the BPS spectrum at such points is rather tricky since these points are exactly \emph{on} instability walls separating different BPS chambers; the mathematically stable modules correspond to the
BPS states which remain stable \emph{on all sides} of the walls.   
This subtlety is important in our present context since we are studying $\cn=2$ models at points having $\mathbb{G}$ symmetry which are typically  very non--generic. One should be careful in
stating the (possibly subtle) physical interpretation of   the straightforward mathematical results.

\subsection{A baby example: the covering $N_f=2\to N_f=0$}\label{sss;babyex} 
Consider the quiver $Q^{(2)}$ in the left hand side of
figure \eqref{quileftright}. 
Its quantum torus algebra $\mathbb{T}_{Q^{(2)}}(q)$
has a very simple structure: the two operators
\begin{equation}
Y_{\bullet_1}Y_{\bullet_2}^{-1}\qquad\text{and}\qquad Y_{\circ_1}Y_{\circ_2}^{-1}
\end{equation}
commute with everything and hence are just
$c$--numbers in any irreducible representation.
Hence, for some complex numbers
$t$, $s$
\begin{equation}\label{tststs}
Y_{\bullet_2}=t\,Y_{\bullet_1},\qquad Y_{\circ_2}=s\,Y_{\circ_1}.
\end{equation}
Since the validity of the WCF is independent of the chosen representation of the torus algebra, we may
take $s=t=-1$. Then the generators
of $\mathbb{T}_{Q^{(2)}}$ are odd with respect to
the Galois group $\Z_2\equiv \{1,\iota\}$, which then acts on the
quantum torus algebra as
\begin{equation}\label{xxxqer}
\iota\cdot Y_{\gamma}\equiv Y_{\iota\gamma}= (-1)^{|\gamma|}\, Y_{\gamma},\qquad \text{where }|\gamma|\equiv\left|\sum_i n_i e_i\right|=\sum_i n_i.
\end{equation}
If $\gamma$ is a Schur root of $\widehat{A}(2,2)$, that is, the dimension vector of a brick of the associated algebra, this is equivalent to
\begin{equation}
\iota\cdot Y_{\gamma}\equiv Y_{\iota\gamma}= (-1)^{m(\gamma)}\, Y_{\gamma},
\end{equation}
where $m(\gamma)$ is the magnetic component of the charge vector $\gamma$.

Let  $N_f=2$
SQCD be endowed with a pull back
central charge\footnote{ This is not a loss of generality: the quantum monodromy and hence
the functional relation between quantum monodromies of different $\cn=2$ theories are
\emph{independent} of the chosen central charge.} as in \S.\,\ref{pullbackZ},
so that $\arg Z(\iota\cdot\gamma_j)=\arg Z(\gamma_j)$. Let $\gamma_j$ be the
charge of a stable BPS state which is not the
$W$ boson; then (cfr.\! table \eqref{ttable20})
$\iota\cdot\gamma_j$ is the charge of a \emph{distinct} BPS particle which is also stable and has the same phase $\arg Z_{\gamma_j}$ and spin $s_j$. Their two 
quantum dilogs come together in the ordered product
for $\mathbb{M}(q)$
\begin{equation}
\mathbb{M}(q)=\cdots\; \Psi\!\big(q^{s_j}\,Y_{\gamma_j};q\big)^{(-1)^{2s_j}}\,\Psi\!\big(q^{s_j}\,Y_{\iota\cdot\gamma_j};q\big)^{(-1)^{2s_j}}\;\cdots 
\end{equation} 
If $m(\gamma_j)\neq 0$ (which implies $s_j=0$,
cfr.\! eqn.\eqref{ttable20}) the product of
two dilogs simplifies to a single dilog
of nome $q^2$ 
\begin{equation}\label{assz76}
\mathbb{M}(q)=\cdots\; \Psi(Y_{2\gamma_j};q^2)\;\cdots 
\end{equation} 
The change of nome, $q\mapsto q^2$,
reflects the fact that the subalgebra of $\mathbb{S}(q)\subset\mathbb{T}_{Q^{(2)}}(q)$ generated by the two elements
\begin{equation}\label{sub2ele}
Y_\bullet =Y_{\bullet_1}Y_{\bullet_2},
\qquad
Y_\circ =Y_{\circ_1}Y_{\circ_2}
\end{equation}
(and their inverse)
is identified with the torus algebra of
$N_f=0$ SQCD, $\mathbb{T}_{Q^{(0)}}(q)$
up to the redefinition of the nome $q\mapsto q^2$
\begin{equation}\label{idtoral}
\mathbb{S}(q)=\mathbb{T}_{Q^{(0)}}(q^2). 
\end{equation}
Then the factor $\Psi(Y_{2\gamma_j};q^2)$
in eqn.\eqref{assz76} is identified with the
element of $\mathbb{T}_{Q^{(0)}}(q^2)$
corresponding to the monodromy factor of the (single) hypermultiplet in the quotient theory having charge the push down of $\gamma_j$.

It remains to consider the factors in $\mathbb{M}(q)$ from the BPS states of the cover $N_f=2$ SQCD with zero magnetic charge. They have all the same\footnote{ Indeed, if $X$ is a brick with $m(X)=0$ it is either in the $W$ family or $F_\lambda X$ is in the $W$ family.} phase
$\arg Z(\gamma)$. So they contribute 
to the monodromy the factor\footnote{ The order of the six factors is irrelevant since they commute.}
\begin{multline}
\Psi\!\big(q^{-1/2}Y_{\bullet_1}Y_{\circ_1};q\big)\,
\Psi\!\big(q^{-1/2}Y_{\bullet_2}Y_{\circ_2};q\big)\,
\Psi\!\big(q^{-1/2}Y_{\bullet_1}Y_{\circ_2};q\big)\,
\Psi\!\big(q^{-1/2}Y_{\bullet_2}Y_{\circ_1};q\big)\;
\times\\
\times\;
\Psi\!\big(q^{-3/2}Y_{\bullet_1}Y_{\circ_1}Y_{\bullet_2}
Y_{\circ_2};q\big)^{-1}\,
\Psi\!\big(q^{-1/2}Y_{\bullet_1}Y_{\circ_1}\mathsf{X}_{\bullet_2}
Y_{\circ_2};q\big)^{-1}=\\
=\Psi\!\big((q^{-1/2}Y_{\bullet_1}Y_{\circ_1})^2; q^2\big)^2\, \Psi\!\big(q^{-1/2}
(q^{-1/2}Y_{\bullet_1}Y_{\circ_1})^2;q)^{-1}\,
\Psi\!\big(q^{1/2}
(q^{-1/2}Y_{\bullet_1}Y_{\circ_1})^2;q)^{-1}
\end{multline}
where we used eqn.\eqref{tststs}; 
this product belongs to the subalgebra $\mathbb{S}(q)$ and is identified with an element of the quantum torus algebra of the quotient theory
i.e.\! pure SYM.
Explicitly,
\begin{equation}
\begin{split}
\frac{\Psi(Y_{\bullet+\circ}; q^2)^2}{\Psi(q^{-1/2}Y_{\bullet+\circ};q)\;
\Psi(q^{1/2}Y_{\bullet+\circ};q)}&=\prod_{n\geq 0}\frac{(1-q^nY_{\bullet+\circ})\;
(1-q^{n+1}Y_{\bullet+\circ})}{(1-q^{2n+1}Y_{\bullet+\circ})^2}=\\
&=\prod_{n\geq 0}(1-q^{2n}Y_{\bullet+\circ})
\prod_{n\geq 0}(1-q^{2n+2}Y_{\bullet+\circ})=\\
&=\Psi\big((q^2)^{-1/2}Y_{\bullet+\circ};q^2\big)^{-1}\;\Psi\big((q^2)^{1/2}Y_{\bullet+\circ};q^2\big)^{-1},
\end{split}
\end{equation}
which is precisely the contribution from a single
vector multiplet of charge $\delta=e_\bullet+e_\circ$, that is, the usual $W$ boson of $SU(2)$ SYM.
Thus, in all (pull back) BPS chambers, the quantum monodromy of the Galois cover $N_f=2$
theory is equal --- factor by factor --- to the quantum  monodromy of the quotient $N_f=0$ model
up the redefinition $q\mapsto q^2$.
Therefore, the WCF for the 
Galois cover implies the WCF for 
the quotient, and we get the functional identity
\begin{equation}
\mathbb{M}^{(N_f=0)}(q^2)\doteq \mathbb{M}^{(N_f=2)}(q),
\end{equation}
where $\doteq$ means equality 
of actions in the $\Z_2$--invariant subsector
of a representation of the covering quantum torus algebra satisfying \eqref{xxxqer}.

\subsection{Galois coverings, cluster mutations, and finite BPS chambers}

The description of the quantum monodromy $\mathbb{M}(q)$ in \cite{CNV} was based on
quantum cluster algebras \cite{gon1,gon2}
and the mutation algorithm \cite{ACCERV1,ACCERV2} (for the mathematical side see \cite{kell2}).
 
 To compare with the standard conventions\footnote{ We adopt Keller's conventions \cite{kell2}. The change of sign is required to make the \emph{positivity property} of cluster algebras manifest (unfortunately it makes the physics less transparent).} in cluster algebra theory, we should interchange the two square roots of
 $q$, $q^{1/2}\leftrightarrow-q^{1/2}$;
this amounts to replacing the quantum dilog $\Psi(y;q)$ with the function $\bee(y)\equiv \Psi(-y;q)$ \cite{kell2}. The  (initial) $Y$--seed of the quantum cluster algebra is identified with set of generators
$Y_i$ of the quantum torus algebra $\mathbb{T}_Q(q)$.
A quantum cluster mutation is the composition of
elementary mutations at single nodes of the quiver $Q$. The elementary quantum mutation at node $k\in Q$ of the $Y$--seed
acts by
\begin{itemize}
\item[\textit{i)}] an elementary mutation of 
the $2$--acyclic quiver with superpotential $Q\to \mu_k(Q)$ \cite{zele};
\item[\textit{ii)}] an operation on the $Y$--seed
$\cq_k\colon \mathbb{T}_Q\to \mathbb{T}_{\mu_k(Q)}$ given by the `right intertwinner' \cite{gon1,gon2,kell2}
\begin{equation}\label{nnnz98q}
\cq_k\cdot Y_i= \bee(Y_k) \phi_k(Y_i)\,\bee(Y_k)^{-1}
\end{equation}
where $\phi_k$ is defined in eqn.\eqref{xxxyyy}.
\end{itemize}
Explicitly (here $B_{ij}$ is the exchange matrix of $Q$)
\begin{equation}
\cq_k\cdot Y_i= \begin{cases}
Y_k^{-1} & \text{if } i=k\\
Y_i\prod_{s=1}^{B_{ki}}(1+q^{-1/2+s}Y_k) & \text{if  $B_{ki}\geq 0$ in $\mu_k(Q)$}\\
q^{-B_{ik}^2/2}\, Y_i Y_k^{B_{ik}}\prod_{s=1}^{B_{ik}}(1+q^{1/2-s}Y_k)^{-1} &
\text{if  $B_{ki}\leq 0$ in $\mu_k(Q)$}.\end{cases}
\end{equation}

It follows from the arguments of \cite{CNV,ACCERV1,ACCERV2}
that if there exists a sequence $k_1,k_2,\cdots, k_s$
such that 
\begin{equation}\label{qqqrwq}
\phi_{k_1}\phi_{k_2}\cdots\phi_{k_s}(Y_i)=Y_{\pi(i)}^{-1},\end{equation}
 for some permutation $\pi$
of the nodes of $Q$, then the quantum monodromy is $\mathbb{M}(q)= \mathbb{K}(q)^2$, where the half--monodromy operator
$\mathbb{K}(q)$ is defined\footnote{ More precisely: \textit{defined up to the commutant of the quantum torus algebra}; see \cite{CCV} for more details.} by its adjoint action on $\mathbb{T}_Q(q)$
\begin{equation}\label{hallamona}
\mathrm{Ad}\big(\mathbb{K}(q)\big)=\Pi^{-1}\cq_{k_1}\cq_{k_2}\cdots\cq_{k_s},
\end{equation}
where the operator $\Pi$ implements the automorphism
$Y_i\to Y_{\pi(i)}$. We call a sequence of mutations with the property \eqref{qqqrwq} a \emph{green sequence for} $Q$.

Comparing with eqn.\eqref{nnnz98q}, we see that $\mathbb{M}(q)$ is automatically a finite product of quantum dilogarihms $\mathbb{E}(Y_\gamma)$ \cite{CNV}. Accordingly
to \S.\ref{zzzuuq}, each such dilogarithm is associated with a BPS hypermultiplet; then the BPS spectrum is \emph{finite} consisting precisely of $s$ hypermultiplets. Conversely, \emph{all \underline{generic}} finite chambers arise this way \cite{ACCERV1,ACCERV2}. The combinatorics of quantum cluster mutations,
which act as products of quantum dilogarithms, describes quantum monodromies which
satisfy the \textbf{(genericity condition)} of eqn.\eqref{gencon}.  
In particular, the quantum mutations in eqn.\eqref{hallamona} are \emph{automatically}
ordered according to the phase of the central charge of the corresponding BPS particle.

\subsubsection{Mutations of $2$--acyclic quivers with freely acting automorphisms}

Now suppose the $2$--acyclic finite quiver with superpotential 
$(Q,\cw)$ has a group $\mathbb{G}$ of automorphisms, acting freely on the nodes,
such that the $\mathbb{G}$--orbit category of its Jacobian category is again a bounded Jacobian category of a $2$--acyclic quiver. Then, for all $i\in Q$ and
$g\in\mathbb{G}$, $B_{i\,gi}=0$; moreover, $Q$ has   no
subquivers of the form $i \rightarrow j \rightarrow gi$.
This means that the elementary quantum cluster mutations $\cq_i$ and $\cq_{gi}$ commute for all
$g\in\mathbb{G}$, and we may unambiguously define
\begin{equation}\label{oooq1}
\mu_{\mathbb{G}i}=\prod_{g\in\mathbb{G}}\mu_{g i},\quad \phi_{\mathbb{G}i}=\prod_{g\in\mathbb{G}}\phi_{g i},\quad \cq_{\mathbb{G}i}=\prod_{g\in\mathbb{G}}\cq_{g i},
\end{equation}
where all factors are distinct since $\mathbb{G}$ acts freely. The mutations $\mu_{\mathbb{G}i}$
transform a quiver which is $\mathbb{G}$--symmetric into a quiver which is $\mathbb{G}$--symmetric; at the level of the quotient quiver,
the mutation $\mu_{\mathbb{G}i}$ induces the elementary mutation at the node $\mathbb{G}i$
(which we also denote as $\mu_{\mathbb{G}i}$).
The same remark applies to the associated change of basis $\phi_{\mathbb{G}i}$. \smallskip

Suppose the covering $\cn=2$ theory has a \emph{finite} BPS chamber which contains a $\mathbb{G}$--invariant point at which condition \eqref{gencon} holds.
At the $\mathbb{G}$--invariant point, the BPS states form complete $\mathbb{G}$--orbits, while all BPS states in a $\mathbb{G}$--orbit have the same phase. Since the product
\eqref{qqqrwq}  is automatically phase--ordered,
all factors in the same $\mathbb{G}$--orbit should
appear grouped together. Then we have
\begin{equation}\label{rrrrbbbbm93}
\phi_{k_1}\phi_{k_2}\cdots\phi_{k_s}=
\phi_{\mathbb{G}i_1}\phi_{\mathbb{G}i_2}\cdots
\phi_{\mathbb{G}i_t},\qquad t=s/|\mathbb{G}|.
\end{equation}  
But the \textsc{rhs} is canonically interpreted
as a green sequence for the quotient quiver, and hence as the BPS spectrum in a corresponding  \emph{finite} chamber of the quotient theory having $s/|\mathbb{G}|$ hypers.

Hence the generic finite chambers behave well under
\emph{strict--sense} Galois covers of $\cn=2$ theories. 
We recall an elementary example
of eqn.\eqref{rrrrbbbbm93} which reproduce (for finite chambers) the results of \S.\ref{su2bpscov} and \S.\ref{rrrrr6519} in terms of cluster algebra combinatorics. For $k\mid \gcd(p,q)$,
consider the generalized SYM cover with simply--laced gauge group $G$, i.e. 
\be\label{covrfffxxx}
\C \widehat{A}(p,q)\boxtimes G \longrightarrow \C \widehat{A}(p/k,q/k)\boxtimes G
\ee
whose green sequences (and hence finite BPS chambers) are given in appendix B of ref.\cite{CDZG}. From their explicit form, it is pretty obvious\footnote{ 
 The finite chamber of $\C \widehat{A}(p,q)\boxtimes G$ consists
of $(p+q)$ copies of the maximal chamber spectrum of Argyres--Douglas of type $G$ \cite{CDZG}, which is trivially a $k$--fold cover of the
$(p+q)/k$ copies in the finite chamber of $\C \widehat{A}(p/k,q/k)\boxtimes G$.} that green functions for covering and quotient models are related as in eqn.\eqref{rrrrbbbbm93}.

\subsubsection{Example: BPS spectra of branched covers of genus zero $\cs[A_1]$ models}\label{091cb}

As a new example, we consider the $\Z_k$ Galois covers of $SU(2)^2$ SYM coupled to a bifundamental described by the quiver in figure
\ref{zerogenus}. We note that all mutations at
white nodes $\circ$ (resp.\! black nodes $\bullet$)
commute. We then write $\phi_\circ$, $\phi_\bullet$
and $\phi_\ast$ for the product of all mutations at nodes denoted by the same symbol in figure \ref{zerogenus}; they are products of mutations 
of the form $\phi_{\Z_k i}$. We have
\begin{equation}\label{xyaz}
\phi_\circ\phi_\bullet\phi_\ast\phi_\circ(Y_i)=Y^{-1}_{\pi(i)},
\end{equation}
where $\pi$ is the permutation which fixes the nodes $\ast$ and interchanges the black/white
pairs $\bullet,\circ$ which are connected by an arrow belonging to an oriented triangle.

Eqn.\eqref{xyaz} gives a finite chamber of the
covering theory with $7k$ hypermultiplets which is a
$k$--fold cover of the seven hyper  finite chamber of 
$SU(2)^2$ with a bifundamental.

\subsection{$\Z_k$ coverings of $SU(2)$ SQCD: friezes and $SL_2$--tilings}\label{tilings123z}

Of course, we are more interested in infinite chambers.
The example of \S.\,\ref{sss;babyex} was particular simple since the quantum monodromies
of the covering and quotient theories could be seen as acting on the same Hilbert space and hence
compared directly. 
We wish to generalize 
the relation between the cover and quotient monodromies
to the $\Z_k$ covering
of $SU(2)$ SYM coupled to arbitrary
Argyres--Douglas systems of types $D_{p/k}$ and $D_{q/k}$ by SYM coupled to AD systems of types $D_p$, and $D_q$,
which corresponds to the cover of $\C$--categories
\begin{equation}\label{covvvo}
\C \widehat{A}(p,q)\longrightarrow \C\widehat{A}(p/k,q/k),\qquad\text{where } k\mid \gcd(p,q).
\end{equation}  
To this end, we take advantage of the description of the quantum monodromy $\mathbb{M}(q)$ in terms of quantum cluster 
algebras \cite{CNV,gon1,gon2,kell2,ysyst},
and of the existence of linear recursions
 for cluster variables of affine quivers \cite{kell-lin}
 which, in the present context, may be
 phrased in the suggestive language of
 \emph{friezes} and $SL_2$--tilings of the plane \cite{frises}.

Although the results below will be valid for all
covers \eqref{covvvo}, to simplify the presentation
 we first assume $p=q$. We consider the
quiver $\widehat{A}(p,p)$ where sources and sinks alternate; we denote the sources as $\circ_a$ and the sinks as $\bullet_a$ and we take the index $a\in\Z$ with the mod $p$ identifications
$\circ_{a+p}\sim \circ_a$, $\bullet_{a+p}\sim \bullet_a$. The labels are chosen as follows
\begin{equation}\label{aaaqqqcc}
\xymatrix{\cdots \ar[r] & \bullet_a & \circ_{a+1}\ar[l]\ar[r] & \bullet_{a+1} &\circ_{a+2}\ar[l]\ar[r] & \bullet_{a+2} &\circ_{a+3}\ar[l]\ar[r] & \bullet_{a+3} & \ar[l]\cdots}
\end{equation}
Mutations at nodes ${\bullet_a}$ (resp.\! ${\circ_a}$) commute between themselves.
Following eqn.\eqref{oooq1}, we write
\begin{equation*}
\mu_\bullet=\prod_{a=1}^p\mu_{\bullet_a},\quad 
\mu_\circ=\prod_{a=1}^p\mu_{\circ_a},
\quad
\phi_\bullet=\prod_{a=1}^p\phi_{\bullet_a},\quad 
\phi_\circ=\prod_{a=1}^p\phi_{\circ_a},
\quad\cq_\bullet=\prod_{a=1}^p\cq_{\bullet_a},\quad \cq_\circ=\prod_{a=1}^p\cq_{\circ_a}.
\end{equation*}
One has $\mu_\bullet(Q)=\mu_\circ(Q)=Q^\mathrm{op}$ and $\mu_\circ\mu_\bullet(Q)=Q$. Then
\begin{equation}
\phi_\bullet\!\begin{pmatrix}Y_{\circ_a}\\
Y_{\bullet_a}\end{pmatrix}=\begin{pmatrix}Y_{\circ_a}\\
Y_{\bullet_a}^{-1}\end{pmatrix},\qquad
\phi_\circ\phi_\bullet\!\begin{pmatrix}Y_{\circ_a}\\
Y_{\bullet_a}\end{pmatrix}=\begin{pmatrix}Y_{\circ_a}^{-1}\\
Y_{\bullet_a}^{-1}\end{pmatrix},
\end{equation}
so that the half--monodromy is $\mathrm{Ad}(\mathbb{K}(q))=\cq_\circ\cq_\bullet$. Then
\begin{equation}\label{muqwer}
\cq_\bullet\!\begin{pmatrix}Y_{\circ_a}\\
Y_{\bullet_a}\end{pmatrix}=_Q
\begin{pmatrix}Y_{\circ_a}\prod\limits_{\circ_a\to\bullet_b\atop \text{in }Q}\big(1+q^{1/2}Y_{\bullet_b})\\
Y_{\bullet_a}^{-1}\end{pmatrix},
\quad
\cq_\circ\!\begin{pmatrix}Y_{\circ_a}\\
Y_{\bullet_a}\end{pmatrix}=_{\mu_\bullet(Q)}
\begin{pmatrix}Y_{\circ_a}^{-1}\\
Y_{\bullet_a}\prod\limits_{\circ_b\to\bullet_a\atop \text{in }Q}\big(1+q^{1/2}Y_{\circ_b})\end{pmatrix}
\end{equation}
We promote each operator $Y_i$ to
a sequence of operators $\{\tilde Y_{i,n}\}_{n\in\Z}$ by setting $\tilde Y_{\bullet_a,0}\equiv Y_{\bullet_a}$, $\tilde Y_{\circ_a,0}\equiv Y_{\bullet_a}$ and
\begin{equation}
\begin{gathered}
\tilde Y_{\bullet_a,2n+1}=\cq_\bullet \tilde Y_{\bullet_a,2n},\qquad
\tilde Y_{\bullet_a,2n+2}=\cq_\circ \tilde Y_{\bullet_a,2n+1},\\
\tilde Y_{\circ_a,2n+1}=\cq_\bullet \tilde Y_{\circ_a,2n},\qquad
\tilde Y_{\circ_a,2n+2}=\cq_\circ \tilde Y_{\circ_a,2n+1}
\end{gathered}
\end{equation}
One has $\tilde Y_{\bullet_a,2n+1}=
\tilde Y_{\bullet_a,2n}^{-1}$ and
$\tilde Y_{\circ_a,2n}=
\tilde Y_{\circ_a,2n-1}^{-1}$, so
half the operators are redundant, and we
 simplify the notation by writing
\begin{equation}
Y_{\bullet_a,n}\equiv \tilde Y_{\bullet_a,2n},
\qquad Y_{\circ_a,n}=\tilde Y_{\circ_a,2n+1}.
\end{equation}
In terms of the sequence of operators $\{Y_{\bullet_a,n},Y_{\circ_a,n}\}_{a, n\in\Z}$
the action of the quantum monodromy is given by shifts of the index $n$
\begin{equation}\label{monomiawe}
\mathrm{Ad}(\mathbb{M}(q)^m)Y_{i,n}=Y_{i,n+2m},\qquad i=\bullet_a, \circ_a.
\end{equation} 
From eqn.\eqref{muqwer}, the sequence of operators $Y_{i,n}$ satisfies 
the quantum recursion relation
\begin{equation}\label{eee41m}
Y_{i,n}\,Y_{i,n+1}= \prod_{i\to j}\big(1+q^{1/2} Y_{j,n+1}\big)\prod_{j\to i}\big(1+q^{1/2} Y_{j,n}\big),
\end{equation}
where all
arrows are now referred to the original form of the quiver $Q$. Hence the action of
the quantum monodromy $\mathbb{M}(q)$ on the quantum torus algebra $\mathbb{T}_Q(q)$
is described by the solutions to the recursion relations
\eqref{eee41m}, see eqn.\eqref{monomiawe}.
\medskip

Our goal is to show that the action of
the monodromy for the quiver
$\widehat{A}(p,p)$ is a $k$--fold cover of the
action of the monodromy for the quiver $\widehat{A}(p/k,p/k)$. This is the same as saying that the solutions to the
recursion relations \eqref{eee41m} for
$\widehat{A}(p,p)$ are `$k$--fold covers' of
the solutions for $\widehat{A}(p/k,p/k)$.
(We could be more general, and consider the $k$--fold cover $\widehat{A}(p,q)$ of $\widehat{A}(p/k,qk)$; we shall comment on such generalization momentarily). 
To show this property of \eqref{eee41m}, we take advantage
from the fact that its quantum 
solutions are determined, up to
the correct prescription
for the operator ordering, by the corresponding classical solutions, i.e.\! by their limit
as $q\to 1$ \cite{CNV}. In this limit, the half--monodromy action
for $\widehat{A}(p,p)$, $Y_{i,n}\to Y_{i,n+2}$ is given by a rational map $\C^{2p}\xrightarrow{K(p)} \C^{2p}$
and likewise for $\widehat{A}(p/k,p/k)$
by a rational map $\C^{2p/k}\xrightarrow{K(p/k)} \C^{2p/k}$. Consider the diagonal map 
\begin{equation}
\Delta\colon \C^{2p/k}\longrightarrow \overbrace{\C^{2p/k}\oplus \C^{2p/k}\oplus \cdots\oplus \C^{2p/k}}^{k\ \text{factors}}\cong \C^{2p}
\end{equation}
which embeds $\C^{2p/k}$ into the $\Z_k$ fixed subset of $\C^{2p}$. To prove the covering property of the quantum monodromy is equivalent to show that
the following diagram of rational maps is commutative
\begin{equation}\label{zzz123v}
\begin{gathered}
\xymatrix{\C^{(p+q)/k}\ar[d]_\Delta\ar[rr]^{K(p/k,q/k)}&& \C^{(p+q)/k}\ar[d]^\Delta\\
\C^{p+q}\ar[rr]_{K(p,q)} && \C^{p+q}}
\end{gathered}
\end{equation}
This is already evident from the form of eqn.\eqref{eee41m};
indeed the recursion relation is invariant
under translation of the first index 
\begin{equation}
i\equiv \bullet_a,\circ_a\longrightarrow \bullet_{a+\ell}, \circ_{a+\ell} \equiv i[\ell]
\end{equation} for all $\ell\in\mathbb{N}$.
Then, if the initial values $\{Y_{\bullet_a,0}, Y_{\circ_a,0}\}$ are chosen to be
periodic in $a$ of period $\ell$, the solution
$\{Y_{\bullet_a,n}, Y_{\circ_a,n}\}_{n\in\Z}$ is also periodic in $a$ with the same period. By definition, $\{Y_{\bullet_a,0}, Y_{\circ_a,0}\}$
for the $\widehat{A}(p,q)$ case is periodic of period $p+q$, while for $\widehat{A}(p/k,p/k)$ of period $(p+q)/k$. Eqn.\eqref{zzz123v} then says that, if we
make a \emph{special} choice for the $(p+q)$ periodic initial datum of the $\widehat{A}(p,q)$ model which is actually $(p+q)/k$ periodic, we get a $(p+q)/k$ periodic solution which we may identify with a solution of the recursion for the $\widehat{A}(p/k,q/k)$ quiver.
This is confirmed by the explicit solution of the recursion \eqref{eee41m} that we now describe.

\subsubsection{Solutions of the recursion: $SL(2,\C)$--tiling of the plane}
It is convenient to 
parametrize the $Y_{i,n}$ in terms of
new variables $X_{i,n}$ (in the cluster algebraic
language, we are re--expressing the $Y$--seed in terms of the $X$--seed \cite{fomzele}) as
\begin{equation}
Y_{i,n}= \prod_{i\to j} X_{j,n}\prod_{j\to i} X_{j,n-1}.
\end{equation} 
In terms of the $X_{i,n}$ the recursion becomes
\begin{equation}\label{ffrise}
X_{i,n+1}X_{i,n}=1+\prod_{i\to j}X_{j,n+1}\prod_{j\to i}X_{j,n}.
\end{equation}
Written in this form, the recursion is well known: it is the \emph{frieze} of the opposite quiver
$Q^\mathrm{op}$ \cite{kell-lin,frises}. Friezes exist for all acyclic quivers; they are defined by a source sequence \cite{kell-lin}.
In the example of the alternating $\widehat{A}(p,p)$  the source sequence is $(\bullet_1,\cdots,\bullet_p,\circ_1,\cdots\,
\circ_p)$. Written in terms of friezes, our
formulae hold for all $\widehat{A}(p,q)$ covers.

For $\widehat{A}(p,q)$
the solutions to the frieze recursion \eqref{ffrise} may be written in terms
of $SL_2$--tilings of the two dimensional square lattice $\Z^2$ \cite{frises}. Let $R$ be a commutative ring with unit:
an \emph{$SL(2,R)$--tiling of the plane} is an assignment of an element $r(i,j)\in R$ to each vertex $(i,j)\in\Z^2$ of the square lattice such that each plaquette
in the lattice, seen as a $2\times 2$ matrix,
belongs to $SL(2,R)$, that is,
\begin{equation}
\left|\;\begin{matrix}r(i,j) & r(i+1,j)\\
r(i,j-1)& r(i+1,j-1)\end{matrix}\;\right|\equiv r(i,j)\,
r(i+1,j-1)-r(i+1,j)\,r(i,j-1)=1.
\end{equation}

\begin{figure}
 \begin{tiny}
 \begin{equation*}
 \xymatrix{\ar@{..}[dddddd]\ar@{..}[rrrrrrrrrr] &\ar@{..}[dddddd]&\ar@{..}[dddddd]&\ar@{..}[dddddd]&\ar@{..}[dddddd]&\ar@{..}[dddddd]&\ar@{..}[dddddd]&\ar@{..}[dddddd]&\ar@{..}[dddddd]&\ar@{..}[dddddd]&\ar@{..}[dddddd]\\
 \ar@{..}[rrrrrrrrrr] &&&&&&(2,2)\ar@{=>}[r]^{m_5}&(2,2)\ar@{<=}[u]_{m_6} 
 &&&\\
 \ar@{..}[rrrrrrrrrr] &&&&&(1,1)\ar@{=>}[r]^{m_3}&(2,1)\ar@{<=}[u]_{m_4}&&&&\\
 \ar@{..}[rrrrrrrrrr] &&&&(0,0)\ar@{=>}[r]^{m_1}&(1,0)\ar@{<=}[u]_{m_2}&&&&&\\
 \ar@{..}[rrrrrrrrrr] &&&(-1,-1)\ar@{=>}[r]^{m_{-1}}&(0,-1)\ar@{<=}[u]_{m_{0}}&&&\bigstar\ar@{--}[lll]\ar@{--}[uuu]&&&\\ 
  \ar@{..}[rrrrrrrrrr] &&\ar@{=>}[r]^{\text{\begin{large}$\phantom{\big|}L$\end{large}}}&\ar@{<=}[u]_{m_{-2}}&&&&&&&\\
   \ar@{..}[rrrrrrrrrr] &\ar@{=>}[r]&\ar@{<=}[u]&&&&&&&&}
 \end{equation*}
 \end{tiny}
 \caption{\label{frontier}The frontier $L$ in $\Z^2$ corresponding to
 the quiver $\widehat{A}(p,p)$ with the alternating orientation for any $p$. E.g.\! to the point $\star$ are associated the sequences of initial values $X_{(-1,0)},X_{(0,0)},X_{(1,0)},X_{(2,0)},X_{(3,0)},X_{(4,0)},X_{(5,0)}$
 and of matrices $m_0,
 m_1,m_2,m_3,m_4,m_5$.}
 \end{figure}
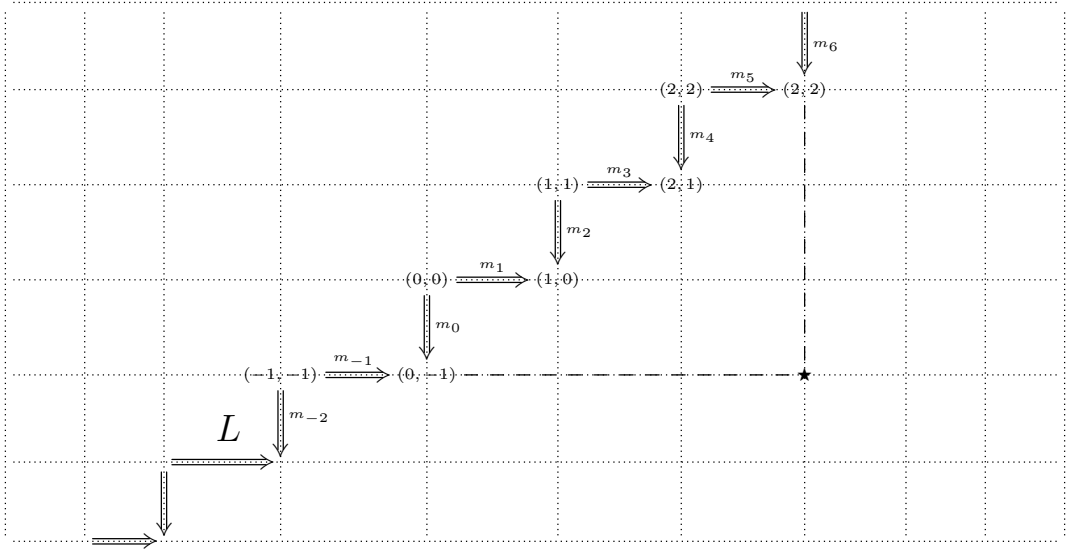

The rule to express the solutions to the frieze \eqref{ffrise} for a quiver $\widehat{A}(p,q)$
in terms of $SL(2,\C)$--tilings of the plane is as follows \cite{frises}. As in figure \eqref{aaaqqqcc},
we consider the \emph{infinite cover} $\widehat{A}(p,q)^\infty$
of the acyclic quiver $\widehat{A}(p,q)$ obtained
by numbering its arrows by an integer mod $p+q$,
and then forgetting the identifications $i\sim i+p+q$ of the nodes. $\widehat{A}(p,q)^\infty$ is then 
an infinite sequence, labelled by integers
$k\in\Z$, of direct $\rightarrow$
and inverse arrows $\leftarrow$
which repeat in $k$ with period $p+q$.
We number nodes in $\widehat{A}(p,q)^\infty$
so that the $k$ node lays between the $(k-1)$--th and $k$--th arrow.
We embed $\widehat{A}(p,q)^\infty$ as a continuous path $L$ along the links of the square plane lattice according to the rule that a direct arrow
corresponds to a horizontal link and an inverse arrow to a vertical link, see figure \ref{frontier} for the example of
 $\widehat{A}(p,p)$ with the alternating orientation.
The path $L$ so obtained is called the \emph{frontier}
 of the $SL_2$--tiling. Sources/sinks of $\widehat{A}(p,q)^\infty$ are mapped to the corners of $L$.
 
The initial condition of the recursion relation
is specified by assigning to the vertex $k$ of the frontier $L$ the complex number $X_{k,0}$,
which repeats with periodicity $p+q$ in $k$, i.e.\!
\begin{equation}
X_{k+p+q,0}=X_{k,0}
\end{equation}
To the $k$--th link
 of the frontier $L$ we associate the
$2\times 2$ matrix $m_k$ which also repeats with
period $p+q$, i.e.\! $m_{k+p+q}=m_k$,
\begin{equation}\label{qqqqxz12}
m_k=\begin{cases}
\begin{pmatrix} X_{k-1,0} & 1\\
0 & X_{k,0}\end{pmatrix} &\text{the $k$--th arrow is direct ($\equiv$ horizontal link in $L$)}\\
\begin{pmatrix} X_{k,0} & 0\\
1 & X_{k-1,0}\end{pmatrix} &\text{the $k$--th arrow is inverse ($\equiv$ vertical link in $L$).}
\end{cases}
\end{equation}

The $SL(2,\C)$--tiling with boundary condition
$X_{i,0}$ is then constructed by assigning to each 
vertex $(i,j)$ below the frontier\footnote{ The pattern above the frontier is symmetric, so we may limit ourselves to the pattern below it \cite{frises}.} the complex number $c(i,j)$ determined as follows. One connects the given point $(i,j)$ (denoted by $\star$ in figure \ref{frontier}) to the frontier $L$ by drawing the vertical and horizontal segments; the intersection points determine a finite
segment $L_{(i,j)}$
 in the frontier $L$, hence a finite sequence of 
 points with attached values
 $X_{k_0},\;X_{k_0+1}\;\cdots,\;X_{k_0+\ell+1}$,
 as well as of links with associated matrices $m_{k_0+1},\;m_{k_0+2},\;
\cdots\;m_{k_0+\ell+1}$, see figure \ref{frontier} for an example.
Consider the subsequence obtained by forgetting the first and the last matrices, and let $M(i,j)$
the product of the remaining ones
\begin{equation}
M(i,j)=m_{k_0+1}m_{k_0+2}\cdots \m_{k_0+\ell}.
\end{equation}
Then 
\begin{equation}\label{12qwe}
c(i,j)=\frac{1}{\prod_{s=1}^\ell X_{k_0+s,0}}\begin{pmatrix}1 & X_{k_0,0}\end{pmatrix} M(i,j) \begin{pmatrix}1 \\ X_{k_0+\ell+1}\end{pmatrix},
\end{equation}
is the unique $SL(2,\C)$--tiling on the plane
which reduces to the assigned values $X_{k,0}$
on the frontier $L$ \cite{frises}. This is obvious
for the alternating case, and shown to be true in general in ref.\cite{frises}.

Let $(i_k,j_k)\in\Z^2$ be the point on the frontier $L$
which corresponds to the node $k\in\widehat{A}(p,q)^\infty$. Then the explicit solution of the frieze
\eqref{ffrise}, and hence to the (classical)
$Y$--seed recursion \eqref{eee41m} is \cite{frises}
\begin{equation}
X_{k,n}= c(i_k+n,j_k-n),\qquad n\in\Z.
\end{equation}
The solution is periodic in $k$,
$X_{k+p+q,n}=X_{k,n}$ since the boundary condition was periodic. Again, the statement is obvious for the alternating $\widehat{A}(p,p)$
and true in general by ref.\!\cite{frises}.

From the $SL_2$--tiling viewpoint,
the Galois covering property of the action of the monodromy is just the self--similarity of
the $SL_2$--tiling under `period rescaling' 
$(p,q)\to(p\ell, q\ell)$ which is an immediate consequence of the identity of infinite quivers
\begin{equation}\label{00012cx}
\widehat{A}(p,q)^\infty\equiv \widehat{A}(p\ell,q\ell)^\infty.
\end{equation} 
Note that from the explicit solution
\eqref{12qwe} it follows that the variables $X_{k,n}$
(and hence $Y_{k,n}$), as a function of $n$ satisfy a linear recursion relation of finite length \cite{frises}
(see also \cite{kell-lin}). 
Comparing with eqn.\eqref{monomiawe},
we see that the monodromy acts by linear recursions, a fact whose physical meaning is discussed in \cite{ysyst}.

\subsubsection{Extension to general class $\cs[A_1]$ theories}

The previous explicit construction of the (classical limit of the) action of the monodromy $\mathbb{M}(q)$ in terms of friezes may be extended to all class $\cs[A_1]$ theories; the extension is particularly easy when all punctures on the Gaiotto surface $\Sigma_G$ are irregular.

Above we have thought of the Euclidean algebra 
$\C\widehat{A}(p,q)$ as a hereditary algebra;
but it is also a string algebra and hence 
all its indecomposable modules may be described in the string/band language (see appendix \ref{stringbands}).
The points of the lattice $\Z^2$ below the frontier
(cfr.\! figure \ref{frontier}) are naturally identified (modulo periodicity) with the \emph{string modules}
of $\C\widehat{A}(p,q)$: to each point $(i,j)$
we associate the string given by the sequence of direct/inverse arrows in the corresponding
segment of frontier $L_{(i,j)}$ (see figure \ref{frontier}).
Thus we see that the map
\begin{equation}
\text{(string module)}\longmapsto \text{(frieze variable)}\equiv \text{$X$--seed}
\end{equation}
is given by replacing each direct (resp.\! inverse) arrow in the string by  a $2\times 2$ matrix $m_k$
having the form in the upper line (resp.\! bottom line) of eqn.\eqref{qqqqxz12}, and multiplying by the appropriate factors for the two end points of the string.
That this prescription works in general (in the string case) is proven in ref.\!\cite{string frieze}.
From the sequence if $X$--seeds we get the sequence of $Y$--seeds in the standard way \cite{fomzele}; then the classical action of $\mathbb{M}(q)$ is given by a shift of the
$Y$--seed sequence.

In particular, this shows the covering property for the action of the monodromy of the \emph{dessins d'enfants} examples of section 4, whose algebras are string algebras.

\subsection{Covers of $\cn=2$ gauge theories with gauge group $G$}

The above analysis may be extended
to the $\Z_k$ coverings of $G$ gauge theories 
\begin{equation}\label{sssza}
\C \widehat{A}(p,q)\boxtimes G\to
\C \widehat{A}(p/k,q/k)\boxtimes G
\end{equation}
 discussed in \S.\ref{kkk12qn}. The monodromy of pure SYM also
acts by linear recursions (called $Q$--systems \cite{qsyst1,qsyst2}), and the situation looks rather similar to the one
discussed above as a consequence of asymptotic freedom of the associated $\cn=2$ QFTs \cite{ysyst}. For simplicity, we consider the case $p=q$ and use the square form of the
quiver $\widehat{A}(p,p)\,\square\,G$. 
Again, as in \S.\ref{tilings123z} we replace the alternating affine quiver $\widehat{A}(p,p)$ by its  
infinite cover $\widehat{A}(p,p)^\infty$,
so that we work with the infinite
cover product quiver $\widehat{A}(p,p)^\infty\,\square\,G$.
By standard manipulations \cite{CNV,keller}, 
similar to the ones
in \S.\ref{tilings123z}, we reduce the action of the quantum monodromy in the classical limit
to the recursion relation of the 
$(A_\infty,G)$ $Y$--system
\begin{gather}
Y_{k,a,n-1}Y_{k,a,n+1}= \frac{(1+Y_{k-1,a,n})(1+Y_{k+1,a,n})}{\prod_{b\in G} (1+Y^{-1}_{k,b,n})^{A_{ab}}}, \quad k\in\Z,\ a\in G,\ n\in \Z,
\end{gather} 
where the matrix $A_{ab}$ is related to the  Cartan matrix of the (simply--laced) group $G$ by
\begin{equation} C_{ab}=2\,\delta_{ab}-A_{ab}.
\end{equation}
In this notation the monodromy acts by a shift in $n$
\begin{equation}
Y_{k,a,n}\xrightarrow{\ \ \mathbb{M}\ \ }Y_{k,a,n+2 h(G)}
\end{equation}
where $h(G)$ is the Coxeter number of $G$.

One reduces the $Y$--system for the infinite cover
$\widehat{A}(p,p)^\infty\,\square\,G$ to the $Q$--system for the finite quiver $\widehat{A}(p,p)\,\square\,G$
just by taking the initial values of the recursion\footnote{ In the physical problem only half the data $\{Y_{k,a,0}, Y_{k,a,1}\}$ are independent.},
$Y_{k,a,0}$ and $Y_{k,a,1}$, to be periodic in $k$
of period $2p$. Again, the identity \eqref{00012cx}
(with $p=q$) guarantees the covering
property for the action of the monodromies
in the $\Z_k$ covering of $G$ gauge theories
\eqref{sssza} (i.e.\! the analogue of the diagram \eqref{zzz123v} is commutative).

\section{BPS spectra of genus one class $\cs[A_1]$ theories}\label{genusone}

In this section we illustrate the Galois cover of BPS spectra in the context of the class $\cs[A_1]$ theories \cite{GMN:2009,Gaiotto} defined by a meromorphic quadratic differential
$\phi_2$ over a genus one curve $E$.
The techniques of section 2 will allow to describe the BPS spectra both explicitly and elegantly.

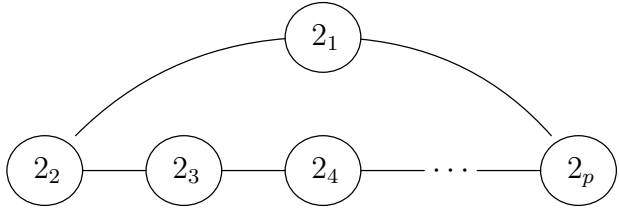
\begin{figure}
 \begin{equation*}
  \begin{gathered}\xymatrix{&& *++[o][F-]{\phantom{\big|}2_1\,}\ar@{-}@/_1.2pc/[lld]\ar@{-}@/^1.2pc/[rrd]\\
 *++[o][F-]{\phantom{\big|}2_2\,}\ar@{-}[r]& *++[o][F-]{\phantom{\big|}2_3\,}\ar@{-}[r]& *++[o][F-]{\phantom{\big|}2_4\,}\ar@{-}[r]&\cdots\ar@{-}[r]& *++[o][F-]{\phantom{\big|}2_p\,}}
  \end{gathered}
 \end{equation*}
\caption{The (gauge) quiver representations of the $T_{1,\,p}$ theory. Each circle represents a $SU(2)$ $\cn=2$ SYM sector, while each link stands for a bifundamental quark hypermultiplet. }\label{quiverg1pp}
\end{figure}

We start by considering the theory $T_{1,\,p}$ defined by a quadratic differential with $p\geq 1$ second order poles (regular 
punctures).
In some corner of its parameter space, $T_{1,\,p}$ is the quiver gauge theory in figure \ref{quiverg1pp}
with gauge groups $SU(2)$ on the nodes and
bifundamental quarks on the links.  

Let $d\mid p$. We claim that (at special points in parameter space) the $\cn=2$ theory $T_{1,\,p}$ is a 
Galois $\Z_d$--cover of the $T_{1,\,p/d}$ theory. In particular,
taking $d=p$, $T_{1,\,p}$ is a $\Z_p$--cover of
$T_{1,\,1}$, which is $\cn=2^*$ SYM with gauge group $SU(2)$. Thus we may construct the $T_{1,\,p}$ BPS spectrum
(at $\Z_p$ symmetric points) by pulling back
the known spectrum of $\cn=2^*$ which has
a simple Lie theoretic description.

The $\Z_p$--cover may be described in terms of Gaiotto geometries:
the elliptic curves $E_p$, $E_{p/d}$ associated, respectively, to $T_{1,\,p}$
and $T_{1,\,p/d}$ are isogenous, 
\begin{equation}\label{isogeny}
E_p\xrightarrow{\ \phi\ } E_{p/d},\qquad \deg \phi=d.
\end{equation}
$\phi$, being an isogeny, is
a (holomorphic) covering map and an Abelian group
homomorphism; its kernel is a $\Z_d$ subgroup of $E_p$, which is the Galois group of the cover. The curve $E_{p/d}$ has $p/d$ regular punctures at $z_1,\cdots, z_{p/d}$ (which we assume to be in general position), and their pre--images in $E_p$
form a set of $p$ regular punctures invariant
under the covering group $\Z_d$.
The $T_{1,\,p}$ theory at a $\Z_d$ symmetric point is specified by a quadratic differential of the form 
\begin{equation}
\phi_2(E_p)=\phi^*\phi_2(E_{p/d})
\end{equation} for some quadratic differential $\phi_2(E_{p/d})$ on $E_{p/d}$ having second order poles at the $z_i$.
For instance, in the special case $d=p$, the quadratic
differential which specifies the $T_{1,\,p}$ theory
takes the form $\phi^*(\lambda\,\wp(z, \tau)\,dz^2+\mu\, dz^2)$
for some constants $\lambda,\mu$.

In the $d=p$ case we may see the Gaiotto geometry as a \emph{pair} $(E_p, C)$ where $E_p$ is an elliptic curve and $C\subset E_p$ is a $\Z_p$ subgroup. As it is well known \cite{elliptic}, the isomorphism classes of such pairs, called \emph{enhanced elliptic curves for $\Gamma_0(p)$}, are parametrized
 by the upper half plane  $\ch$ modulo the congruence modular subgroup $\Gamma_0(p)$
 \begin{gather}
 \tau \sim \frac{a\tau+b}{c\tau+d},\qquad \tau\in\ch,\ \intertext{where}
 \begin{pmatrix}a & b\\c& d
 \end{pmatrix} \in\Gamma_0(p) \equiv \left\{\begin{pmatrix}a & b\\c& d
 \end{pmatrix}\in SL(2,\Z)\; \Bigg|\; c=0 \mod p\right\}.
 \end{gather}
 A given point $\tau$ in the modular curve
 $Y_0(p)\equiv \ch/\Gamma_0(p)$ corresponds to the (isomorphism class) of the $\Z_p$ invariant
 Gaiotto geometry given by a complex torus of
 periods $(1,\tau)$ with regular punctures at the $p$ points
 $z_k=k/p$ ($k=0,1,\dots, p-1$). 
 The cyclic covering map $\phi$ in eqn.\eqref{isogeny}  then is simply the map between the tori
 of periods $(1,\tau)$ and $(1,p\tau)$ given by multiplication by the integer $p$, $z\mapsto p\,z$. 
 
% Related considerations from the viewpoint of the $tt^*$ geometry of the corresponding 2d (2,2) models are presented in appendix \ref{s:tts}.

\subsection{Categorical covers and duality frames}\label{Sframes}
Let us consider the particular quiver $Q_p$ in the
mutation class of the $T_{1,\,p}$ model
\begin{equation}\label{specialpytttaquiH}
\begin{gathered}\xymatrix{
&& \circ_{a-1}\ar@<0.4ex>[dd]\ar@<-0.4ex>[dd] & & \circ_{a}\ar@<0.4ex>[dd]^{B_a}
\ar@<-0.4ex>[dd]_{A_a} & && \circ_{a+k}\ar@<0.4ex>[dd]\ar@<-0.4ex>[dd] &\\
{\ }&{\;\ast_{a-2}}\ar@{..}[l]|{\cdots\cdots\cdots} \ar[ru] & &\ast_{a-1}\ar[ul]\ar[ur]^{\xi_a} & &\ast_{a} \ar[ul]_{\eta_a}\ar@{..}[r]|{\cdots\cdots\cdots}&{\;\ast_{a+k-1}}\ar[ur]& &{\ast_{a+k}\;}\ar[ul]\ar@{..}[r]|{\cdots\cdots\cdots}&\\
&& \bullet_1\ar[ul]\ar[ur] & &\bullet_{a}\ar[ul]^{\xi_a^*}\ar[ur]_{\eta_a^*} & &&\bullet_{a+k}\ar[ur]\ar[ul]&}\end{gathered}
\end{equation}
where the 
index $a$ in the node labels takes value in $\Z$ with  periodic 
identification
mod $p$
\begin{equation}
\ast_{a+p}\sim \ast_a,\qquad \circ_{a+p}\sim \circ_a,\qquad \bullet_{a+p}\sim\bullet_a.
\end{equation}
In the usual Lagrangian regime, where $T_{1,\,p}$ is the weakly coupled gauge theory in figure \ref{quiverg1pp}, we identify each Kronecker subquiver $\circ_a\rightrightarrows \bullet_a$ of $Q_p$ with a vertex of the gauge quiver, and the node $\ast_a$ with the
flavor symmetry associated to the $a$--th bifundamental link. This is \emph{not} the situation we wish to study: 
in this paper we are mainly interested in other `strongly coupled' regimes of the $T_{1,\,p}$ model which have a nicer Galois cover interpretation.\footnote{ Of course, the ordinary weak coupling may also be seen as a Galois cover
of $T_{1,\,1}$.}

The quiver $Q_p$ has a $\Z_p$ automorphism group, acting freely on the nodes, given by
\begin{equation}
\Z_p\ni k\colon (\circ_a,\bullet_a,\ast_a)\longmapsto (\circ_{a+k},\bullet_{a+k},\ast_{a+k}).
\end{equation}
Without loss of generality\footnote{ The BPS spectrum is independent of the $\lambda_a$ as long as they are not zero.} we may choose the free superpotential parameters $\lambda_a\in\C^\times$ associated to the punctures (see \cite{LF1,ACCERV1}) to be all equal,
$\lambda_a=\lambda\neq0$; then the superpotential 
\begin{equation}\label{supesuper}
\cw_p=\sum_{a\in\Z_p} \Big(\xi_a^* A_a \xi_a+\eta_a^* B_a\eta_a+\lambda\, \eta_a^*A_a\eta_a\,\xi_{a+1}^*B_{a+1}\xi_{a+1}\Big)\end{equation}
is also $\Z_p$--invariant, and the group $\Z_p$ acts by automorphisms of
the (bounded) Jacobian category $\cj_p\equiv\C Q_p/(\partial\cw_p)$. This group acts freely on
the objects of $\cj_p$.
If $d\mid p$, the category $\cj_p$
is a Galois cover of the category $\cj_{p/d}$
with Galois group $\Z_d\simeq \Z_p/\Z_{p/d}$.
In particular, $\cj_p$ is a $\Z_p$--cover of the
category $\cj_1$ corresponding to 
 $SU(2)$ $\cn=2^*$ whose (unique) quiver is the Markov quiver $Q_1$
\begin{equation}
\begin{gathered}
\xymatrix{\circ\ar@<0.4ex>[dd]\ar@<-0.4ex>[dd]\\
&& \ast\ar@<0.4ex>[ull]\ar@<-0.4ex>[ull]\\
\bullet\ar@<0.4ex>[urr]\ar@<-0.4ex>[urr]}
\end{gathered}
\end{equation}
endowed with the superpotential $\cw_1$.
The three nodes $\circ$, $\bullet$, $\ast$ of $Q_1$
correspond, respectively, to the three orbits of $\Z_p$ on the nodes of $Q_p$ i.e.\! $\{\circ_a\}$, $\{\bullet_a\}$, and $\{\ast_a\}$.

A stability function on $Q_1$ is specified by three points in the upper half--plane, $Z_\circ$, $Z_\bullet$, and $Z_\ast$. The pulled back central charge
on $Q_p$ is then
\begin{equation}\label{ZZp}
Z_{\circ_a}=Z_\circ,\quad Z_{\bullet_a}=Z_\bullet,\quad Z_{\ast_a}=Z_\ast\quad\text{for all }a=1,2,\dots,p.
\end{equation}
Of course, this central charge is very \emph{non}--generic; the corresponding BPS spectrum is much simpler than the one in generic BPS chambers. The choice \eqref{ZZp} is nevertheless technically natural \cite{'tHooft:1979bh}, since it corresponds to an enhancement of symmetry by $\Z_p$. 

For the
$\cn=2^*$ theory all BPS chambers
 are essentially equivalent, up to the action of the $\Z_3$ automorphism of the category $\cj_1$
and the $SL(2,\Z)$ $S$--duality group.
However these equivalent regimes pull back
to physically \emph{inequivalent} chambers for the
$T_{1,p}$ theory, so the covering trick allows us to explore several distinct physical situations by a single computation.

A $SL(2,\Z)$--frame for $\cn=2^*$ is specified by the vector multiplet which plays the role of weakly coupled $W$ boson, or, equivalently,
by declaring which $\mathbb{P}^1$ family of bricks  of $\mathsf{mod}\,\cj_1$ is associated to the $W$ boson. The most common choice (the `minimal' $W$) is the family of regular bricks with support on one of the three Kronecker subquivers of the Markoff quiver $Q_1$. However this is not the most canonical choice \cite{cattoy};  we shall work with the `canonical' choice of $W$ i.e.\! the family of modules $W(\mu:\lambda)$
\begin{equation}
W(\mu:\lambda)\colon\quad\begin{gathered}
\xymatrix{\C\ar@<0.4ex>[dd]^{0\choose1}\ar@<-0.4ex>[dd]_{1\choose 0}
\\
&& \C\ar@<0.4ex>[ull]^0\ar@<-0.4ex>[ull]_0\\
\C^2\ar@<0.4ex>[urr]^{(\mu\  0)}\ar@<-0.4ex>[urr]_{(0\ \lambda)}}
\end{gathered},\qquad(\mu:\lambda)\in\mathbb{P}^1.
\end{equation}
In terms of the central charge, the stability of $W(\mu:\lambda)$ requires
\begin{equation}\label{whichZ}\arg Z_\ast<\arg Z_\bullet <\arg Z_\circ.\end{equation}
In this $S$--duality frame the magnetic charge of a module $X$ is
\begin{equation}
m(X)=\dim X_\circ-\dim X_\ast.
\end{equation}

In the `minimal' $W$ frame, the $\Z_p$--cover 
produces the weakly coupled Lagrangian
regime of the $T_{1,\,p}$ theory: we have
$p$ light\footnote{ By a \emph{light} state in the covering theory we mean  a particle whose mass is bounded as the coupling constant $g^2_\mathrm{YM}$ of the quotient $\cn=2^*$ theory goes to zero (with respect to the chosen $S$--duality frame). }, mutually local, $W$ bosons associated to the $p$ Kronecker subquivers of $Q_p$, and the only other states which are local relatively to all $W$ bosons are $p$ bifundamental hypermultiplets.
All other BPS particles are magnetically charged.

On the contrary,  the $\Z_p$--cover 
of the `canonical' $W$ frame
leads to a \emph{strongly coupled} version of $T_{1,\,p}$
with a \emph{single} light (i.e.\! weakly coupled) $W$ boson which is magnetically charged with respect to all the original Lagrangian $SU(2)$ groups. The stable $W$ boson is the pull back of the one for $\cn=2^*$ \footnote{ This property fails for the `minimal' $W$ frame. The point is that the isotropy subgroup of the
cover light $W$'s in the `minimal' frame, $\mathbb{G}_{W_\mathrm{min}}$, is trivial, whereas for the canonical case $\mathbb{G}_{W_\mathrm{can}}\equiv\mathbb{G}$.}.  
The charge vector of the unique light $W$ is
\begin{equation}
\sum_{a\in\Z_p}\Big(e_{\circ_a}+2\,e_{\bullet_a}+e_{\ast_a}\Big),
\end{equation}
and the magnetic charge with respect to the
associated weakly coupled $SU(2)$ gauge group
\begin{equation}
m(X)=\sum_{a\in\Z_p}\Big(\dim X_{\circ_a}-\dim X_{\ast_a}\Big).
\end{equation}
The BPS spectrum contains only finitely many hypermultiplets
 which are local with respect to this $W$ boson;
 in fact the stable BPS states with zero cover magnetic charge are in one--to--one correspondence with the BPS states of two copies of the Argyres--Douglas system of type $D_{2p}$
 (the correspondence does not preserve the charge vectors and the Dirac pairing, so the theory does \emph{not} look, in this frame, $SU(2)$ SYM coupled to two $D_{2p}$ AD systems).

 Geometrically, the pair $(E_p, C)$ which corresponds
 to this `strongly coupled' regime is given,
 modulo the congruence modular group $\Gamma_0(p)$,
  by a $\tau\in Y_0(p)$ in a neighborhood of the cusp at infinity, $\tau=i\infty$. The usual weak coupling is the cusp at the origin $\tau=0$.
  \medskip

For $\cn=2^*$ with the central charge \eqref{whichZ},
all stable modules have vanishing arrows $\ast\rightarrow \circ$, i.e.\! are modules of the triangular factor $\cj_1^\mathrm{tr}$ of $\cj_1$ specified by the bi--quiver\begin{equation}
\begin{gathered}
\xymatrix{\circ\ar@<0.4ex>[dd]\ar@<-0.4ex>[dd]\\
&& \ast\ar@{..>}@<0.4ex>[ull]\ar@{..>}@<-0.4ex>[ull]\\
\bullet\ar@<0.4ex>[urr]\ar@<-0.4ex>[urr]}
\end{gathered}
\end{equation}
Thus we have a triangular chamber, and the BPS spectrum is controlled by its Tits form. In particular, the 
spectrum may be described in Lie theoretic terms:
the charge vectors of BPS particles are roots of the toroidal Lie algebra $A_1^{(1,1)}$.

The pull back of this triangular chamber for $\cn=2^*$
produces a triangular chamber for $T_{1,\,p}$.
Note that acting with a $\Z_3$ automorphism of 
the category $\cj_1$ and then pulling back to
$\cj_p$, we get non--isomorphic triangular
factor of $\cj_p$ (however their Tits forms are  $\Z$--equivalent). 

We claim that by pulling back the description of the $\cn=2^*$ spectrum in terms of the root system $A^{(1,1)}_1$,
we get a description of the BPS spectrum of $T_{1,\,p}$ (in our chamber) in terms of 
root systems of \emph{extended affine
Lie algebras.}
To show the claim, we begin by checking the triangularity of the
$\Z_p$--symmetric chamber of $T_{1,\,p}$.

\subsection{Triangularity of the $\Z_p$--symmetric BPS spectra}\label{tttttrqgj}

The first observation is that we may
replace the
Jacobian algebra $\cj_p$ of the
superpotential in eqn.\eqref{supesuper} with
the Jacobian algebra\footnote{ The algebra $\cj^g_p$ is
not finite dimensional. However the relevant modules are in facts modules of a finite--dimensional \emph{string} algebra \cite{cattoy}, as the argument in the text shows. }  $\cj_p^g$
obtained by forgetting the order six terms in $\cw_p$.
This is an instance of the gentling trick
of appendix \ref{ap:gentling}:
each node $\ast_a$ corresponds to a flavor $SU(2)$ which is gaugeable. We may
weakly gauge all the $SU(2)^p$ flavor symmetry
by coupling $p$ spectator $SU(2)$ SYM sectors and compute the BPS spectrum of the fully gauged theory. In the limit of vanishing spectator gauge coupling, $g_\mathrm{spec}\to 0$, the BPS spectrum  consists of heavy states carrying
spectator magnetic charge, which decouple at
$g_\mathrm{spec}= 0$,  and states of zero spectator magnetic charge which survive the decoupling limit. 
The surviving states consist of the BPS spectra of the several decoupled sectors,
i.e.\! the original model with $SU(2)^p$ flavor symmetry plus the $p$ spectator $W$ bosons.

At the level of the quiver with superpotential
the spectator gauging of the $a$--th flavor $SU(2)$ modifies the quiver
\eqref{specialpytttaquiH} at the node
$\ast_a$ by splitting it into two nodes connected by
an arrow $\circledast_a \xrightarrow{\ K_a\ }\ast_a$\begin{equation}
\begin{gathered}
\xymatrix{
\circ_{a}\ar@<0.4ex>[dd]^{B_a}\ar@<-0.4ex>[dd]_{A_a} && \circ_{a+1}\ar@<0.4ex>[dd]^{B_{a+1}}\ar@<-0.4ex>[dd]_{A_{a+1}}\\
&\ast_a\ar[ul]_{\eta_a}\ar[ur]^{\xi_{a+1}} &\\
\bullet_a\ar[ur]_{\eta_a^*} & &\bullet_{a+1}\ar[ul]^{\xi^*_{a+1}}}
\end{gathered}\xrightarrow{\,\text{ spectactor gauging }\,}\begin{gathered}
\xymatrix{
\circ_{a}\ar@<0.4ex>[dd]^{B_a}\ar@<-0.4ex>[dd]_{A_a} &&& \circ_{a+1}\ar@<0.4ex>[dd]^{B_{a+1}}\ar@<-0.4ex>[dd]_{A_{a+1}}\\
&\circledast_a\ar[ul]_{\eta_a}\ar[r]^{K_a} &\ast_a\ar[ur]^{\xi_{a+1}} &\\
\bullet_a\ar[ur]_{\eta_a^*} && &\bullet_{a+1}\ar[ul]^{\xi^*_{a+1}}}
\end{gathered}
\end{equation}   
while the new superpotential is given
by eqn.\eqref{supesuper} with the order six terms
omitted. The BPS states of the fully gauged theory
which have bounded masses as $g_\mathrm{spec}\to 0$ correspond to the stable modules
of the Jacobian algebra of the modified quiver
such that all new arrows $K_a$ are isomorphisms.
Being an isomorphism, $K_a$ identifies
nodes $\circledast_a$ and $\ast_a$, so that
the relevant modules may be seen as
stable $\cj_p^g$--modules, i.e.\! modules of the original quiver with the purely cubic superpotential. This light sector of the fully gauged theory should contain the BPS spectrum of the original model plus $p$ spectator $W$ bosons.

In $\cj_p^g$ the relations take a very simple
form
\begin{equation}\label{genrelations}
A_a\xi_a= \xi_a\xi_a^*=\xi_a^*A_a=B_a\eta_a=\eta_a\eta_a^*=\eta_a^*B_a=0\qquad \forall\;a.
\end{equation}

Let $X$ be a stable module of $\cj_p^g$
which is not a spectator $W$. We write
$X|_{\mathsf{Kr}_a}$ for the $\mathsf{Kr}$--module obtained by restriction to the $a$--th
Kronecker subquiver. By eqn.\eqref{ringel} we have
\begin{equation}
X|_{\mathsf{Kr}_a}=p_a\oplus t_a\oplus q_a.
\end{equation} 
We claim that
\begin{equation}\label{claim1}
\eta^*_a\big|_{q_a}=\xi^*_a\big|_{q_a}=0,\qquad \text{and dually}\qquad
\mathrm{Im}\,\eta_a,\; \mathrm{Im}\,\xi_a\subset t_a\oplus q_a.
\end{equation}
This is a consequence of the
relations \eqref{genrelations} and the fact that, restricted to $q_a$ (resp.\! $p_a$) the arrows
$A_a$, $B_a$ are surjective (resp.\! injective).
From eqn.\eqref{claim1} it follows that each
$q_a$ (resp.\! $p_a$) is a submodule
(resp.\! quotient) of $X$. Since $X$ is stable
\begin{equation}\label{eeeqqr}
\arg Z(q_a) < \arg Z(X) < \arg Z(p_a),
\end{equation}
which is impossible in view of \eqref{whichZ}
unless, for each $a$, either $q_a$ or $p_a$ vanishes. The pulled back
central charge \eqref{whichZ} corresponds to a special $\Z_p$
symmetric point such that 
\begin{equation}\label{rrr5621}
\arg Z(p_a)<\arg Z(q_b)\quad \text{for all $a$, $b$.}
\end{equation}
  Then, if $p_a\neq 0$ (resp.\! $q_a\neq 0$) for one $a$ then $q_b=0$ (resp.\! $p_b\neq 0$) for \emph{all} $b\in\Z_p$. In the next subsection we shall relax the condition \eqref{rrr5621}
  which is too strong from a physical perspective.

From \S.\ref{purereview} we have $t_a=\oplus_s R_{n_s}(\lambda_s)$. Comparing the relations \eqref{genrelations} with the structure of $R_{n_s}(\lambda_s)$, we conclude that if $t_a$ contains a direct summand $R_{n_s}(\lambda_s)$
with $\lambda_s\neq 0,\infty$ then $X$ belongs to the $\mathbb{P}^1$ family of a magnetically charged vector multiplet with support on the $a$--th Kronecker subquiver\footnote{ In the weak coupling Lagrangian regime, a state with these quantum numbers would be interpreted as the $W$ boson of the $a$--th gauge group.} $\mathsf{Kr}_a$.
For all other stable modules $\lambda_s$ is equal to $0$ or $\infty$.
Since $X$ is a brick,
there are no direct summands $R_{n_s}(0)$,
$R_{n_s}(\infty)$ with $n_s>1$. Indeed, let $\pi$ and $(\cdot)|$ be the projection and restriction to the direct summand $R_{n\geq2}(\infty)$. Taking into account the
relations \eqref{genrelations}, in presence of
such a summand we have the situation
\begin{equation}\label{erreenne}
\begin{gathered}
  \xymatrix{& \C^{n\geq 2}\ar@<0.4ex>[dd]^{\text{Id}}
  \ar@<-0.4ex>[dd]_{A_a\neq 0}\\
\ast_{a-1}\ar[ur]^{\pi\xi_a} && \ast_a\ar[ul]_0\\
& \C^{n\geq 2}\ar[ul]^{\xi^*_a|}\ar[ur]_0}
 \end{gathered}\quad\text{with }A_a\xi_a=\xi_a^* A_a=0,\ \text{and $A_a$ nilpotent,}
\end{equation}
which shows that multiplication by $A_a$ at the nodes of the $a$--th Kronecker subquiver and by $0$ elsewhere
 is a non--zero nilpotent endomorphism.
Then $t_a$ is the direct sum of copies of $R_1(0)$ and $R_1(\infty)$. In view of \eqref{genrelations}, for each $R_1(\infty)$
summand we have a situation as in \eqref{erreenne}
with $n=1$ and $A_a=0$ (for $R_1(0)$ the situation is symmetric with $\xi_a\leftrightarrow \eta_a$,
$\xi_a^*\leftrightarrow \eta_a^*$). 
We distinguish three possibilities
\begin{align}
&a)\ R^q_1(\infty):\ \xi_a^*|=0, && b)\ R^p_1(\infty):\ \pi\xi_a=0, && c)\ R^r_1(\infty):\ \pi\xi_a,\;\xi_a^*|\neq0.
\end{align}
In appendix \ref{nor1} we show that no stable representation (which is not a spectator $W$)
contains summands of the form $c)$.
$R_1^q(\infty)$, $R_1^q(0)$ are subrepresentations of $X$ while $R_1^p(\infty)$, $R_1^p(0)$ are quotients.
Arguing as in eqn.\eqref{eeeqqr}, we get a contradiction if $X$ contains (even at different Kroneckers)
summands of both kinds, $R_1^q$ and $R_1^p$;  we also get a contraction if 
we have non--trivial summands 
of the form $R_1^q$ and $p_a$ or $R_1^p$ and $q_a$.
Hence, for all $a\in\Z_p$, 
$X|_{\mathsf{Kr}_a}$ is either: \textit{1)} a direct sum of
a preinjective module $q_a$ and regular modules
of the form $R^q_1(\infty)$, $R^q_1(0)$,
or \textit{2)}
a direct sum of
a preprojective module $p_a$ and regular modules
of the form $R^p_1(\infty)$, $R^p_1(0)$.

Consider first the possibility \textit{1)}. Suppose that $X_{\ast_a}\neq 0$ for some $a$.
 $X$ has a quotient $Y$
with support on $\ast_a$; since $\arg Z_\ast \leq \arg Z(M)$ for all indecomposable modules $M$, with equality iff $M$ is the simple with support on $\ast_a$, we deduce that $X$, being stable,
is such a simple module. On the other hand, if
$X_{\ast_a}=0$ for all $a$, $X$, being indecomposable, has support
on a single Kronecker subquiver. In both cases the
arrows $\xi_a$, $\eta_a$ vanish for all $a$.

In the possibility \textit{2)} the arrows $\xi_a$, $\eta_a$ vanish by definition. Then, with our present choices, we get a triangular chamber with respect to the triangular factor algebra defined by the bi--quiver
\begin{equation}\label{specialpytttaqui}
\begin{gathered}\xymatrix{
& \circ_1\ar@<0.4ex>[dd]\ar@<-0.4ex>[dd] & & \circ_{2}\ar@<0.4ex>[dd]^{B_2}
\ar@<-0.4ex>[dd]_{A_2} & &&& \circ_{p}\ar@<0.4ex>[dd]\ar@<-0.4ex>[dd] &\\
\ast_{p} \ar@{..>}[ru] & &\ast_1\ar@{..>}[ul]\ar@{..>}[ur]^{\xi_2} & &\ast_{2} \ar@{..>}[ul]_{\eta_2}&\cdots &\ast_{p-1}\ar@{..>}[ur]& &\ast_{p}\ar@{..>}[ul]\\
& \bullet_1\ar[ul]\ar[ur] & &\bullet_{2}\ar[ul]^{\xi_2^*}\ar[ur]_{\eta_2^*} & &&&\bullet_{p}\ar[ur]\ar[ul]&}\end{gathered}
\end{equation}
The covering BPS spectrum is then 
controlled by the Tits form of the above 
triangular algebra which happens to be the Tits form of an extended affine Lie algebra of nullity $p+1$.
We describe these Lie theoretical aspects in the  subsection \ref{extended}. Before going to that,
we have to discuss the physical significance of the 
above computation.

\subsection{Physical interpretation: perturbing away from the $\Z_p$--point}

Although the above analysis of the $\Z_p$--symmetric point is mathematically robust, its physical significance may be questioned on the grounds that, at such a point, there are stable  modules with the same BPS phase, $\arg Z(X)$,
which are \emph{not} mutually local. For the present class of models, this is an automatic consequence of $\Z_p$--symmetry, which implies
special alignments of the BPS phases in the $Z$--plane. 

To get the physically sound interpretation of the mathematical result, we have to slightly perturb this alignment in the $Z$--plane to a situation in which any two stable representations, $X_1, X_2$
with $\arg Z(X_1)=\arg Z(X_2)$ are mutually local,
i.e.\! $\langle \mathbf{dim}\,X_1,\mathbf{dim}\,X_2\rangle_D=0$. Of course, there are many possible directions in which we may perturb, which correspond to different BPS spectra, because the $\Z_p$ points lay at the intersection of several walls of marginal stability. 

The perturbations which lead to BPS spectra which are simple to describe are the ones
which keep the chamber triangular (in the broader sense). We start from these perturbations, which are rather special. First of all,
we require the phases to satisfy the order
\begin{equation}
\arg Z_{\ast_a}< \arg Z_{\bullet_b}< \arg Z_{\circ_c},\quad \text{for all }a,b,c.
\end{equation}
In addition to this phase--order condition, to ensure that no $R_1^r$ summand appears, we need a few special alignments in $Z$--plane (cfr.\! appendix \ref{nor1})
\begin{equation}\label{xxxxz123n}
\arg(Z_{\circ_{a-1}}+Z_{\bullet_{a-1}}+Z_{\ast_a})=
\arg(Z_{\circ_{a}}+Z_{\bullet_{a}}+Z_{\ast_a}),\quad a=1,2\dots, p.
\end{equation} 
Calling $\epsilon$ the overall size of the perturbation, we remain free to choose the complex numbers $Z_{\ast_a}=Z_\ast+O(\epsilon)$, $Z_{\bullet_a}=Z_\bullet+O(\epsilon)$, and
real numbers $t_a=1+O(\epsilon)$ ($a=2,3,\dots, p$), such that
\begin{equation}
Z_{\circ_a}=t_a\,Z_{\circ_{a-1}}+(t_a\,Z_{\bullet_{a-1}}-Z_{\bullet_a})+(t_a-1)Z_{\ast_a}\ \ \text{for }a\geq 2,\quad \text{while }Z_{\circ_1}=Z_\circ.
\end{equation}
In this perturbed chamber, the relation \eqref{rrr5621} does not hold in general;
it may be violated for those pairs $a$, $b$
such that 
\begin{equation}
\arg(Z_{\circ_a}+Z_{\bullet_a})> \arg(Z_{\circ_b}+Z_{\bullet_b}),
\end{equation}
and in this case only if the dimensions 
of both summands $p_a$ and $q_b$ are \emph{parametrically large} of order $O(1/\epsilon)$.  At any rate, it remains true (by construction) that we have no $R_1^r$ summands, and hence, for all $a$,
$X|_{\mathsf{Kr}_a}$ is either: \textit{i)} a direct sum of $p_a$ and copies of $R_1^p$, or \textit{ii)}  a direct sum of $q_a$ and copies of $R_1^q$. We attach a sign $\epsilon_a(X)=\pm 1$ to the $a$--th Kronecker subquiver to distinguish the two possibilities: for case \textit{i)} we set $\epsilon_a(X)=+1$, while for case \textit{ii)} we set $\epsilon_a(X)=-1$. Then if $\epsilon(X)_a=+1$ (resp.\! $\epsilon_a(X)=-1$) $X|_{\mathsf{Kr}_a}$ is a quotient (resp.\! submodule) of $X$, and the arrows $\xi_a,\eta_a$ (resp.\! $\xi^*_a,\eta_a^*$) vanish. So all (non--spectator) stable brick $X$ is triangular with
respect to a $X$--dependent triangular factor $\ct[\epsilon_a(X)]$ of $\mathcal{J}$
whose Tits form $q(\boldsymbol{x};\{\epsilon_a(X)\})$ is $\Z$--equivalent to the Tits form $q(\boldsymbol{x})$ of the bi--quiver
\eqref{specialpytttaqui}
\begin{equation}\label{eeeeeevzq}
q(x_{\circ_a},x_{\bullet_a}, x_{\ast_a};\{\epsilon_a\})=q(\epsilon_a\,x_{\circ_a},\epsilon_a\,x_{\bullet_a}, x_{\ast_a}).
\end{equation}
However, stable non--spectator modules $X$ having $\epsilon_a(X)=-1$ for some $a$, i.e.\! which are not modules of the original triangular factor \eqref{specialpytttaqui}, have charges, hence masses, parametrically large of order $O(1/\epsilon)$. Note that these very heavy states do not form
full $\Z_p$--multiplets as do states with masses which remain bounded as $\epsilon\to 0$. On the other hand, after the perturbation, the stable states with the same BPS phase are mutually local; indeed, choosing the $Z_i$ to be $\mathbb{Q}$--linearly independent, the only alignments are those in eqn.\eqref{eeeeeevzq}
which correspond to mutually local BPS states.

In conclusion, in this class of perturbed chambers, BPS spectrum is still controlled by the Tits form $q(\boldsymbol{x})$, and hence described by the root system of an extended affine Lie algebra, and it consists of two distinct sectors:
\begin{itemize}
\item[1)] states with masses $O(1)$ which form $\Z_p$--multiplets (with $O(\epsilon)$ mass splittings inside the $\Z_p$--multiplets). They have $\epsilon_a(X)\equiv +1$ for all $a$;
\item[2)] parametrically heavy states, $m=O(1/\epsilon)$ which do not form complete $\Z_p$--multiplets. In this case some (but not all)
signs $\{\epsilon_a(X)\}_{a=1}^p$ must be equal $-1$.
\end{itemize}
 The BPS spectrum at the $\Z_p$--point computed in \S.\ref{tttttrqgj}
 corresponds to the sector 1) of particles whose mass remains bounded as we switch off the perturbation.  For $\epsilon\sim 0$,
the physics of the perturbed chamber is the same as the one obtained by taking our covering computation at the $\Z_p$--symmetric point at its face level (up to $O(\epsilon)$ corrections); that computation is exact asymptotically as $\epsilon\to 0$.  This clarifies the precise physical meaning of the spectrum computed at the $\Z_p$--symmetric
 point.
 
 If we consider a more general perturbation of order $\epsilon$, where the equalities \eqref{xxxxz123n} hold only up to $O(\epsilon)$ corrections,
 it is not longer true that we get a triangular chamber, and hence the full spectrum cannot be simply described in terms of root systems of Lie algebras. Nevertheless, it is easy to see that
 it remains true that the BPS spectrum decomposes in two sectors: a triangular sector with masses of order $O(1)$
 as $\epsilon\to 0$, which form representations of
 $\Z_p$ and coincides with our `mathematical' spectrum at the $\Z_p$--point, and a second sector,
 whose detailed structure is rather involved, but which contains only particles with parametrically large masses $O(1/\epsilon)$ which do not form $\Z_p$ representations. Besides being the result of the analysis of module stability,
 this picture is physically very natural, and it is expected to hold quite in general.
 
 The situation is analogue\footnote{ In fact this is more than just an analogy.} to weak coupling in 
 $\cn=2$ gauge theories; also there as 
 $\epsilon\equiv g^2_\mathrm{YM}\to 0$
 the spectrum decomposes in two parts:
 states with bounded masses which form complete representations of the gauge group (which is interpreted as a flavor symmetry in the limit)
 and a complicated dyonic sector whose
 masses are of order $O(1/\epsilon)$. 
 
 The physical conclusion, is that we may trust
 the BPS spectrum computed at a $\Z_p$--point by pulling back the spectrum of the quotient theory.
In any nearby BPS chamber the physical spectrum is given by the pull back one plus essentially decoupled super--heavy states. The decoupling of the heavy junk is exact asymptotically as $\epsilon\to 0$.

\subsection{Relation with extended affine and Kac--Moody Lie algebras}\label{extended}

 Let us look at the Tits form $q$ of the triangular algebra associated to the bi--quiver
 \eqref{specialpytttaqui}. From \S.\ref{classA1tits},  $q$ is
 the integral quadratic form on $\Gamma\simeq \Z^{3p}$
 \begin{equation}\label{rrr1234}
 q(x_{\circ_a},x_{\bullet_a}, x_{\ast_a})=\frac{1}{2}\,\sum_{a=1}^p\bigg(\big(x_{\bullet_a}-x_{\circ_a}-x_{\ast_a}\big)^2+\big(x_{\bullet_a}-x_{\circ_a}-x_{\ast_{a-1}}\big)^2\bigg).
 \end{equation}

Being a sum of squares, the Tits forms is
manifestly positive semi--definite. By \S.\ref{emalas},
a positive semi--definite Tits form is uniquely 
identified, up to $\Z$--equivalence, by its type $\mathfrak{g}_r$ and nullity $\kappa$. 
The form in eqn.\eqref{rrr1234} has type
$A_{2p-1}$ and nullity $\kappa=p+1$. 
To see this, note that there is an element
of $\mathrm{rad}\,q$ associated to each
Kronecker, i.e. $x_{\circ_a}=x_{\bullet_a}=1$ and zero elsewhere, so, working $\mathrm{mod}\,\mathrm{rank}\,q$, we may set $x_{\circ_a}=0$ for all $a$.
$q$ reduces to the Tits form of the Kac--Moody
algebra whose Dynkin graph is obtained by deleting all $\circ_a$ nodes in the quivers: 
\eqref{specialpytttaqui} reduces to the affine Dynkin graph $A^{(1)}_{2p-1}$.
Being affine this Lie algebra has Tits form of nullity 1
(and type $A_{2p-1}$).
Hence the nullity of the Tits form $q$ is one more the number of Kronecker subquivers; $\mathrm{rad}\,q$ is generated by the dimension vectors $\delta_a=e_{\circ_a}+e_{\bullet_a}$ of the $p$ $W$ bosons of the usual weak coupling plus the
dimension vector $\sum_a(e_{\ast_a}+2e_{\bullet_a}+e_{\circ_a})$ of the $W$ boson which is light in our pulled--back triangular regime. 

This result is most conveniently stated in terms of Lie algebras. 
The Tits form in eqns.\eqref{rrr1234} is
 identified with the Tits forms of the extended affine Lie algebra $A^{[p+1]}_{2p-1}$.
$\boldsymbol{x}\in \mathbb{Z}^{\kappa+r}$ is a \emph{root} 
 iff $q(\boldsymbol{x})\leq 1$ and its support is connected.  A root $\boldsymbol{x}$ is \emph{real} iff $q(\boldsymbol{x})=1$, 
and \emph{imaginary} iff $q(\boldsymbol{x})=0$. 
The imaginary roots form the sublattice $\mathrm{rad}\,q$.
From the recursive construction, eqn.\eqref{recconst}, it is clear that the real roots have the form $\alpha+\delta$ where $\alpha$ is a root
 of the Lie algebra $\mathfrak{g}_r$  and $\delta$ is an element of the lattice of imaginary roots $\mathrm{rad}\,q$.
Equivalently, we may separate the generator $\delta\in \mathrm{rad}\,q$ associated to the light $W$,
$\delta=\sum_a(e_{\ast_a}+2e_{\bullet_a}+e_{\circ_a})$, and write the roots in the 
form $\hat \alpha_n+ \rho$ where $\hat \alpha_n=\alpha+n\delta$ is a root of the affine
Lie algebra $\mathfrak{g}_r^{(1)}$ and $\rho\in \mathrm{rad}\, q/\Z\delta$. 

\subsubsection{The GIM Lie algebra of the $g=1$ $\cs[A_1]$ models}
We write explicitly the Chevalley generators of the \emph{extended affine Lie algebra} $A^{[p+1]}_{2p-1}$ associated to the triangular factor \eqref{specialpytttaqui}. We write $\alpha_i$ for the simple roots of $A_{2p-1}$, $\alpha_\theta\equiv \sum_i\alpha_i$ for its maximal root, and $\{E_i,F_i,H_i\}_{i=1}^{2p-1}$ for its Chevalley generators. The Chevalley generators of
$\mathfrak{L}_p$ are
\begin{equation}\label{generators}
\begin{aligned}
&e_{\circ_a}= E_{2a-1}, && f_{\circ_a}= F_{2a-1}, &&h_{\circ_a}=H_{2a-1}&& a=1,2,\dots, p \\
&e_{\ast_a}=E_{2a},&&f_{\ast_a}=F_{2a}, &&h_{\ast_a}=H_{2a}, &&a=1,2,\cdots,p-1\\
&e_{\ast_p}=F_\theta\, t_0, && f_{\ast_p}=E_\theta\,t_0^{-1}, && h_{\ast_p}=-H_\theta+\text{central}\\
 &e_{\bullet_a}=F_{2a-1}\,t_a, && f_{\bullet_a}=E_{2a-1}\,t_a^{-1}, &&h_{\bullet_a}=-H_{2a-1}+\text{central}&&a=1,2,\dots, p,
\end{aligned}
\end{equation}
where $t_0, t_1,\dots, t_p$ are coordinates on the torus $(\C^*)^{p+1}$. One checks that all the Serre--Slodowy relations \eqref{serre1}--\eqref{serren} following from the Dynkin bigraph  \eqref{specialpytttaqui} are satisfied.

\subsubsection{Dyonic towers, Kac--Moody algebras, and Schur roots}

From the discussion around eqn.\eqref{spintits} and the above analysis it follows that the stable
$\cj_p$-modules $X$  are controlled by the Tits form $q$. Hence, at weak coupling, the charge vectors $\mathbf{dim}\,X$ of the BPS particles satisfies
\begin{equation}
q\!\big(\dim X_{\circ_a},\dim_{\bullet_a}, \dim X_{\ast_a}\big)= \begin{cases} 1 & \text{for hypermultiplets}\\
0 & \text{vector multiplets,}
\end{cases}
\end{equation}
and there are no higher spin BPS states since $q$ is semi--definite. In view of \S.\ref{extended}, the quantum numbers of the hypermultiplets (resp.\! vector multiplets)
are real roots (resp.\! imaginary roots) of the extended affine Lie algebra
$A^{[d+1]}_{2d-1}$.
More precisely,
the charge vectors should be \emph{Schur roots} for the appropriate orientation  of the bi--quiver.

Let $\delta_a=e_{\bullet_a}+e_{\circ_a}$. $\delta_a\in \mathrm{rad}\,q$, so 
$\Gamma_\mathrm{el}\equiv \bigoplus_a \Z\delta_a$
is a sublattice of $\mathrm{rad}\,q$; in the Lagrangian regime, $\Gamma_\mathrm{el}$
 would be identified with the lattice
of the gauge electric charge.
The Tits form $q$ on $\Gamma$ induces a
quadratic form $\overline{q}$ on the quotient
$\Gamma/\Gamma_\mathrm{el}$ which is the Tits form of the Kac--Moody algebra $A^{(1)}_{2p-1}$
whose Dynkin graph is obtained deleting 
the $\circ_a$ nodes.

Let $\boldsymbol{x}\in\Gamma$ the charge vector of a BPS particle. Forgetting its Lagrangian electric charge, amounts to considering the residue class $\overline{\boldsymbol{x}}\in\Gamma/\Gamma_\mathrm{el}$. 
The reduced charge vector $\overline{x}$
of a stable BPS particle (in our pulled back chamber)
is necessarily a
 \emph{Schur}\footnote{ The notion of Schur root depends on the orientation of the Dynkin graph of $A^{(1)}_{2p-1}$.
Here the orientation is the one of the Dynkin quiver obtained from the original quiver by collapsing the Kronecker subquivers to a single node $\bullet_a$ and deleting the dashed arrows.} root of $A^{(1)}_{2p-1}$, real roots being hypermultiplets and imaginary ones vector multiplets.
Considering the reduced charge vector $\overline{\boldsymbol{x}}$, instead of $\boldsymbol{x}$, amounts to forgetting the information about the (Lagrangian)
electric charge of the state, while keeping the
information about its (Lagrangian) magnetic and flavor charges,
$(m_a, f_i)$. This reduced information is encoded in the Schur roots of $A^{(1)}_{2p-1}$.
The condition of the roos being
Schur gives restrictions that we now discuss.

The reduced charge vector is
a Schur root of $A^{(1)}_{2p-1}$. Let 
 $\alpha_i$, $i=0,1,\dots,2p-1$ be the simple roots of the Kac--Moody algebra $A^{(1)}_{2p-1}$. The element of the affine root lattice
\begin{equation}\label{ppppokm}
\begin{split}
\gamma(m_a,f_a):=&
\sum_{a=0}^{p-1}\Big(\big(\dim X_{\bullet_a}-\dim X_{\circ_a}\big)\alpha_{2a}+\dim X_{\ast_a}\,\alpha_{2a+1}\Big)\equiv \\
&\equiv\sum_{a=0}^{p-1}\Big( -m_a\,\alpha_{2a}+f_a\,\alpha_{2a+1}\Big),\end{split}\end{equation} 
 is a positive root of
 the Kac--Moody algebra $A^{(1)}_{2p-1}$. In \eqref{ppppokm} we use the
 nomenclature for the different charges which is
 standard at the ordinary weakly coupled Lagrangian point, i.e.\!
 $m_a$ is the $a$'th Lagrangian magnetic charge and $f_a$ is the Lagrangian $a$--th quark flavor charge.
More precisely, $\gamma(m_a,f_a)$ is a \emph{Schur root} for the orientation of the $\widehat{A}(p,p)$ quiver in which the 
$2a$--th node is a \emph{source}. The Ringel defect \cite{cattoy} of this affine quiver is
\begin{equation}
\mu(m_a,f_a)= -\sum_a\big(f_a+m_a\big),
\end{equation}
and the list of the Schur roots is \cite{cbrq}:
\begin{itemize}
\item[\textit{i)}] the minimal imaginary root $\boldsymbol{\delta}=\sum_i\alpha_i$ which corresponds to a dyonic vector--multiplet;
\item[\textit{ii)}]  the real positive roots with $\mu(m_a,f_a)\neq 0$;
\item[\textit{iii)}]  the real positive roots with $\mu(m_a,f_a)= 0$ and $|m_a|\leq 1$, $f_a\leq 1$ for all $a$.
\end{itemize} 
The above are \emph{necessary} conditions for a reduced charge vector $\gamma(m_a,f_a)$ to correspond to quantum numbers of stable BPS dyons. For each $\gamma(m_a,f_a)$ satisfying these conditions, we need to determine the sets $\{e_a\}$ of Lagrangian electric charges such that
$(e_a,m_a,f_a)$ are the quantum numbers of a stable BPS state. A convenient way of working is to reduce to the stability problem for a representation of the effective $\widehat{A}(p,p)$ with the orientation specified by the $\epsilon_a(X)$'s  and an effective central charge $Z^\text{eff}(\{e_a\})$ which depends on the electric charges $\{e_a\}$.
A representation of  $\widehat{A}(p,p)_{\{\epsilon_a\}}$ with dimension vector $(|m_a|, f_a)$ which 
is stable for $Z^\text{eff}(\{e_a\})$ corresponds to a $(e_a,|m_a|, f_a)$ BPS state.

\subsection{Irregular punctures}

The analysis may be easily extended to class $\cs[A_1]$ theories with $p$ \emph{irregular} punctures corresponding to poles of degree $s+1$ of the quadratic differential $\phi_2$ where $s>1$ \cite{CV11}. 
In one `weakly coupled' corner of its parameter space, this model is
$SU(2)^{2p}$ $\cn=2$ SYM coupled to $p$ half tri--fundamentals and $p$ copies of the Argyres--Douglas systems of type $D_{s+1}$ \cite{CV11}. Here we discuss a different regime which has a nice Galois cover interpretation. 

The $\Z_p$--symmetric form of the quiver, $Q_{p,s}$, is obtained from $Q_p$ by replacing each
node $\ast_a$ with a $A_s$ subquiver $\ast_{a,1}\xrightarrow{K_{a,1}}\ast_{a,2}\xrightarrow{K_{a,2}}\cdots\xrightarrow{K_{a,s-2}} \ast_{a,s-1}\xrightarrow{K_{a,s-1}}\ast_{a,s}$, i.e.
\begin{equation}
\begin{gathered}
\xymatrix{\cdots &&&& \circ_{a+1}\ar@<0.4ex>[dd]^{B_{a+1}}\ar@<-0.4ex>[dd]_{A_{a+1}} &&&& \circ_{a+2}\ar@<0.4ex>[dd]\ar@<-0.4ex>[dd]\\
&\ast_{a,1}\ar[ul]\ar[r]^{K_{a,1}}  &\cdots\ar[r]^{K_{a,s-1}} &\ast_{a,s}\ar[ur]^{\xi_{a+1}} && \ast_{a+1,1}\ar[ul]_{\eta_{a+1}}\ar[r] &\cdots\ar[r] & \ast_{a+1,s}\ar[ur]&& \ar[ul]\cdots
\\
\cdots\ar[ur] &&&& \bullet_{a+1}\ar[ur]_{\eta^*_{a+1}}\ar[ul]^{\xi^*_{a+1}} &&&& \bullet_{a+2}\ar[ur]\ar[ul]}
\end{gathered}
\end{equation}
where again the index $a\in\Z$ is periodically identified $a\sim a+p$. $Q_{p,s}$ is endowed with the superpotential
\begin{equation}
\cw_{p,s}=\sum_{p\in\Z_p}\Big(A_a\xi_a\xi^*_a+ B_a\eta_a\eta^*_a\Big).
\end{equation}
Clearly we have a free action on the nodes
of the $\Z_p$ automorphism group of the $Q_{1,s}$ quiver
\begin{equation}
\Z_p\ni k\colon (\circ_a,\bullet_a, \ast_{a,j})\longmapsto (\circ_{a+k}, \bullet_{a+k}, \ast_{a+k,j})\qquad a\in\Z_p,\ j=1,2,\dots,s,
\end{equation}
which induces a $\Z_p$ automorphism of the Jacobian category $\cj_{p,s}=\C Q_{p,s}/(\partial\cw_{p,s})$. The orbit category $\cj_{p,s}/\Z_p$ is the Jacobian category $\C Q_{1,s}/(\partial\cw_{1,s})$, where $Q_{1,s}$ is the quiver
\begin{equation}
\begin{gathered}
\xymatrix{
&\ast_3\cdots\cdots\ar@<0.6ex>@/^1pc/[dr]\\
\ast_2\ar@<0.4ex>@/^1pc/[ur]& \circ\ar@<0.4ex>[dd]\ar@<-0.4ex>[dd] &
\ast_{s-1}\ar@/^0.8pc/[d]\\
\ast_1\ar@/^0.8pc/[u]
\ar[ur] && \ast_s\ar[ul]\\
&\bullet\ar[ul]\ar[ur]}
\end{gathered}
\end{equation}
By comparing with the previous $s=1$ case,
we see that in the regime
\begin{equation}
\arg Z_{\ast_i} < \arg Z_{\bullet} <\arg Z_\circ,
\end{equation}
the arrows with target $\circ$ vanish. Hence we have a triangular chamber with Tits form the one of the toroidal Lie algebra
$A^{(1,1)}_{s}$. 

The $\Z_p$ cover of this chamber
is also triangular. Its BPS spectrum is controlled by the Tits form of the extended affine Lie algebra
$A^{[p+1]}_{p(s+1)-1}$ in agreement with 
\textbf{Fact 2} of \S.\,\ref{classA1tits}.

\section*{Ackowledgements}
The authors thank the Simons Center for Geometry and Physics for hospitality while working on this paper, and thank the organizers of the Simons Summer Workshop and of the Quiver Variety program. SC thanks the Department of Physics of Harvard University for hospitality. 
The work of MDZ is supported by the NSF grant PHY-1067976.

\appendix

\section{Nullity $\kappa$
and type $\mathfrak{g}_r$ of class $\cs[A_1]$ triangular factors}\label{ppprof}

Here we prove \textbf{Fact 2.}
Let $Q$ be a triangulation quiver of a surface $\Sigma$ having genus $g$,
$p$ punctures, $b$ boundaries (with $p+b>0$), and $\ell_i\geq 1$ marks on the $i$--th boundary. The number of nodes of 
$Q$ is 
\begin{equation}
n\equiv 6(g-1)+3(p+b)+\sum_i\ell_i=-3\chi+\sum_i\ell_i,
\end{equation}
where $\chi$ is the Euler characteristic of $\Sigma$. 
We write $n_\mathfrak{b}$ for the number of blocks of Type $\mathfrak{b}$ in the decomposition of the quiver $Q$. 
We consider a \emph{good} biquiver obtained by
dashing some arrows of $Q$ as described in \S.\ref{classA1tits}.

From figure \ref{biblocks}, a {good} bi--quiver block decomposition defines a group homomorphism
\begin{gather}\label{667nnhq}
 \beta\colon \Z^n\rightarrow \mathbb{L}\equiv \Z^{n_0+n_I+n_{II}} \oplus L_2^{(n_{III}+n_{IV})}\oplus \Z^{3n_V}\\
\beta\colon x_i\mapsto (y_a,w_{m,1},w_{m,2}, z_{s,1},z_{s,2},z_{s,3}),
\end{gather}
where $L_2$ is the rank 2 lattice
$L_2=\{(w_1,w_2)\in \Z^2\;|\; w_1=w_2\mod 2\}$,
 such that the Tits form $q(x_i)$ of the bi--quiver is the pull--back
\begin{equation}
q_A= \beta^* \mathsf{q} 
\end{equation}
of the $\mathbb{Q}$--quadratic form
on the lattice $\mathbb{L}$
\begin{equation}\label{Qquad}
 \mathsf{q}= \frac{1}{2}\sum\nolimits_a y_a^2+\frac{1}{4}\sum\nolimits_m \big(w_{m,1}^2+w_{m,2}^2\big)+\frac{1}{2}\sum\nolimits_{s,a} z_{s,a}^2.
\end{equation}
\medskip

\textbf{Lemma.}
\textit{One has} 
\begin{equation}\mathrm{rank}\,\mathbb{L}=\chi+n
\end{equation}
\medskip

\textsc{Proof.}
The number of nodes in $Q$, $n$, is half the total number of white nodes in the several blocks plus the number of black ones
\begin{equation}
 n\equiv -3\chi+\sum\nolimits_i\ell_i= \frac{1}{2}(n_0+2n_I+3n_{II}+n_{III}+2n_{IV}+n_V)+(2n_{III}+2n_{IV}+4n_V),
\end{equation}
 while 
\begin{equation}\label{ttttxr}
 \sum\nolimits_i \ell_i = 2 n_0+n_I+n_{III},
\end{equation}
since (cfr.\! table \ref{puzzle}) a Type $0$ block corresponds to a triangle with two sides on the boundary, a Type $I$ block to a triangle with one side on the boundary, and Type $III$ to a punctured di--gon with 
one side on the boundary, while the other block Types correspond to internal `puzzle pieces' in the ideal triangulation \cite{triangulation1}. Then
\begin{equation}
\begin{split}
 n+\chi&= n-\frac{1}{3}\left(n-\sum\nolimits_i \ell_i\right)=\\
 &= (n_0+n_I+n_{II})+2(n_{III}+n_{IV})+3n_V\equiv \mathrm{rank}\,\mathbb{L}.\qquad\qquad\quad\square
 \end{split}
\end{equation}

Since the $\mathbb{Q}$--quadratic form $\mathsf{q}$ is \emph{positive definite}, the index of nullity of the extended affine Lie algebra having Tits form $q(x_i)$ is
\begin{equation}
 \kappa= n-\mathrm{rank}_\mathbb{Q}\,\beta\equiv -\chi+\dim_\mathbb{Q}\mathrm{coker}\,\beta,
\end{equation}
and $\kappa=-\chi$ iff $\beta$ is surjective (over $\mathbb{Q}$). Let $\beta$ be the $(n+\chi)\times n$ matrix giving the $\mathbb{Q}$--linear map
extending \eqref{667nnhq}. One has
$\dim_\mathbb{Q}\mathrm{coker}\,\beta=
\dim_\mathbb{Q}\mathrm{ker}\,\beta^{\,t}$.
From figure  \ref{biblocks} it is clear that, for a connected bi--quiver,
$\mathrm{ker}\,\beta^{\,t}$ cannot be non zero whenever in the block decomposition of $Q$ there are black nodes, that is, blocks of  Types $III$, $IV$, $V$.

\paragraph{The case 
$n_{III}=n_{IV}=n_V=0$.} In this case a non--zero zero--eigenvector of $\beta^t$, $(y_1,\dots,y_{n+\chi})$ should satisfy $|y_a|=|y_b|$ for all $a,b$. Hence
$\mathrm{coker}\,\beta$ is at most one--dimensional and $\kappa$ at most $-\chi+1$. 
Suppose then that $\kappa=-\chi+1$. By flipping the sign of some $y_a$, if necessary, we reduce to the situation that
\begin{align}
 \mathsf{q}&=\frac{1}{2}\sum\nolimits_a y_a^2\\
\mathrm{im}\,\beta&=\big\{(y_a)\in \mathbb{Q}^{n+\chi}\;\big|\; y_1+y_2+\cdots+y_{n+\chi}=0\big\},
\end{align}
and from the classification of the non--negative forms
we know that for $-\chi\geq 0$ the form $\mathsf{q}|_{\mathrm{im}\,\beta}$ should be isometric to the Tits form of an affine Lie algebra of type either $A$ or $D$,
while for $\chi=1$ to the Tits form of a finite--type Lie algebra of type $A$ or $D$. In both cases this condition is satisfied by the type $A$ algebra with isometries
\begin{align}
 &-\chi\geq 0 && y_i=x_i-x_{i+1},\quad x_{i+n+\chi}\equiv x_i\\
&-\chi=-1 && x_i=y_i-y_{i+1},\quad i=1,\dots,n,
\end{align}
but not by $D$ type. We conclude that $\kappa=-\chi+1$ implies $A$ type. Conversely, $\kappa=-\chi$ implies $D$--type.

\section{Quick review of strings and bands}\label{stringbands}

\subsection{Indecomposable modules of string algebras}

A finite--dimensional algebra $\mathscr{A}$ is called a \emph{string} algebra \cite{stringmod} iff
it has the form $\C Q /I$ where:
\begin{description}
\item[S1] each node of the quiver $Q$ is the starting point of at most two arrows;
\item[S2] each node of the quiver $Q$ is the ending point of at most two arrows;
\item[S3] for each arrow $\alpha$ in $Q$ there is at most one arrow $\beta$ with $\alpha\beta\not\in I$;
\item[S4] for each arrow $\alpha$ in $Q$ there is at most one arrow $\beta$ with $\beta\alpha\not\in I$;
\item[S5] the ideal $I$ is generated by zero--relations.
 \end{description}

Every string algebra is, in particular, special biserial and tame. Physically, tame means that the BPS spectrum consists only of hypermultiplets and vector--multiplets, higher spin states been forbidden. 
A string algebra has
two kinds of indecomposable representations \cite{stringmod}:
\begin{enumerate}
 \item \emph{string} representations, which without free parameters;
\item \emph{band} representations which come in one--parameter families.
\end{enumerate}

By a \textit{string} $C$  we mean a \emph{walk} (generalized path) in the quiver $Q$ where we are allowed to pass from one node to another one either following an arrow connecting them or going backwards such an arrow. We write a string as a sequence of nodes connected by direct or inverse arrows, {i.e.}\! in the form
\begin{equation}
C\colon\qquad  i_0\xleftarrow{\phi_{\alpha_1}} i_1\xrightarrow{\phi_{\alpha_2}} i_2 \xrightarrow{\phi_{\alpha_3}}\: \cdots\cdots\ i_{n-1}\xleftarrow{\phi_{\alpha_\ell}} i_\ell,
\end{equation}
where the nodes in the sequence are labelled as $\{i_0,i_1,i_2,\cdots, i_\ell\}$ and the direct/inverse arrows as $\{\phi_{\alpha_1},\phi_{\alpha_2},\cdots, \phi_{\alpha_\ell}\}$. 
A string is \underline{not} allowed to contain:
\begin{description}
          \item[C1] adiacent direct/inverse arrow pairs, as $\cdots i\xrightarrow{\phi_a} j\xleftarrow{\phi_a} i \cdots$ or
$\cdots i\xleftarrow{\phi_a} j\xrightarrow{\phi_a} i \cdots$;
\item[C2] adiacent pairs of (direct or inverse) arrows whose composition vanishes by the relations $\partial\cw=0$.
         \end{description}
  
A string $C$ is identified with its inverse $C^{-1}$ (\textit{i.e.}\! the string obtained by reading  $C$ starting from the right instead than from the left). 
Given a string $C$ of length $\ell+1$, the corresponding string module $M(C)$ is obtained in the following way. For each vertex $v$ of $Q$ let
\begin{equation}
 I_v=\{ a \;|\; \text{the }a\text{--th node in the string $C$ is }v \}\subset\{0,1,2,\cdots, \ell\}.
\end{equation}
As $M(C)_v$ we take the vector space of dimension $\#I_v$ with base vectors $z_a$, $a\in I_v$. We define
\begin{equation}\label{kklklkd}
 \begin{cases}\phi_{\alpha_a}(z_a)=z_{a+1} & \phi_{\alpha_a}\ \text{is direct}\\
  \phi_{\alpha_a}(z_{a+1})=z_a & \phi_{\alpha_a}\ \text{is inverse}
 \end{cases}
\end{equation}
Finally, for all arrows $\gamma$ and vectors $z_a$ such that $\gamma(z_a)$ is not of the form in \eqref{kklklkd}, we set $\gamma(z_a)=0$. 

A \emph{band} is a cyclic sequence $C$ of nodes of $Q$ and direct/inverse arrows, again forbidding sequences as in \textbf{C1} and \textbf{C2} above.
Moreover, $C$ is required to be such that all powers $C^n$ are well--defined (cyclic) strings, but $C$ itself is not the power of a   string of smaller length. $C$ is identified up to cyclic rearrangement and overall inversion. 
Given such a band $C$ of length $\ell$, we define a family $M(C,\lambda, n)$ of indecomposable band modules labelled by an integer $n\geq 1$ and a point $\lambda\in \C^*$.
Explicitly, we set $$M(C,\lambda,n)_v=\bigoplus_{a\in I_v} Z_a, \qquad\quad (a=1,2,\dots,\ell),$$ where the $Z_a$'s are copies of $\C^n$, and 
\begin{equation}\label{uuus}
 \begin{cases}
  \phi_{\alpha_a}\Big|_{Z_a}\colon Z_a\xrightarrow{1} Z_{a+1} & \phi_{\alpha_a}\ \text{direct } a\neq \ell\\
  \phi_{\alpha_a}\Big|_{Z_{a+1}}\colon Z_{a+1}\xrightarrow{1} Z_{a} & \phi_{\alpha_a}\ \text{inverse } a\neq \ell\\
  \phi_{\alpha_\ell}\Big|_{Z_\ell}\colon Z_\ell\xrightarrow{J(\lambda)} Z_{1} & \phi_{\alpha_\ell}\ \text{direct }\\ 
  \phi_{\alpha_\ell}\Big|_{Z_{1}}\colon Z_{1}\xrightarrow{J(\lambda)^{-1}} Z_{\ell} & \phi_{\alpha_\ell}\ \text{inverse }
\end{cases}
\end{equation}
where $J(\lambda)$ is the $n\times n$ Jordan block of eigenvalue $\lambda\neq 0$. All other linear maps which are not defined in 
\eqref{uuus} are set to zero.

\textit{If $\mathscr{A}$ is a \emph{string} algebra, the above string and band modules give a complete list of indecomposable (and pairwise non--isomorphic) $\mathscr{A}$--modules} \cite{stringmod}.
Note also that a brick has necessarily $n=1$ since $J(\lambda)\in\mathrm{End}\, M(C,\lambda, n)$ is non trivial for $n\geq 2$.

\subsection{The
Assem--Bru\"stle--Charbonneau--Plamondon theorem}

A \emph{gentle} algebra is a string algebra where the ideal $I$ may be generated by zero--relations of lenght two (\textit{i.e.}\! involving just two arrows).
All Jacobian algebras of class $\cs[A_1]$
theories with only irregular punctures are gentle:
\vglue 12pt

\textbf{Theorem} \cite{ABCJP}. \textit{Let $(Q,\cw)$ be a quiver with superpotential corresponding to a Gaiotto $A_1$--theory with only irregular punctures and at least one such puncture. Then the Jacobian algebra $\mathscr{A}\equiv \C Q/(\partial \cw)$ is gentle.}
\vglue 12pt

Thus for such an `irregular' Gaiotto theory the BPS states are stable modules of a gentle algebra
which may be explicitly constructed in terms of strings/bands.
A string module corresponds to an open WKB trajectory \cite{GMN:2009} on the Gaiotto surface of the corresponding class $\cs[A_1]$ theory, see \cite{ABCJP,LF2} for the explicit construction of the string module out of the WKB geodesic and \textit{viceversa}. Likewise,
a family $M(C,\lambda,1)$ of representations a closed WKB geodesic associated to a primitive\footnote{\ By primitive homotopy class we mean a class which is not a non--trivial integral multiple of an integral class. Note also that the identification $C\leftrightarrow C^{-1}$ corresponds to the fact that the WKB geodesics have no orientation due to the $\Z_2$--monodromy of the differential $\sqrt{\phi_2}$.} homotopy class of  closed WKB gedesics on $\Sigma$. Hence the theorem of ref.\!\cite{ABCJP} just says that the WKB analysis
of the BPS spectra is exact in the present context. 

\subsection{Gentling class $\cs[A_1]$ algebras}\label{ap:gentling}

The
Assem--Bru\"stle--Charbonneau--Plamondon theorem may sound a little odd to a physicist, since WKB is exact for all class $\cs[A_1]$
theory whether or not the Gaiotto surface $\Sigma_G$ has regular punctures. Therefore,
\textit{for all class $\cs[A_1]$ category, it should be possible to construct all relevant modules in terms of strings/bands.}

We consider a general $\cs[A_1]$ theory based on a ultraviolet curve $\Sigma$ of genus $g$ with $p$ regular punctures (double poles of the quadratic differential $\phi_2$) and a number $b$ of
 irregular punctures (poles of order $c_i\geq 3$) with $p+b>0$. 
For $p\not=0$ the algebra is not gentle. However, to study the stable representations of the corresponding algebras $\C Q/(\partial\cw)$ 
we can effectively replace the algebra by a gentle one. We proceed as follows.
Each puncture carries a flavor $SU(2)$ group, so that our theory has a flavor symmetry $SU(2)^p$. We couple $p$ `spectator' copies of  $SU(2)$ SYM, 
each copy weakly gauging the flavor $SU(2)$ of one puncture. 
In the limit of very small spectator YM coupling, $g_\mathrm{spect}\rightarrow 0$, we get a BPS spectrum consisting of the BPS spectrum of the original theory, 
together with $p$ spectator $W$ bosons, as well as particles having $O(1/g_\mathrm{spect}^2)$ masses which are magnetically charged under the spectator gauge groups.

On the other hand, the weakly gauged model corresponds to the $\cs[A_1]$ theory based on the same surface $\Sigma$ with the $p$ ordinary punctures replaced by \emph{cubic} irregular ones. The weakly coupled model, having no more regular punctures, corresponds to a gentle algebra. We may compute the BPS spectrum of the gauged model using the simple Representation Theory of this gentle algebra and then recover the BPS spectrum of the original theory by taking $g_\mathrm{spec}\to 0$, that is, by projecting out states of non--zero spectator magnetic charge as well as the spectator $W$ bosons. We shall refer to this procedure
as gentling.

%
%\begin{figure}
%\centering
%\includegraphics[width=0.35\textwidth, height=0.21\textwidth]{gentling.pdf}  
%\caption{An instance of the Gaiotto glueing procedure: weakly gauging the flavor of a puncture with a spectator $SU(2)$ gauge field 
%is equivalent to the above surgery of Riemann surfaces \cite{Gaiotto, CV11}. The grey shaded region correspond to a boundary with a marked point.}
% \label{gentlingfig}
%\end{figure}
%

\paragraph{The original \textit{vs.}\!  the gentled algebra.}

\begin{figure}
\centering
\includegraphics[width=0.35\textwidth]{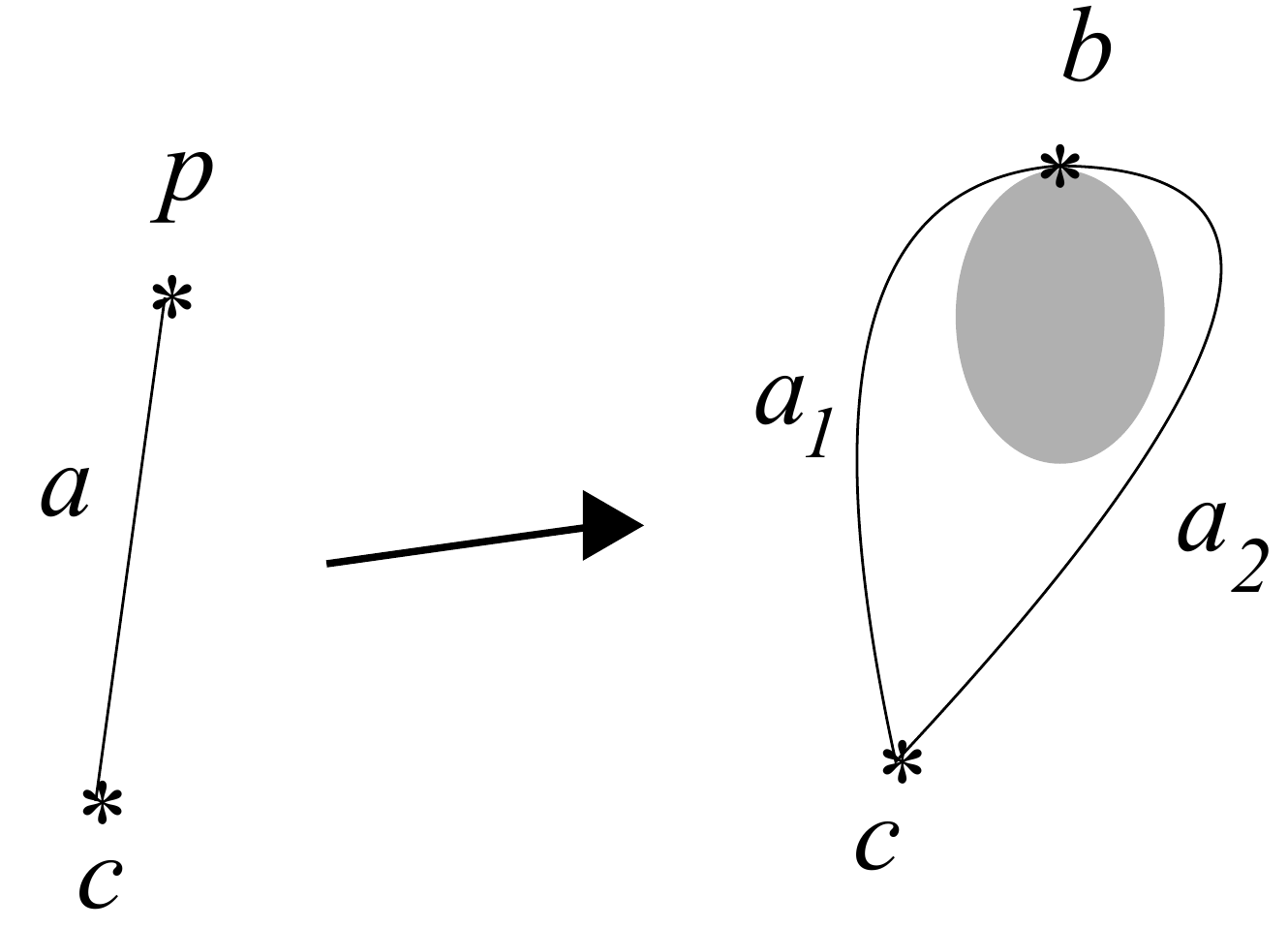}  
\caption{Gauging the $SU(2)$ symmetry associated to a puncture $p$ replaces one arc $a$ of the triangulation
 with a pair of arcs $a1, a2$ with the same starting point ($c$) forming a triangle with the boundary (grey).}
 \label{gentlingfig0}
\end{figure}

We wish to describe more in detail the effect of gentling on the Jacobian algebra $\C Q/(\partial \cw)$. 
To fully specify the Jacobian algebra, we fix one quiver $Q$ in the mutation class of the original theory,
 corresponding to a particular ideal triangulation of its Gaiotto surface $\Sigma$.
From the point of view of $\Sigma$, the gentling procedure may be seen as a \emph{local} operation at each 
of the $p$ punctures\footnote{ For simplicity, here we assume the puncture is not the internal vertex of a self--folded
 triangle \cite{triangulation1}. This is not a restriction: all surfaces have ideal triangulations without self--folded triangles.}:
 we cut a small disk near the puncture in such a way that the puncture becomes a marked point on the disk boundary (figure \ref{gentlingfig0}).  One can always choose a given ideal triangulation of the surface so obtained that matches the previous one up to the doubling one of the ideal arcs originating from the puncture, $a$, in two copies, $a_1,a_2$, 
which form the two sides of a triangle whose third side is the disk boundary (see figure \ref{gentlingfig0}).
 At the level of the quiver, this replaces the node corresponding to the ideal arc  $a$ with two nodes associated 
to the two copies of that arc, $a_1$, $a_2$, forming a Type $I$ block  $a_1\xrightarrow{\ \alpha\ } a_2$ \cite{triangulation1}. 

The procedure being local at each puncture, the global geometry of $\Sigma$ is irrelevant,
 and to analyze the effect of gentling on the Representation Theory of the corresponding BPS quiver, we may replace the surface $\Sigma$ with a small disk around the given puncture $p_0$. Let $r$ be the number of segments of ideal arcs incident in $p_0$; locally in the small disk around $p_0$ the geometry is the same as in the `radial' ideal triangulation of the punctured disk with $r$ marked points on its boundary whose quiver is the \emph{cyclic} orientation of the affine Dynkin graph $A^{(1)}_{r-1}$, which belongs to the mutation class of the $D_r$ AD model. Thus all gentling operation is modeled on the gauging of the $SU(2)$ flavor symmetry of a $D$--type
AD model in a region of its parameter space covered by the cyclic form of its quiver.  Gentling just replaces\footnote{ The discussion
in the text
is valid under the assumption that the arc $a$ does not start and end in the same puncture $p_0$. When it does,
the segment $a_1\xrightarrow{\ \alpha\ }a_2$ appears twice in the covering affine quiver which is then of the form $\widehat{A}(r,2)$.}
 the cyclic orientation of $A^{(1)}_{r-1}$ by the acyclic affine quiver
$\widehat{A}(r,1)$
\begin{equation}\label{gensubquiv}
\begin{gathered}
\xymatrix{& \bullet\ar[dl] &a \ar[l] &\\
\bullet\ar@{.}[d] &&&\bullet\ar[ul]\\
\bullet\ar[dr] &&&\bullet\ar[u]\\
& \bullet\ar[r] &\bullet\ar[ur]}
\end{gathered}\quad\xrightarrow{\ \ \text{gentling}\ \ }\quad
\begin{gathered}
\xymatrix{& \bullet\ar[dl] &a_1 \ar[l]\ar[r]^\alpha & a_2\\
\bullet\ar@{.}[d] &&&\bullet\ar[u]\\
\bullet\ar[dr] &&&\bullet\ar[u]\\
& \bullet\ar[r] &\bullet\ar[ur]}
\end{gathered}
\end{equation}
We stress that this does not necessarily mean that the original quiver had a cyclic $A^{(1)}_{r-1}$ subquiver which is replaced with a $\widehat{A}(r,1)$ subquiver by gentling;
indeed two segments incident in $p$ may correspond to the two ends of the \emph{same} arc, and some nodes and arrows in eqn.\eqref{gensubquiv} should be identified in pair. The above description is valid for a suitable \emph{cover} of the quiver. Indeed, the representation corresponding to the spectator $W$ boson for the gentling of the puncture $p_0$ is precisely the band module of the gentled algebra given by the RHS of figure \eqref{gensubquiv}. In what follows, by a slight abuse of notation $W_\mathrm{spec}$ denotes either the spectator $W$ boson, or the corresponding representation of the gentled algebra, or the corresponding band, according to the context.

\paragraph{Properties of the representations.} If $X$ is a representation of the gentled quiver, the spectator magnetic charge of the corresponding particle is \cite{cattoy} 
\begin{equation}m(X)_\mathrm{spec}\equiv \frac{1}{2} \langle \dim X , \dim W_\mathrm{spec} \rangle_D = \dim X_{a_1}-\dim X_{a_2}.\end{equation}
Let $X$ be an indecomposable representation of
the gentled algebra and $X|_p$ be its restriction
to the $\widehat{A}(r_p,1)$ cover subquiver associated to the gentled $p$ puncture (cfr. \eqref{gensubquiv}).
One has $X|_p= \oplus_s Y_s$ where $Y_s$ are indecomposables of the Euclidean algebra $\C \widehat{A}(r_p,1)$. The following statement is a standard property of the \emph{light category}
\cite{cattoy}
\medskip

\noindent\textbf{Lemma.} \textit{Suppose $X$ has zero spectator magnetic charge, $m(X)_\mathrm{spec}=\dim X_{a_1}-\dim X_{a_2}=0$ and that it is stable for parametrically small spectator coupling $g^2_\mathrm{spec}$. Then $m(Y_s)_\mathrm{spec}\equiv \dim (Y_s)_{a_1}-\dim (Y_s)_{a_2}=0$, 
for all $s$.}
\medskip

From the properties of the light category \cite{cattoy} we also know that, for 
a representation $X$ of zero spectator magnetic charge which is stable as $g^2_\mathrm{spec}\to0$,
 the extra arrow $\alpha$ is either an isomorphism or 
$\dim X$ is $n$ times the charge vector of $W_\mathrm{spec}$ (equal to the image under the cover map of the minimal imaginary root of $\widehat{A}(r,1)$). In the second case, only $n=1$ gives a brick, which belongs to the family of the spectator $W$.  
For all other such bricks $X$ of the gentled algebra, $\alpha$ is an isomorphism, and $X$ is canonically identified with a representation of the original quiver. 

\section{Proof of Fact 4}\label{proofpi}

In this appendix we prove \textbf{Fact 4}.

\subsection{The associated topological cover}

Assume we have a Galois
cover $\mathscr{C}\to \mathscr{J}$ where 
$\mathscr{C}$ and $\mathscr{J}$ are the $\C$--categories of two
class $\cs[A_1]$ QFTs.
Then $\mathscr{C}=\C Q^\prime/(\partial\cw^\prime)$ for a connected triangulation quiver $Q^\prime$
and corresponding superpotential $\cw^\prime$.
Let us first look at the conditions for $Q^\prime$ to cover the quiver $Q$ of $\mathscr{J}$; later we 
shall check under which conditions the covering map $Q^\prime\to Q$ is a covering map of
the full Jacobian categories.

For simplicity, we assume that $Q^\prime$ has a unique block decomposition.\footnote{ Again, this is not a substantial limitation since the
remark in footnote \ref{fool} applies.} The blocks of given Type $b$
then form orbits under $\mathbb{G}$.
We form a collection $\mathfrak{C}$ of blocks by taking one
chosen representative for each orbit.  
Let $\{\circ_{b,i}\}$ be the set of white nodes of the representative
block $b\in\mathfrak{C}$. Node $\circ_{b,i}$ is glued in $Q^\prime$ to node $\circ_{gc,j}$
for some $c\in\mathfrak{C}$ and $g\in\mathbb{G}$.
Suppose $b=c$; 
since $b\neq gc$, one has $g\neq 1$.
Consider the subquiver of $Q_{b,g}^\prime\subset Q^\prime$
 obtained by gluing
the two same--Type blocks $b$ and $gb$
at the node $\circ_{b,i}\equiv \circ_{gb,j}$.
Restricted to $Q_{b,g}^\prime$, $g$ is an automorphism acting freely on nodes. By inspection of the 10 possible gluings of two same--Type blocks, such an automorphism
exists only in four cases:
\begin{itemize}
\item[\textit{i)}] both nodes of two Type I blocks are glued together with arrows in opposite directions to give a pair of opposite
arrows which correspond to two disconnected nodes. This is the quiver of the Argyres--Douglas theory of type $D_2$;
\item[\textit{ii)}] all nodes of two Type II blocks are glued together with \emph{opposite} orientations to form the
(non $2$--acyclic \cite{ACCERV1,ACCERV2}) quiver of the $T_2$ theory;
\item[\textit{iii)}] all nodes of two Type II blocks are glued together with the same orientation to form the
Markov quiver of the $SU(2)$ $\cn=2^*$ theory;
\item[\textit{iv)}]  all white nodes of two
Type IV blocks are glued with opposite orientations to form the $(\Z_2)^3$--symmetric form of the 
quiver of $SU(2)$ SQCD with $N_f=4$. 
\end{itemize}
In these 4 cases, all white nodes have being
already paired up so, since $Q^\prime$ is connected, $Q^\prime\equiv Q_{b,g}^\prime$.
We conclude that, for all quivers $Q^\prime$ (admitting a unique decomposition) which are not in the list \textit{i)}--
\textit{iv)}, we have $b\neq c$, that is, two blocks in the decomposition of $Q^\prime$ which share a
white node belong to \emph{distinct} $\mathbb{G}$--orbits.

Let $Q^\prime$ with a freely acting automorphism group $\mathbb{G}$, which has a unique decomposition and is not in the above list of four.
The quiver $Q=Q^\prime/\mathbb{G}$ is then constructed as follows. We have the collection
$\mathfrak{C}$ of chosen representative of $\mathbb{G}$ orbits of blocks, which we identity with the collection of blocks for $Q^\prime/\mathbb{G}$.
Two blocks $b,c\in\mathfrak{C}$ are glued together
by identifying the white nodes $\circ_{b,i}
\sim \circ_{c,j}$ iff there is a $g\in\mathbb{G}$
(necessarily unique) such that $\circ_{b,i}$
and $\circ_{gc,j}$ are identified in $Q^\prime$.
By the previous argument, the white nodes
which get identified belong to distinct blocks,
and hence form an admissible block decomposition of the $Q^\prime/\mathbb{G}$ quiver, which is then an ideal triangulation quiver of some
surface $\Sigma_\mathrm{quot}$ \emph{endowed with a preferred
block decomposition $\mathfrak{C}$.}

Given a block decomposition $\mathfrak{C}$ of a triangulation quiver $Q$, we construct 
a surface with punctures and marks on the boundaries $\Sigma_\mathfrak{C}$. This is done simply by taking for each block in $\mathfrak{C}$
the corresponding $n$--gon from table
\ref{puzzle} and gluing them together
by identifying in pairs their sides according to the rule given by
their correspondence with glued pairs of white nodes of the blocks in $\mathfrak{C}$. 

Hence starting with a surface $\Sigma^\prime$
and one of its triangulation quivers $Q^\prime$
(which is not one of the exceptions above)
having a freely acting automorphism group
$\mathbb{G}$, by the  procedure above we
uniquely construct a surface (with punctures and marks)
$\Sigma_\mathfrak{C}$. Moreover, the procedure gives us a covering map
\begin{equation}
\gamma\colon \Sigma^\prime\to \Sigma_\mathfrak{C},
\end{equation} 
given in the following way. 
For each block Type $\mathfrak{b}\in\mathfrak{B}$
let $S_\mathfrak{b}$ be the $n$--gon in table
\ref{puzzle}.
We write $S_\mathfrak{b}(b)$ for the $n$--gon 
associated to the 
 block
$b$ in the decomposition of $Q^\prime$.
From each chosen block representative $b\in\mathfrak{C}$ and $g\in\mathbb{G}$ 
we have a homeomorphism 
\begin{equation}
\gamma_{b,g}\colon S_\mathfrak{b}(gb)\to S_\mathfrak{b}(b),
\end{equation}
which sends the segments of the boundary (and the arcs of the internal triangulation)
into corresponding segments (and arcs).
The family of maps $\{\gamma_{b,g}\;|\;b\in\mathfrak{C},g\in\mathbb{G}\}$ glue into
a continuous map 
\begin{equation}
\gamma\colon \Sigma^\prime\to \Sigma_\mathfrak{C},
\end{equation}
which is a covering map which may be
branched only at the base points of the
ideal triangulation (i.e.\! punctures and marks)
which are not the inner puncture of a self--folded triangle. This is the map to be constructed.

A covering map of (punctured and marked)
surfaces $\gamma\colon
\Sigma^\prime\to \Sigma$,
branched only at punctures, defines
a quotient map between the corresponding
quivers. For instance, the $\Z_{pq}$--symmetric quiver of the punctured disk
with $pq$ marks on the boundary, which is an oriented $pq$ cycle, is a $\Z_p$--cover
(branched over the puncture) of the 
punctured disk with $q$ marks. However, this 
covering of quivers do not extend to
a cover of Jacobian categories, since 
the superpotential on the quotient quiver is not the quotient superpotential. This is the general case;
the term in the superpotential \cite{LF1}
associated to a puncture (not inside a self--folded triangle) has degree equal to the number $s$ of ideal arcs incident in the puncture; under a cover branched on this puncture with local branch number $m$, this number reduced from $s$ to $s/m$, and agreement is possible only if $m=1$ i.e.\! there is no branching.
\medskip

The conclusion is: \textit{(with the mentioned exceptions) the non--trivial Galois covers,
$F\colon \mathscr{C}\to\mathscr{J}$
between class $\cs[A_1]$ categories
$\mathscr{C}$, $\mathscr{J}$ are
in one--to--one to unbranched covers between the corresponding bordered surfaces 
$f\colon \Sigma(\mathscr{C})\to\Sigma(\mathscr{J})$ where the punctures (resp.\! marks) on
$\Sigma(\mathscr{C})$ are taken to be the pre--images of the punctures (marks) on $\Sigma(\mathscr{J})$.}

\paragraph{Gaiotto viewpoint.}
We replace the bordered surfaces
$\Sigma(\mathscr{C})$, $\Sigma(\mathscr{J})$
by the corresponding Gaiotto curves obtained by replacing the $i$--th
boundary component with $\ell_i\geq 1$ marks
with an irregular puncture at which the 
quadratic differential has a higher pole of the form
\begin{equation}
\phi_2\Big|_{z\sim z_i}\sim \mathrm{const.}\;(z-z_i)^{-\ell_i}\, \left(\frac{d(z-z_i)}{z-z_i}\right)^{\!\!2}.
\end{equation} 
Locally around a branch point of order
$m$, the cover looks like $z-z_i= w^m+\cdots$,
so
\begin{equation}
w^*\phi_2=\mathrm{const.}^\prime\;
w^{-m\ell_i} \left(\frac{dw}{w}\right)^{\!\!2},
\end{equation}
which correctly corresponds to a boundary component with a number of marks which is the original one, $\ell_i$ times the local degree of the covering map, $m$.

\subsection{Exceptional cases}
There are two sources of exceptions:
\begin{itemize}
\item[A)] the list of four special $\Sigma^\prime$
\textit{i)}--\textit{iv)}. The quotient quivers of \textit{iii)} and \textit{iv)} are not $2$--acyclic
(this does \emph{not} mean they are not useful covers). The quiver of \textit{ii)} is not
$2$--acyclic from the start. It remains
\textit{i)}: the $D_2$ Argyres--Douglas model
(a doublet of free hypermutliplet) double covers
the $A_1$ Argyres--Douglas model (i.e.\! a free hyper). Of course, the statement that the BPS spectrum of two copies of the free hyper makes
two copies of the BPS spectrum of one free hyper, it by no means a surprise. 
\item[B)] The quiver $Q^\prime$ has more than one block decomposition. Suppose we have a Galois cover of triangulation quivers $Q^\prime\to Q^\prime/\mathbb{G}$.
Fix a particular block decomposition of the quotient
quiver; the pull back of the this block decomposition is a particular block decomposition of $Q^\prime$ with the property that $\mathbb{G}$ acts by permutation of same--Type blocks.
This is enough to make the previous argument work. It is possible that $Q^\prime/\mathbb{G}$
is the triangulation quiver of several topologically distinct surfaces; fixing a block decomposition implicitly chooses a topological surface in this set, and we describe the surface cover as a cover of this particular surface; thus, in this case, we may describe in many ways as covering of surfaces
the same covering of class $\cs[A_1]$ categories.
\end{itemize}

\section{$g=1$: no $R_1^r$ summands at 
$\Z_p$--symmetric points}\label{nor1}

Here we show that no $R_1^r$ appears in a module of the Jacobian algebra $\cj_p^g$ which is stable in our pulled back chamber and is not a spectator $W$. 

\subsection{Strings and bands}
The model with the flavor symmetry fully gauged has a gentle Jacobian algebra, whose indecomposable modules may be constructed using strings and bands (see \cite{cattoy} for a review). Hence the
relevant modules of $\cj^g_p$ may be described in the same language. 
We write $\overline{Q}$ for the (two--side) ideal in 
$\cj_p^g\equiv \C Q_p/(\partial\cw_\mathrm{cubic})$ which is the image of the arrow ideal of $\C Q_p$. We say that a $\cj_p^g$--module $M$ has order $n\in\mathbb{N}$ iff
\begin{equation}
\overline{Q}^{\,n}\cdot M\neq 0,\qquad\quad \overline{Q}^{\,n+1}\cdot M=0.
\end{equation}

\noindent\textbf{Lemma.} \textit{Let $X\in\mathsf{mod}\,\cj_p^g$ be a brick which is \underline{not} a spectator $W$. Then its order is at most 5.
Then $X$ is canonically identified with 
a brick of the \emph{string} algebra $\cj_p^g/\overline{Q}^{\,6}$.}
\medskip

\noindent\textsc{Proof.}
For $a=1,2,\dots, p$ consider the families $\Lambda^{(a)}=\{\lambda^{(a)}_{\circ_b}, \lambda^{(a)}_{\bullet_b},\lambda^{(a)}_{\ast_b}\}_{b=1}^p$ of linear maps defined by
\begin{equation}\label{eeeq}
\begin{gathered}
\lambda^{(a)}_{\circ_a}=\eta_a\xi^*_{a+1}B_{a+1}\xi_{a+1}\eta_a^*A_a,\quad \lambda^{(a)}_{\circ_{a+1}}=
\xi_{a+1}\eta_a^*A_a\eta_a\xi^*_{a+1}B_{a+1},\quad
\lambda^{(a)}_{\circ_b}=0\ \text{for }b\neq a, a+1\\
\lambda^{(a)}_{\bullet_a}=A_a\eta_a\xi^*_{a+1}B_{a+1}\xi_{a+1}\eta_a^*,\quad \lambda^{(a)}_{\bullet_{a+1}}=
B_{a+1}\xi_{a+1}\eta_a^*A_a\eta_a\xi^*_{a+1},\quad
\lambda^{(a)}_{\bullet_b}=0\ \text{for }b\neq a, a+1\\
\lambda^{(a)}_{\ast_a}=
\eta_a^*A_a\eta_a\xi^*_{a+1}B_{a+1}\xi_{a+1}+
\xi_{a+1}^*B_{a+1}\xi_{a+1}\eta_a^*A_a\eta_a,\hskip0.92cm 
\lambda^{(a)}_{\ast_b}=0\ \text{for }b\neq a;
\end{gathered}
\end{equation}
one checks that $\Lambda^{(a)}\in \mathrm{End}_{\mathsf{mod}\,\cj_p^g}X$. So, if $X$ is a ${\mathsf{mod}\,\cj_p^g}$--brick, $(\Lambda^{(a)})_i=\lambda^{(a)}\in\C$ for all $a$ and all nodes $i$
in the support of $X$. We lift $X$ to
a $\mathsf{mod}\,\cj_p^\mathrm{fg}$--module by reinserting the
isomorphisms $K_b$ at all nodes $\ast_b\in\mathrm{supp}\,X$. Since $\mathsf{mod}\,\cj_p^\mathrm{fg}$
is gentle the brick $X$ may be explicitly constructed as the string module of a string $C$ or as a band module of the form
$M(C,1,\lambda)$. To produce a non zero $(\Lambda^{(a)})_i$ the string/band module should
contain a substring given, up to cyclic reordering,
by the sequence of arrows (direct or inverse) $C_a$ making the
band of the $a$--th spectator boson. Essentially by definition, the resulting $(\Lambda^{(a)})_i=\lambda^{(a)}\in \C^\times$
if and only it \emph{is} the band module $M(C_a,1,\lambda^{(a)})$, which corresponds to the $a$--th spectator $W$ boson. All other bricks of $\mathsf{mod}\,\cj_p^\mathrm{fg}$, such that $K_a$ are isomorphisms,
must have $(\Lambda^{(a)})_i= 0$ for all $a$,
which implies that the corresponding
strings/bands do not contain
any of the $C_a$ as a substring. Identifying such bricks of $\mathsf{mod}\,\cj_p^\mathrm{fg}$ with bricks of 
$\mathsf{mod}\,\cj_p^g$, we conclude that 
 all products of six arrows appearing in
eqn.\eqref{eeeq} (including the ones in $\lambda^{(a)}_{\star_a}$) vanish. By diagram chasing 
one checks that these six arrows products generate the ideal $\overline{Q}^{\,6}$. The last statement follows from the fact
that $\cj_p^g$ failed to be a string algebra \cite{stringmod}
only because it was not finitely dimensional. \hfill $\square$

\subsection{No $R_1^r$}

We draw only the relevant piece of the quiver:
\begin{gather}\label{yyyyym}
\begin{gathered}
  \xymatrix{&\circ_{a-1}\ar@<0.4ex>[dd]
  \ar@<-0.4ex>[dd]_{A_{a-1}} && \circ_a\ar@<0.4ex>[dd]^{B_a}
  \ar@<-0.4ex>[dd]\\
\ast_{a-2}\ar[ur]^{\xi_{a-1}}&&
\ast_{a-1}\ar[ur]^{\xi_a} \ar[ul]_{\eta_{a-1}}&& \ast_{a}\ar[ul]\\
&\bullet_{a-1}\ar[ur]_{\eta^*_{a-1}}\ar[ul]^{\xi^*_{a-1}} && \bullet_{a}\ar[ul]^{\xi^*_a}\ar[ur]}
 \end{gathered}
\end{gather}
We write $R_1^{(a)}$ for a $R_1$ summand of the restriction of $X$ to the $a$--th Kronecker subquiver $\mathsf{Kr}_a$.
If $C$ is the string (band) corresponding to $X$,
each $R_1^{(a)}$ is represented by a maximal\footnote{ Here and below \emph{maximality} is meant with respect to inclusion in the $a$--th Kronecker $Kr_{a-1}$, i.e. any strictly larger segment of $C$ contains arrows different from $\xleftarrow{A_{a}}$ and 
$\xrightarrow{B_{a}}$.} segment of $C$ in $\mathsf{Kr}_a$ consisting 
of a single arrow (direct or inverse).
 We focus on the case in which this arrow is  the 
\emph{second} arrow of the $a$--th Kronecker, $B_a$, the other case being related by an automorphism of the quiver. 
In the $R_1^r$ case $R_1^{(a)}$ appears in $C$ inside the substring $\cdots\ \xleftarrow{\xi^*_a}\ 
\xleftarrow{B_a}\ \xleftarrow{\xi_a}\ \cdots$. 
Note that if $\xleftarrow{\xi_a}$ is the last arrow in the string, or if it is followed by $\xrightarrow{\eta_{a-1}}$, we have a destabilizing quotient with
support at the node $\ast_{a-1}$. Hence $C$ should contain the substring 
\begin{equation}\label{SS3}
\cdots\ \xleftarrow{\xi^*_a}\ 
\xleftarrow{B_a}\ \xleftarrow{\xi_a}\ 
\xleftarrow{\eta^*_{a-1}}\cdots.\end{equation}

Consider the part of the string which precedes
$\xleftarrow{\xi^*_a}$. We have three possibilities:
$\xleftarrow{\xi^*_a}$ is the first arrow of the string; or it is preceded by $\cdots\;\xrightarrow{\eta^*_{a-1}}$; or it is preceded by $\cdots\;\xleftarrow{\eta_{a-1}}$.
In the last case, $\xleftarrow{\eta_{a-1}}$ cannot be the first arrow, nor cannot be preceded by
$\xrightarrow{\xi_{a-1}}$, since in both cases we have a destabilizing submodule with support at
$\circ_{a-1}$. Thus in the third case we must have
\begin{equation}
\cdots\ \xleftarrow{A_{a-1}}\  \xleftarrow{\eta_{a-1}}\ \xleftarrow{\xi^*_a}\ 
\xleftarrow{B_a}\ \xleftarrow{\xi_a}\ 
\xleftarrow{\eta^*_{a-1}}\cdots
\end{equation}
But any string/band module $X$ which contains this substring has $\overline{Q}^6\cdot X\neq 0$, and hence cannot be a non--spectator brick.
We remain with the first two cases, where we have a \emph{subrepresentation} $Y$ of $X$
\begin{equation}
\cdots\ \overbrace{\xleftarrow{\xi^*_a}\ 
\xleftarrow{B_a}}^{Y\ \text{subrepr.}}\ \xleftarrow{\xi_a}\ 
\xleftarrow{\eta^*_{a-1}}\cdots
\end{equation}
with central charge $Z_\circ+Z_\bullet+Z_\ast$.

The arrow $\xleftarrow{\eta^*_{a-1}}$
is followed by a maximal string segment  in $\mathsf{Kr}_{a-1}$ which corresponds to either a $R_1^{(a-1), r}$, or a $R_1^{(a-1),p}$, or a $P_n^{(a)}$ direct
summand of $X|_{a-1}$. In the first case
$\overline{Q}^6\cdot X\neq 0$ and hence $X$ is not
a non--spectator brick.
In the second and third cases we have, respectively, the substrings
\begin{align}
&\cdots\ \overbrace{\xleftarrow{\xi^*_a}\ 
\xleftarrow{B_a}}^{Y\ \text{subrepr.}}\ \xleftarrow{\xi_a}\ 
\overbrace{\xleftarrow{\eta^*_{a-1}}\ \xleftarrow{A_{a-1}}}^{V\ \text{quotient}}
\\
&\cdots\ \overbrace{\xleftarrow{\xi^*_a}\ 
\xleftarrow{B_a}}^{Y\ \text{subrepr.}}\ \xleftarrow{\xi_a}\ 
\overbrace{\xleftarrow{\eta^*_{a-1}}\ \big(\xleftarrow{A_{a-1}}\ \xrightarrow{B_{a-1}}\big)^n}^{U_n\ \text{quotient}}\ \cdots\qquad n=0,1,2,\dots
\end{align}
which contain, respectively, a quotient of $X$
of the form $V$ or, respectively, $U_n$ whose central charges are
\begin{equation}
Z(V)=Z_\circ+Z_\bullet+Z_\ast,\qquad Z(U_n)=
n Z_\circ+(n+1)Z_\bullet+Z_\ast.
\end{equation}
Since $Z(Y)=Z(V)$ a module of the first form is never stable, while the stability of
a module of the second form
requires (in particular)
\begin{equation}
\arg Z(U_n)\equiv \arg\!\big(n Z_\circ+(n+1)Z_\bullet+Z_\ast\big)> \arg(Z_\circ+Z_\bullet+Z_\ast)\equiv Z(Y)
\end{equation}
i.e.
\begin{equation}
\arg\!\big(n (Z_\circ+Z_\bullet)+Z_\bullet\big)> \arg(Z_\circ+Z_\bullet), \qquad n\geq 0
\end{equation}
which is impossible since $\arg Z_\bullet < \arg Z_\circ$.


\newpage

\begin{thebibliography}{83}


\bibitem{KS1} M.~Kontsevich and Y.~Soibelman, ``Stability structures, motivic Donaldson-Thomas invariants and cluster transformations'', \arXiv{0811.2435}

\bibitem{GMN:2008} 
D. Gaiotto, G. W. Moore, A. Neitzke, 
``Four-dimensional wall-crossing via three-dimensional field theory'', 
Commun. Math. Phys. {\bf 299} (2010) 163-224, \arXiv{0807.4723}

\bibitem{GMN:2009}
D.~Gaiotto, G.~W.~Moore, and A.~Neitzke, ``Wall--crossing, Hitchin Systems, and the WKB Approximation'', \arXiv{0907.3987}

\bibitem{Gaiotto}
D.~Gaiotto, ``{N=2 dualities}'',  JHEP {\bf 1208} (2012) 034 \arXiv{0904.2715}

\bibitem{CNV} S.~Cecotti, A.~Neitzke and C.~Vafa, ``$R$--Twisting and $4d/2d$ correspondences'',  \arXiv{1006.3435}

\bibitem{ttstar} S.~Cecotti and C.~Vafa, ``Topological--anti--topological fusion'', 
Nucl. Phys. {\bf B367} (1991) 359-461.

\bibitem{CV92} S.~Cecotti and C.~Vafa, 
``On classification of {$\N=2$} supersymmetric theories'', 
Commun. Math. Phys. {\bf 158} (1993) 569-644, 
\arXiv{hep-th/9211097}

\bibitem{CV11} S. Cecotti and C. Vafa, ``Classification of complete $\cn=2$ supersymmetric theories in $4$ dimensions'', Surveys in differential geometry, vol {\bf 18} (2013) \arXiv{1103.5832}

%\cite{Cecotti:1992qh}
\bibitem{Cecotti:1992qh} 
  S.~Cecotti, P.~Fendley, K.~A.~Intriligator and C.~Vafa,
  ``A New supersymmetric index,''
  Nucl.\ Phys.\ B {\bf 386}, 405 (1992)
  [hep-th/9204102].
  
  
 \bibitem{branes} 
   K. Hori, A. Iqbal and C. Vafa, 
  ``D-branes and mirror symmetry'', 
   hep-th/0005247.
 
%\cite{Gaiotto:2010be}
\bibitem{Gaiotto:2010be} 
  D.~Gaiotto, G.~W.~Moore and A.~Neitzke,
  ``Framed BPS States,''
  Adv.\ Theor.\ Math.\ Phys.\  {\bf 17}, 241 (2013)
  [arXiv:1006.0146 [hep-th]].
  
  %\cite{Gaiotto:2011tf}
\bibitem{Gaiotto:2011tf} 
  D.~Gaiotto, G.~W.~Moore and A.~Neitzke,
  ``Wall-Crossing in Coupled 2d-4d Systems,''
  JHEP {\bf 1212}, 082 (2012)
  [arXiv:1103.2598 [hep-th]].

%\cite{Gaiotto:2012rg}
\bibitem{Gaiotto:2012rg} 
  D.~Gaiotto, G.~W.~Moore and A.~Neitzke,
  ``Spectral networks,''
  Annales Henri Poincare {\bf 14}, 1643 (2013)
  [arXiv:1204.4824 [hep-th]].

%\cite{Gaiotto:2012db}
\bibitem{Gaiotto:2012db} 
  D.~Gaiotto, G.~W.~Moore and A.~Neitzke,
  ``Spectral Networks and Snakes,''
  Annales Henri Poincare {\bf 15}, 61 (2014)
  [arXiv:1209.0866 [hep-th]].
  
  
  \bibitem{infty} S. Cecotti, and M. Del Zotto, ``Infinitely many N=2 SCFT with ADE flavor symmetry'', JHEP \textbf{1301}(2013)191, \arXiv{1210.2886}

\bibitem{CDZG} 
S. Cecotti, M. Del Zotto, and S. Giacomelli, 
``More on the N=2 superconformal systems of type $D_p(G)$'', 
JHEP {\bf 1304} (2013)153, \arXiv{1303.3149}

\bibitem{gal1}
P. Gabriel, ``The universal cover of a
representation--finite algebra'',
In \emph{Representation of algebras,}
\textsc{Lectures Notes in Mathematics}
\textbf{903}, Springer (1981), pp 68--105.

\bibitem{gal2}
K. Bongartz and P. Gabriel,
``Covering spaces in Representation
Theory'', Inven. Math. \textbf{65} (1982)
331--378.


\bibitem{ACCERV1} M.~{Alim}, S.~{Cecotti}, C.~{Cordova}, S.~{Espahbodi}, A.~{Rastogi}, and
  C.~{Vafa}, ``{BPS Quivers and Spectra of Complete N=2 Quantum Field
  Theories},'' Commun. Math. Phys. {\bf 323} (2013) 1185-1227, \arXiv{1109.4941}

\bibitem{ACCERV2} M.~{Alim}, S.~{Cecotti}, C.~{Cordova}, S.~{Espahbodi}, A.~{Rastogi}, and C.~{Vafa}, ``$N=2$ Quantum Field Theories and their BPS Quivers'', 
Advances in Theor. Math. Phys. \textbf{18} (2014) 27--127, \arXiv{1112.3984}

\bibitem{cattoy} S.~Cecotti, ``Categorical tinkertoys for $N=2$ gauge theories'', Int. J. Mod. Phys. {\bf A28} (2013) 1330006 \arXiv{1203.6743}


%\cite{'tHooft:1979bh}
\bibitem{'tHooft:1979bh} 
  G.~'t Hooft,
  ``Naturalness, chiral symmetry, and spontaneous chiral symmetry breaking,''
  NATO Sci.\ Ser.\ B {\bf 59}, 135 (1980).
  %151 citations counted in INSPIRE as of 16 mar 2015


\bibitem{MN1}
J. Minahan and D. Nemeschansky,
``An N=2 Superconformal Fixed Point with E6
Global Symmetry,''
 Nucl. Phys. B 482 (1996) 142?152 [hep-th/9608047].
 
 \bibitem{MN2}
J. Minahan and D. Nemeschansky,
 ``Superconformal Fixed Points with En Global Symmetry,''
  Nucl. Phys. B 489 (1997) 24--46 [hep-th/96010076].
  
  \bibitem{wallcrossingtopstrings}
S. Cecotti and C. Vafa,
"{BPS wall crossing and topological strings}",
 (2009),
  \arXiv{0910.2615} [hep-th].
  
 \bibitem{guk1}
 T. Dimofte and S. Gukov,
 ``Refined, motivic, and quantum,''
  Lett. Math. Phys. \textbf{91}, 1 (2010) \texttt{[arXiv:0904.1420 [hep-th]].}
  
 \bibitem{guk2}
  T. Dimofte, S. Gukov and Y. Soibelman, 
  ``Quantum wall crossing in N=2 gauge theories,'' Lett. Math. Phys. \textbf{95}, 1 (2011) 
  \texttt{[arXiv:0912.1346 [hep-th]].}

\bibitem{Denef} F.~Denef, 
``Quantum quivers and Hall/Holes Halos'',  \arXiv{hep-th/0206072}


\bibitem{DZS} M. Del Zotto, and A. Sen, ``About the Absence of Exotics and the Coulomb Branch Formula'', \arXiv{1409.5442} [hep-th].

\bibitem{zele}
H.~Derksen, J.~Wyman, and A.~Zelevinsky,
``Quivers with potentials and their representations I: Mutations,''
Selecta Mathematica \textbf{14} (2008) 59--119.

\bibitem{cbrq}
W. Crawley--Boevey, ``Lectures on Representations of
Quivers,'' available on line at
{\tt http://www1.maths.leeds.ac.uk/pmtwc/quivlecs.pdf}.

\bibitem{ringel}
C.M.~Ringel,
\textit{Tame Algebras and Integral Quadratic Forms,}
\textsc{Lecture Notes in Mathematics} \textbf{1099}, Springer, Berlin, (1984).


\bibitem{bluebook2}
D.~Simson and A.~Skowro\'nski,
\textit{Elements of the Representation Theory of
Associative Algebras.} Volume 2: ``Tubes and Concealed Algebras of Euclidean Type'',
London Mathematical Society Student Text \textbf{71}, Cambridge University Press (2007). 

\bibitem{Kac1}  V. Kac, ``Infinite root systems, representations of graphs, and invariant theory'',
Invent. Math. \textbf{56} (1980) 57--92.

\bibitem{Kac2}  V. Kac, ``Infinite root systems, representations of graphs, and invariant theory. II'',
J. Algebra \textbf{78} (1982) 141--162.

\bibitem{Kac3} V. Kac, 
``Root systems, representations of quivers and invariant theory'', 
(Montecatini, 1982). Lecture Notes in Mathemathics \textbf{996} Springer (1983) pages 74--108.

\bibitem{dPlects} J. de la Pe\~na, ``Integral quadratic forms and the representation type of an algebra'', 2006 ICTP lectures. Available on line at
{\tt http://webusers.imj-prg.fr/ bernhard.keller/ictp2006/lecturenotes/delapena-all.pdf}

\bibitem{Arnold1} S. Cecotti and M. Del Zotto,
``On Arnold's 14 `exceptional' $\cn=2$ superconformal gauge theories,''
JHEP \textbf{1110} (2011) 099, \arXiv{1107.5747}

\bibitem{Arnold2}  M. Del Zotto,
``More Arnold's $\cn=2$ superconformal gauge theories,''
JHEP \textbf{1111} (2011) 115, \arXiv{1110.3826}

\bibitem{MN3} S. Cecotti, and M. Del Zotto, ``The BPS spectrum of the 4d N=2 SCFT's $H_1, H_2, D_4, E_6, E_7, E_8$'', JHEP {\bf 1306}(2013)75, \arXiv{1304.0614}

\bibitem{KacMoody2} V. Kac, 
\textit{Infinite dimensional Lie algebras,} 3rd edition, Cambridge University Press (1990).

\bibitem{Slodowy1} P. Slodowy, ``Beyond Kac--Moody algebras and inside'', Can. Math. Soc. Conf. Proc. 5(1986), 361--371, CMP 18:10.

\bibitem{Slodowy2} P. Slodowy, ``Kac--Moody algebras, assoziiert Gruppen und Verallgemeinerungen'', Habilitationsschrift, Universit\"at Bonn, 1984.

\bibitem{gim3} B. Allison, S. Azam, S. Berman, Y. Gao, A. Pianizola, ``Extended affine Lie Algebras and their root systems'', Mem. Amer. Math. Soc. {\bf 126} (1997)

\bibitem{gim4} E. Neher, ``Lectures on Extended Affine Lie algebras'', \arXiv{1003.2352} [math.RA]

\bibitem{toroidal}
R. Moody, S. Rao, and T. Yokonuma, 
``Toroidal Lie algebras and vertex representations'', Geom. Dedicata \textbf{35} (1990), no. 1-3, 283--307.

\bibitem{triangulation1}
S.~{Fomin}, M.~{Shapiro}, and D.~{Thurston}, ``{Cluster algebras and
  triangulated surfaces. Part I: Cluster complexes},'' Acta Math. \textbf{201} (2008), 83-146.
  
  \bibitem{LF1}
D. Labardini--Fragoso,
``Quivers with potentials associated to triangulated surfaces'',
{\tt arXiv:0803.1328 [math.RT]}.

\bibitem{LF2}
D. Labardini--Fragoso,
``Quivers with potentials associated to triangulated surfaces. Part II: Arc representations'',
{\tt arXiv:0909.4100 [math.RT]}.

\bibitem{GLBS} C. Geiss, D. Labardini-Fragoso, J. Schroer, ``The representation type of Jacobian algebras'', \arXiv{1308.0478}

\bibitem{gal3}
R. Martinez-Villa and J.A. de la Pe\~na,
``The universal cover of a quiver with relations'',J. Pure and Appl. Algebra \textbf{30} (1983) 277--292.

\bibitem{gal4}
P. Dowbor and A. Skrowro\'nski,
``Galois coverings of representation infinite
algebras'', Comment. Math. Helv. \textbf{62} (1987) 311--337.

\bibitem{gal5}
A. Skrowro\'nski,
``Periodicity in representation theory of algebras'',
ICTP Lecture Notes 2006 available on--line at
{\tt http://webusers.imj-prg.fr/\\
bernhard.keller/ictp2006/lecturenotes/skowronski.pdf}.

\bibitem{galsurvery}
J.A. de la Pe\~na,
``Group actions on algebras and module
categories'',
Rev. Un. Mat. Argen. \textbf{48}
(2007) 1--20.

\bibitem{stringmod}
M.C.R. Butler and C.M. Ringel, 
``Auslander--Reiten sequences with few middle terms and applications to string algebras'',
 Comm. in Algebra \textbf{15} (1987) 145--179.
 
\bibitem{ABCJP} I. Assem, T. Br\"ustle, G. Charbonneau--Jodoin, P.-G. Plamondon, 
``Gentle algebras arising from surface triangulations'', \arXiv{0903.3347}

\bibitem{assem1}
I. Assem, D.~Simson and A.~Skowro\'nski,
\textit{Elements of the Representation Theory of
Associative Algebras.} Volume 1: ``Techniques of Representation Theory'',
London Mathematical Society Student Text \textbf{65}, Cambridge University Press (2006). 

\bibitem{hurwitz}
A. Hurwitz,
``Ueber Riemann'sche Fl\"aachen mit gegebenen
Verzweigungspunkten'',
Math. Ann. \textbf{39} (1891) 1--60.

\bibitem{belyi}
G.V. Belyi,
``On Galois extensions of a maximal cyclotomic field,'' Math. USSR Izvestija \textbf{14} (1980)
247--265.


\bibitem{elliptic}
F. Diamond and J. Shurman,
\textit{A first course in modular forms,}
\textsc{Graduate Texts in Mathematics} \textbf{228},
Springer (2005).

\bibitem{farkas}
H.M. Farkas and I. Kra,
\textit{Theta constants, Riemann surfaces and the
modular group,} \textsc{Graduate Studies in Mathematics} \textbf{37}, AMS (2001).


\bibitem{coxeterz}
R.  Stekolshchik,
\textit{Notes on Coxeter transformations and the
McKay correspondence,}
\textsc{Springer Monographs in Mathematics,}
(2008).

\bibitem{itzk}
P. Beazley Cohen, C. Itzykson, J. Wolfart,
``Fuchsian triangle groups and Grothendieck dessins.
Variations on a theme of Belyi'',
Comm. Math. Phys. \textbf{163} 605--627 (1994).

\bibitem{groth}
A. Grothendieck,
``Esquisse d'un programme,'' pp. 5--48
in L. Schneps, P. Lochak (eds.)
\textit{Geometric Galois Actions. 1,}
\textsc{London Math. Lecture Note Series}
\textbf{242} Cambridge University Press
(1997).

\bibitem{wol}
J. Wolfart,
``ABC for polynomials, dessins d'enfants,
and uniformization --- a survey'',
in \textit{Elementare und analytische Zahlentheories},
Schr. Wiss. Ges. Johann Wolfgang Goethe
Univ. Frankfurt am Main \textbf{20},
Stuttgart (2006).

\bibitem{wol2}
J. Wolfart,
``Traingle groups and Jacobians of CM type'',
available on line at {\tt http://www.math.uni-frankfurt.de/wolfart}.

\bibitem{zvon}
A.K. Zvonkin,
``Belyi functions: examples, properties,
and applications'',
available on line at {\tt https://www.labri.fr/perso/zvonkin/Research/belyi.pdf}.

\bibitem{magot}
N. Magot and A.K. Zvonkin,
``Belyi functions for Archimedean solids'',
Discrete Math. \textbf{217} 249--271 (2000).

\bibitem{petronio}
E. Pervova and C. Petronio,
``On the existence of branched coverings between
surfaces with prescribed branch data. I,''
Alg. and Geom. Topology \textbf{6}, 1957--1985 (2006).

\bibitem{tack}
Y. Tachikawa and S. Terashima, 
``Seiberg--Witten geometries revisited,''
\texttt{arXiv:1108.2315 [hep-th].}

\bibitem{keller}
B. Keller,
``The periodicity conjecture for pairs of Dynkin diagrams,''
\texttt{arXiv:1001.1531 [math.RT].}

\bibitem{tack2}
Y. Tachikawa, 
``Six--dimensional DN theory and four--dimensional SO--USp quivers'',
 JHEP \textbf{0907}:067 (2009) 
 \texttt{[arXiv:0905.4074].}

\bibitem{gon1}
V.V. Fock and A.B. Goncharov,
``The quantum dilogarithm and 
representations of quantum cluster varieties,''
Inven. Math. \textbf{175} (2009) 223--286.


\bibitem{gon2}
V.V. Fock and A.B. Goncharov,
``Cluster ensembles, quantization and the dilogarithm,''
Ann. scient. de l'ENS \textbf{42}
(2009) 865--930.


\bibitem{kell2}
B. Keller,
``On cluster theory and quantum dilogarithm identities'',
\texttt{arXiv:1102.4148.}


\bibitem{fadd}
L.D. Faddeev and R.M. Kashaev,
``Quantum dilogarithm'',
Modern Phys. Lett. \textbf{A9}
(1994) 427--434.

\bibitem{ysyst}
S.~Cecotti and M.~Del Zotto,
  ``$Y$ systems, $Q$ systems, and 4D $\mathcal{N}=2$ supersymmetric QFT,''
  J.\ Phys.\ A {\bf 47}, no. 47, 474001 (2014)
  [arXiv:1403.7613 [hep-th]].

\bibitem{kell-lin}
B. Keller and S. Scherotzke,
``Linear recurrence relations for cluster
variables of affine quivers'',
\texttt{arXiv:1004.0613.}


\bibitem{frises}
I. Assem, C. Reutenauer, and D. Smith,
``Frises'',
\texttt{arXiv:0906.2026.}

\bibitem{stringfrieze}
I. Assem, G. Dupont, R. Schiffler,  and D. Smith,
``Friezes, strings and cluster variables'',
\texttt{arXiv:1009.3341.}

\bibitem{CCV}
  S.~Cecotti, C.~Cordova and C.~Vafa,
  ``Braids, Walls, and Mirrors,''
  arXiv:1110.2115 [hep-th].

\bibitem{fomzele}
S. Fomin and A. Zelevinsky,
``Cluster algebras IV: Coefficients'',
\texttt{arXiv:math/0602259 [math.RA].}

\bibitem{qsyst1}
R. Kedem,
``$Q$--systems as cluster algebras'',
J. Math. Phys. A: Math. Theor. \textbf{41}
(2008) 1940011, \texttt{arXiv:0712.2695 [math.RT].}

\bibitem{qsyst2}
P. Di Francesco and R. Kedem,
``$Q$--systems, heaps paths and cluster
positivity'',
Comm. Math. Phys. \textbf{293} (2009)
727820, \texttt{arXiv:0811.37027 [math.CO].}

%\bibitem{surprise}
%Y.-H. He and J. Read,
%``Dessins d'enfants in N=2 generalised quiver theories'',
%{\tt arXiv:1503.06418.} 


\end{thebibliography}
\end{document}